\documentclass[10pt,a4paper]{article}
\usepackage[utf8]{inputenc}
\usepackage[english]{babel}
\usepackage{amsmath}
\usepackage{physics}
\usepackage{amsfonts}
\usepackage{amssymb}
\usepackage{mathtools}
\usepackage{graphicx}
\usepackage{slashed}
\usepackage{circuitikz}
\usepackage{comment}
\usepackage[left=2cm,right=2cm,top=2cm,bottom=2cm]{geometry}
\usepackage{caption}
\usepackage{subcaption}
\usepackage{makecell}
\usepackage{bm}
\usepackage{tikz}

\def\ben{\begin{enumerate}}
	\def\een{\end{enumerate}}
\def\bit{\begin{itemize}}
	\def\eit{\end{itemize}}
\def\beq{\begin{equation}}
	\def\eeq{\end{equation}}
\def\bea{\begin{eqnarray}}
	\def\eea{\end{eqnarray}}
\def\bq{\begin{quote}}
	\def\eq{\end{quote}}
\def \lsim{\mathrel{\vcenter
		{\hbox{$<$}\nointerlineskip\hbox{$\sim$}}}}

\def\gappeq{\mathrel{\rlap {\raise.5ex\hbox{$>$}}
		{\lower.5ex\hbox{$\sim$}}}}
\def\lappeq{\mathrel{\rlap{\raise.5ex\hbox{$<$}}
		{\lower.5ex\hbox{$\sim$}}}}

\def\LNP{\Lambda_{NP}}
\def\te{\tau\to e}
\def\tm{\tau \to \mu}
\def\tl{\tau \to \ell}
\def\mue{\mu\to e}

\def\meg{\mu \to e \gamma}

\def\mec{\mu \! \to \! e~ {\rm conversion}}

\def\g{\gamma}

\usepackage[compat=1.1.0]{tikz-feynman}
\usepackage{cite}

\renewcommand{\tl}{ \tau \leftrightarrow l }
\renewcommand{\te}{ \tau \leftrightarrow e }
\renewcommand{\tm}{ \tau \leftrightarrow \mu }
\newcommand{\me}{ \mu \leftrightarrow e }
\renewcommand{\LNP}{ \Lambda_{\rm NP}}

\renewcommand*{\thefootnote}{\fnsymbol{footnote}}

\begin{document}

\def\sd{\color{olive}}
\def\ma{\color{blue}}

\begin{center}
	{\Large {\bf
            The sensitivity of  $\mu\to e$ processes to $\tau$ flavour change
			
	}}
	\vskip 20pt
	{\large Marco Ardu$^1$\footnote{E-mail address: marco.ardu@umontpellier.fr},
		Sacha Davidson$^1$\footnote{E-mail address:
			s.davidson@lupm.in2p3.fr}, and Martin Gorbahn$^2$\footnote{E-mail address: martin.gorbahn@liverpool.ac.uk} }
	
	\vskip 10pt  
	
	{\it $^1$LUPM, CNRS,
		Université Montpellier
		Place Eugene Bataillon, F-34095 Montpellier, Cedex 5, France
	}\\
        {\it $^2$Department of Mathematical Sciences,
          University of Liverpool, Liverpool L69 3BX, United Kingdom
    }\\
\end{center}
\begin{abstract}
\noindent
$~$\\Transforming a $\mu$ to a $\tau$,  then the $\tau$ to to an $e$,  results in $\mu \to e$.
In an EFT framework, we explore the sensitivity of $\mu\to e$ observables to products of $(\mu\to \tau)\times (\tau \to e)$ interactions, and show that   the exceptional sensitivity  of upcoming $\me$ experiments could allow to probe  parameter space  beyond the reach  of 
{ upcoming $\tl$ searches in  Higgs, $\tau$ and $B$ decays.}
We  describe the $\tl$  interactions as dimension six operators in the SM EFT, identify pairs of them giving interesting contributions to $\me$  processes,
and obtain the anomalous dimensions   mixing  those  pairs  into dimension eight $\mu\to e$ operators. 
{   We find  that $\mu \to e$ processes are  sensitive to
  $\tau$ flavour-changing  $B$ decays at rates
  comparable to current $B$ anomalies, but cannot reduce rates
  --- as appropriate in many current $B$ anomalies---
because they do not interfere with the SM.}

\end{abstract}

\renewcommand*{\thefootnote}{\arabic{footnote}}
\setcounter{footnote}{0}

\section{Introduction}

{ The three lepton flavours are accidentally conserved in the Standard Model, if it is defined with  massless neutrinos.} But the non-zero neutrino masses and mixing angles established by the observation of neutrino oscillations clearly demonstrate that leptons change flavour. Extending the Standard Model (SM) with Dirac neutrino  masses generically predicts flavour-changing contact interactions  among the charged leptons
(LFV or CLFV--- for reviews, see, eg,\cite{Kuno:1999jp,Calibbi:2017uvl}), but the branching ratios are GIM-suppressed by small neutrino masses $Br\sim G_F^2m_\nu^4\sim 10^{-50}$\cite{Hernandez-Tome:2018fbq,Blackstone:2019njl}, so beyond any foreseeable experimental reach. Searches for CLFV are thus of great interest, as an observation would be an unambiguous signature of New Physics (NP) { that could shed light on the neutrino mass mechanism. In addition, } null results generally limit the parameter space of Beyond the SM theories, many of which predict sizable LFV rates. In Table \ref{tab:LFVsearches} a subset of LFV processes is listed with the { current experimental bounds on their branching ratios, and the expected sensitivities of  upcoming searches.}

\begin{table}[th]
\begin{center}
\begin{tabular}{|l|l|l|}
\hline
Process & Current bound on BR   & Future Sensitivity    \\
\hline
$\mu\to e \gamma $ & $ < 4.2 \times 10^{-13}$ \cite{MEG:2016leq}
		   & $  { 10^{-14}}$ \cite{MEGII:2018kmf} \\
$\mu\to \bar{e}ee $ & $ < 1.0  \times 10^{-12}$ \cite{SINDRUM:1987nra}
			& $10^{-16}$ \cite{Blondel:2013ia} \\
$\mu A \to e A$ &  $< 7 \times 10^{-13}$ \cite{SINDRUMII:2006dvw}
			&$10^{-16}$ \cite{COMET:2009qeh}\\
\hline
$\tau\to l\gamma$ & $<3.3\times 10^{-8}$ \cite{tau1} & $ 3 \times 10^{-9}(e),  10^{-9}(\mu) $ \\
$\tau\to e\bar{e} e$ & $< 2.7\times 10^{-8}$ \cite{tau2}
			& $  5\times 10^{-9} $\cite{belle2t3l} \\
$\tau\to \mu\bar{\mu} \mu$ & $< 2.1\times 10^{-8}$ \cite{tau2}
			& $ 4\times 10^{-9} $\cite{belle2t3l} \\
$\tau\to \mu \bar{e} e ,e \bar{\mu} \mu$ & $< 1.8, 2.7\times 10^{-8}$ \cite{tau2}
& $ 3, 5 \times 10^{-9} $\cite{belle2t3l} \\
			...&...&...\\
			\hline
$\tau\to \ell\pi^0$ & $<8.0\times 10^{-8}$ \cite{Belle:2007cio} &  $ 4 \times 10^{-9} $ \cite{belle2t3l}\\
$\tau\to \ell\eta$ & $<6.5\times 10^{-8}$ \cite{Belle:2007cio} & $ 7 \times 10^{-9} $ \cite{belle2t3l} \\
 $\tau\to \ell\rho$ & $<1.2\times 10^{-8}$ \cite{Belle:2007cio} &  $  10^{-9}$ \cite{belle2t3l} \\
\hline
$h\to e^\pm\mu^\mp$ & $<6.1\times10^{-5}$ \cite{AtlasHiggs}& $2.1\times10^{-5}$\cite{ILC} \\	
$h\to e^\pm\tau^\mp$ & $<2.2\times 10^{-3}$ \cite{cmsHiggs}& $2.4\times10^{-4}$\cite{ILC} \\
$h\to \tau^\pm\mu^\mp$ & $<1.5\times 10^{-3}$ \cite{cmsHiggs} &  $2.3\times10^{-4}$\cite{ILC}\\
			\hline
		\end{tabular}
	\end{center} 
\caption{Some $\mu \leftrightarrow e$  and  $\tau \leftrightarrow l$ processes  ($l\in \{e,\mu\}$),  with the current experimental bound on the branching ratios.  The last column lists the future sensitivities used in our projections, which correspond to the expected reach  of upcoming or planned experiments (except for
  $\mu\to e \gamma$, where the MEGII experiment at PSI, which starts taking data in 2022, aims to reach $BR \sim 6\times 10^{-14}$). Additional $\tl$ processes involving $b$ quarks are listed in table \ref{tab:tetu}. \label{tab:LFVsearches}} 
\end{table}

{ The current limits on $\mu  \to e$  flavour change are  more retrictive than those on  $\tau\to l$, where $l \in \{e,\mu\}$, due to the possibility of making  intense muon beams. Furthermore, a  significant  gain  in sensitivity  is expected at upcoming   $\mu  \to e$  experiments (see table \ref{tab:LFVsearches}), sometimes  allowing: }
\begin{equation}
	Br(\mu \to e\dots) \lsim Br(\tau\to e\dots)Br(\tau\to \mu\dots)\label{eq:brsquaredtaubrmuon}
\end{equation}
This is  interesting, as the three { $\Delta F=1$ lepton flavour changes
 are related:
$$
\begin{array}{rcl}
&\tau&\\
~~\nearrow\!\!\!\!\!\!\!&\!\!& \!\!\!\!\!\!\searrow\\
e~&\longrightarrow &~\mu
\end{array}~~~
$$
If two lepton flavours are unconserved,
then no symmetry forbids the third to happen, so it could  be generated from the first two  at some order in the perturbative expansion.
Eq.~(\ref{eq:brsquaredtaubrmuon}) tells us that $\mu \to e$  searches are potentially sensitive to the product  of $\mu\to \tau$ and $\tau\to e$ interactions respecting  $\tau$ LFV   constraints.
{ So the aim of this manuscript, is to explore what can be learned about}
$\tau \leftrightarrow l$ interactions, using $\mu \to e$ observables.
We are interested  in the  model-independent aspects of this question,  so we  assume that
the NP responsable  for  LFV is heavy, and use
Effective Field Theory (EFT) \cite{Georgi:1993mps,Buras:1998raa,Manohar} to 
parametrise low energy LFV.

In   this EFT  approach, lepton flavour violation is mediated by  contact interactions among Standard Model particles, which correspond  in the Lagrangian to  higher dimensional operators respecting the appropriate gauge symmetries (Our EFT formalism is presented in more detail in section \ref{sect:techINTRO}).
We will suppose  a New Physics scale $\LNP \geq$ 4 TeV (``beyond the LHC''),
describe  $\tau\to l$ interactions via dimension six operators, and calculate the  log-enhanced }
contributions to dimension eight $\mu \to e$ operator coefficients,
which appear in their  Renormalization Group evolution between
$\LNP$ and $m_W$.
These contributions arise from  the insertion in  loop diagrams  of  both
a $\mu\to \tau$   and a  $\tau\to e$  operator, and 
can be reliably computed in  EFT --- although
they may not be the dominant contributions to $\mu \to e$ processes coming from $\tau \leftrightarrow l$ interactions (see section \ref{ssec:models}).
   We will find that   upcoming $\mu\leftrightarrow e$ searches could be sensitive to
    $\tau\leftrightarrow \ell$ interactions beyond the reach of upcoming
    $\tau$ experiments.

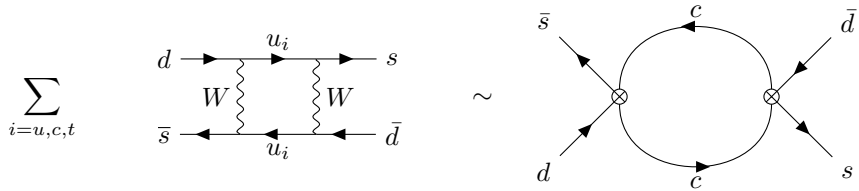
\begin{figure}[t]
	\centering
	$\displaystyle\sum_{i=u,c,t}$
\begin{subfigure}{.3\textwidth}
	\centering
 \begin{tikzpicture}
	\begin{feynman}[small]
		\diagram* [ horizontal=a to b] {
			i1 [particle=\(d\)]
			-- [fermion] a
			-- [fermion, edge label=\(u_i\)] b
			-- [fermion] f1 [particle=\(s\)],
			i2 [particle=\(\overline s\)]
			-- [anti fermion] c
			-- [anti fermion, edge label'=\(u_i\)] d
			-- [anti fermion] f2 [particle=\(\bar{d}\)],
			{ [same layer] a -- [photon, edge label'=\(W\)] c },
			{ [same layer] b -- [photon, edge label=\(W\)] d},
		};
	\end{feynman}
\end{tikzpicture}
\end{subfigure}$\sim$
\begin{subfigure}{.3\textwidth}
	\centering
\begin{tikzpicture}
	\begin{feynman}[small]
		\vertex (fin) at (-1,1) {\(\bar{s}\)};
		\vertex[style=crossed dot] (a)  at (0,0) {};
		\vertex[style=crossed dot] (b) at (2,0) {};
		\vertex (mu) at (-1,-1) {\(d\)};
		\vertex (fout) at (3,1) {\(\bar{d}\)};
		\vertex (e) at (3,-1) {\(s\)};
		\diagram* [inline=(a.base)]{
			(fin) -- [anti fermion] (a),
			(mu) -- [fermion] (a),
			(a) -- [anti fermion, half left, looseness=1.42, edge label=\(c\)] (b) -- [anti fermion, half left, looseness=1.42, edge label=\(c\)] (a),
			(b) -- [fermion] (e),
			(b) -- [anti fermion] (fout),
		};
	\end{feynman}
\end{tikzpicture}
\end{subfigure}
\caption{ The GIM mechanism in $K-\bar{K}$ mixing: the mass-independent dimension six contribution  from the box cancels in the flavour sum because of CKM unitarity.
  Then at ${\cal O}(G_F^2)$, the top contribution is not dominant due to small mixing with the down quark, whereas the dimension eight term $\sim G^2_Fm_c^2$ is relevant, and can be calculated  in the low-energy EFT (Fermi theory) as  the loop contribution  with  two dimension six operators inserted, as illustrated on the right.}\label{fig:GIMexample}
\end{figure}

The paper is organized as follows. In Section \ref{sect:techINTRO} we introduce the formalism  for  the  EFT calculation (notation and operators), and we make several estimates to focus the calculations  on contributions within future $\mu \to e$ experimental sensitivity.
 Our  results  are illustrated  in Section
 \ref{sec:Calculations}, where  the Renormalization Group Equations (RGEs) for dimension eight operators are reviewed,
 we discuss examples of anomalous dimensions calculated from double insertions of dimension six operators, and  give the weak scale matching   of $\mu\to\tau\times \tau\to e$ onto low energy $\mu\to e$ operators.
 The complete results for
 (dimension 6)$^2\to$ dimension 8 mixing can be found in appendix \ref{appendix:AnomalousDimensions}. 
In Section \ref{sec:Pheno} we discuss some phenomenological implications : $\mu\to e$ observables are sensitive to  products of $\tau\leftrightarrow l$ operator coefficients and we compare this sensitivity to  the limits
coming from  searches for $\tl$ processes.

\section{EFT, operators and notation}
\label{sect:techINTRO}

In this  section,  we start by comparing our calculation to the expectations of a few models in subsection \ref{ssec:models},  then   review the EFT framework  in sections \ref{ssec:EFT} to \ref{ssec:EoM}. Finally  in subsection \ref{ssec:estimates}, we  estimate  which $(\mu\to \tau)\times (\tau \to e)$ loop diagrams
could be accessible to future $\mu\to e$ experiments,  making them interesting to calculate.

\subsection{A few models}
\label{ssec:models}

In this subsection,  we  discuss two  models---one being the SM--- in order to
illustrate  the relationships between
$\tau\leftrightarrow l$ and  $\mu\leftrightarrow e$ observables, and
to compare our EFT calculation  with the  expectations of  UV complete models.

First, consider a model where   two  heavy bosons, $M\gg m_W$,  are added  to the SM,
with  flavour diagonal, and  respectively
$\tau \leftrightarrow \mu$ and $\tau \leftrightarrow e$  renormalizable  interactions.  A first source of $\mu \leftrightarrow e$ flavour change
could be additional  renormalizable $\me$ interactions of the heavy bosons  ---  not forbidden by symmetry --- but  these do not interest us, because their  magnitude  depends on the model and  is  independent of  the  $\tau \leftrightarrow l$ interactions.
We are interested in  $\mu\to e$ processes  which    occur  due to diagrams involving  both  the $\mu\to \tau$
  and $\tau \to e$ interactions.
  The part of these amplitudes which  is reproduced by    our EFT  calculation, can be  identified by   matching the model onto EFT at the heavy boson mass scale $M$.
  The model  generates   $\tau \leftrightarrow l$   four-fermion amplitudes at tree level, and  could induce  $\mu \leftrightarrow e$ amplitudes at one loop.  These all
  are expected to match onto  dimension six  operators in the EFT, with coefficients   of  ${\cal O}(\lambda_{\tau l}/M^2)$
  and ${\cal O}(\frac{\lambda^*_{\mu \tau} \lambda_{\tau e}}{16\pi^2 M^2})$.
Our EFT calculation cannot reproduce these model dependent coefficients \footnote{Despite that    
   the UV dependence is also apparent in the loop integrals performed in the  EFT, which  are power divergent \cite{QuadDivergINPREP}.}.
 Instead, the EFT below the heavy boson scale allows to combine the dimension six  $\tau \leftrightarrow e$ and  $\tau \leftrightarrow \mu$  operators into a
 dimension eight  $\mu \leftrightarrow e$ operator,  giving a contribution  to  the 
   $\mu \leftrightarrow e$ amplitude  
 $\lsim {\cal O}(\frac{\lambda^*_{\mu \tau} \lambda_{\tau e}v^2}{16\pi^2 M^4})$ 
 ($v$ is the  vacuum expectation value of the SM Higgs).  By power-counting, this is  subdominant   compared to the model-dependent matching contribution discussed above. So  this model illustrates that   
 $\tau \leftrightarrow e$ and  $\tau \leftrightarrow \mu$  interactions
 could  generically combine into
 larger $\mu\leftrightarrow e$ rates than  the EFT allows to  compute.

 As a  second example, consider $K-\bar{K}$ mixing in the SM, where the
 dominant contribution is computable in the   EFT (Fermi theory).
The  box diagram   in the full SM is illustrated  on the left in figure  \ref{fig:GIMexample};
evaluated with  only    massless  $u$ quarks  in the loop,
it gives an amplitude $\propto (V_{us}^*V_{ud})^2/(16\pi^2 m_W^2)$,
 where $V$ is the CKM matrix.
 This would  match at $m_W$  onto  a dimension six $\Delta F=2$ operator
 in the low-energy theory Fermi theory.
However, due to CKM unitarity,
this   ${\cal O} (\frac{1}{16\pi^2 m_W^2}$)  amplitude
vanishes when summing over all up-type quark flavours
and neglecting their masses. Instead, the amplitude
in the full SM has a GIM dependence on the quark masses 
$\propto (V_{cs}^*V_{cd})^2m_c^2/16\pi^2 m_W^4+(V_{ts}^*V_{td})^2m_t^2/16\pi^2 m_W^4$.
In matching this to the low-energy EFT, the  $m_t^2/16\pi^2 m_W^4$ piece
would match onto a dimension six operator, but is negligeable due to the small mixing between the third and first generation.
And the log-enhanced  part of the amplitude
$\propto m_c^2$  is reproduced in the EFT
by calculating  the diagram with two insertions of
dimension six operators, illustrated on the right of Figure \ref{fig:GIMexample}.  So in the Standard Model,  our calculation can sometimes  reproduce the observed flavour changing rates.

\subsection{EFT for LFV}
\label{ssec:EFT}

If  the
new particles with  lepton flavour changing
interactions are
heavy,
LFV at  lower energies  can be parametrised  via contact interactions,
which appear as 
non-renormalisable  operators in the Lagrangian of an EFT
(see {\it eg} \cite{Buras:1998raa,Manohar} for a review).
{ In this subsection, we sketch the EFT  background of  our calculation, and
introduce some notation.}

Above the weak scale, we use the Lagrangian of the SMEFT, in which  the SM Lagrangian is augmented by operators of higher dimension that respect the $SU(3)\times SU(2) \times U(1)$ gauge symmetry of the SM, and are constructed out of SM fields.  We are interested in LFV operators of dimension  6 or 8,  so we write
\begin{equation}\label{eq:SMEFTLag}
  \mathcal{L}_{\rm SMEFT}=\mathcal{L}_{\rm SM}
  +\left(\sum_{A,\zeta}\frac{C_A^{[6]\zeta} \mathcal{O}_A^{[6]\zeta}}{v^2}
  +\sum_{B,\xi}\frac{C_B^{[8]\xi}\mathcal{O}_B^{[8]\xi}}{v^4} +\rm{h.c}\right)
\end{equation}
where $v = 174$ GeV,  the   operator subscripts indicate the gauge structure and particle content,  and  the superscripts contain  the operator dimension in brackets [suppressed when unneccessary], additional information about the operator structure in parentheses (see  section \ref{sec:Operators} for examples), and the flavour indices.
{ The LFV  operators of interest here are listed in section \ref{sec:Operators}.
In the flavour sums  of  eq. (\ref{eq:SMEFTLag}), each index   runs over  all three  generations.
The  doublet and singlet lepton generations are  the charged lepton mass eigenstates $\{e,\mu,\tau\}$, the singlet quarks are  also labelled by their flavour, and the  quark doublets are in the $u$-type mass basis, with  generation indices that  run  $1\to 3$. 
}

The SM Lagrangian  is in  the notation of \cite{Manohar1}, so the covariant derivative on doublet leptons is
\begin{equation}
(D_\mu \ell)^{I}_{i} = \left( \delta_{IJ}\partial_\mu + i \frac{g}{2} \tau^a_{IJ} W^a_\mu 
+ i \delta_{IJ} g' Y(\ell)
  B_\mu \right) \ell^{J}_{i},
\label{Dl}  
\end{equation}
{where  $\tau^a$  are  Pauli matrices, $I,J$ are SU(2) doublet indices and
  $i$ is a flavour index. }
At all scales, the doublet and singlet leptons are   in
  the low energy  mass eigenstate basis,
 so   the  lepton Yukawa  matrix
  $[y_e]$ can have off-diagonal entries,
  in the presence of the  operator  ${\cal O}_{eH}$ (see equations
\ref{eq:yukdim6} and \ref{me}).  We follow \cite{Ardu:2021koz} in choosing this  basis,
because  it defines lepton flavour in the presence of LFV,  so  it 
simplifies our calculations(as mentioned at the end of  section \ref{ssec:EoM}). The Yukawa matrix eigenvalue of fermion  $f$  is written $y_f$.

The dimension six  operators in eq. (\ref{eq:SMEFTLag}) are in the ``on-shell''  basis  of \cite{BUCHMULLER1986621} 
as pruned  in \cite{Grzadkowski:2010es}, where   ``on-shell'' means that  the equations of motion were used to   reduce  the basis.   Complete bases of on-shell
 dimension eight operators have  appeared recently \cite{Murphy:2020rsh, Li:2020gnx},  and  our  dimension eight operators are in these lists. However in reality, we are only interested in the subset of dimension eight  $\mu\leftrightarrow e$ operators  to which experiments could be sensitive, which was given in  \cite{Ardu:2021koz}.  Finally, some operators in  eq. (\ref{eq:SMEFTLag}) are  hermitian in flavour space ({\it ie} $[{\cal O}^{\bar{i}j\bar{k}l}_A]^\dagger =  {\cal O}^{\bar{j}i\bar{l}k}_A$);  we include these operators  multiplied by an extra $1/2$, as the Hermitian conjugate is included in (\ref{eq:SMEFTLag}) and summing over flavour indices would otherwise lead to double counting with respect to the conventions of \cite{Manohar1}.

 We assume  LFV heavy particles are beyond  the  reach of the LHC in the next decade,  because we are  interested in combining observables from upcoming experiments at  low-energy.  Concretely,   this means  that {  the operator coefficients, or Wilson coefficients (WCs),} satisfy
 $$C_A^{[n]\zeta}\leq \left(\frac{v}{\Lambda_{\rm NP}}\right)^{n-4} ~~~
 ,~~\Lambda_{\rm NP} =  4~ {\rm TeV}~~, (v= 174 ~ {\rm GeV})$$
 and that  we calculate  Renormalisation Group running of
 LFV operators in SMEFT from $\LNP \to m_W$. 
Should new particles with LFV interactions and masses $m_W <M_{NP} <4$ TeV induce larger coefficients, our results would still apply,  but might be incomplete because additional operators and diagrams could contribute.

The WCs  $\{C_A^{[n]\zeta}\}$ function as coupling constants for  LFV interactions. 
Their numerical value can be obtained by matching  the EFT onto a model, for instance  by equating the Greens functions of the model and the EFT  at the new particle mass scale $\sim \Lambda_{\rm NP}$.
The Renormalisation the Group Equations (RGEs) govern the scale dependence of the WCs below $\Lambda_{\rm NP}$.
The solution of these equations resums the logarithms that are generated by the light particle, which propagate as dynamical particles in the EFT.
So in SMEFT, the one-loop RGEs of dimension six SMEFT operators
arise from  decorating a dimension six operator with a loop involving renormalisable interactions   \cite{Manohar1,Manohar2,Manohar3}, and from  loops involving two dimension 5 operators\cite{Davidson:2018zuo}. The  mixing of   a product of dimension five and six operators  into dimension seven  has  also been calculated in SMEFT\cite{Chala:2021juk}, as have some  anomalous dimensions for 
some  operators of dimension eight \cite{Davidson:2019iqh, Chala:2021pll,Silva:2022tln}.

Upon reaching a particle mass scale,  the high scale EFT can be matched onto   another EFT, where the now-heavy particles are removed.  
{ For instance,  in crossing the electroweak scale, SMEFT  Greens functions  are calculated  in the broken SM, with  the Higgs doublet written
 \begin{equation}
       H=\begin{pmatrix}
               G^+\\ 
               v+\frac{1}{\sqrt{2}}(h+iG^0)
       \end{pmatrix}
 \end{equation}
where the $G$s are the Goldstones and $h$ is the SM Higgs boson.
These Greens functions  are then  matched to those   of  a
QED and QCD invariant EFT  (we refer to it as low energy EFT) in which the  non-renormalisable  operators are built out of SM fields  lighter than the $W$ boson\cite{Dekens:2019ept}. }

The running and matching continues from the weak scale  down to the experimental scale, where  rates can be calculated in terms of the WCs and matrix elements of the operators.
For three or four-legged $\mu \to e$ processes which are otherwise flavour diagonal ({\it ie}
$\mu\to e\gamma$ and  $\mu\to e\gamma\gamma$, but not including  $K \to \mu^\pm e^\mp$), the  ``leading'' evolution  between the  experimental scale and the weak scale has been obtained\cite{Crivellin:2017rmk}. This includes the one-loop RGEs for dimension  five  and six  operators, and some large two-loop  anomalous dimensions where the  one loop mixing vanishes\cite{Ciuchini:1993fk}.  
Several branching ratio calculations in the low energy EFT are given in the
$\mu\to e$  review \cite{Kuno:1999jp}, {  and $\mu A \to eA$ conversion rates
can be calculated from \cite{Kitano:2002mt}.}
These results  can be    combined to  calculate the  current and upcoming
sensitivity of $\mu\leftrightarrow e$ experiments to WCs at the weak scale,
and also extrapolated  to give the
sensitivities to the  $\tau\leftrightarrow  l$ WCs considered in this manuscript
\cite{Davidson:2016edt}.

The aim of this manuscript is to calculate the contributions
to $\mu \to e$ observables that arise from combining $\tau \to e$ and $\mu\to \tau$ operators. This could occur in SMEFT running, in matching at the weak scale, and in running below the weak scale. In SMEFT, loop diagrams containing pairs of dimension six operators 
renormalize the Wilson coefficients of dimension eight operators,
such that the RGEs for the latter take the schematic form \cite{Herrlich:1994kh}
\begin{equation}
	(16\pi^2)\frac{d\Vec{C}^{[8]}_A}{d\log M}=\Vec{C}^{[8]}_B\gamma_{BA}+\Vec{C}^{[6]}_X\hat{\gamma}_{XY,A}\Vec{C}^{[6]}_{Y},
\label{eqn4}
\end{equation}
having aligned the operator coefficients in the row vectors $\Vec{C}^{[8]}$, $\Vec{C}^{[6]}$, and where $\gamma$ is the anomalous dimension matrix of dimension eight coefficients while $\hat{\gamma}$ mixes pairs of dimension six into dimension eight. The RGEs of dimension eight operators are currently unknown and only partial calculations have been performed \cite{Davidson:2019iqh, Chala:2021pll}. This manuscript fits into this ongoing effort.

{ We define the anomalous dimensions with a $1/(16\pi^2)$ prefactor, while we unconventionally do not factor out SM couplings. Two insertion of dimension six operators renormalize the dimension eight coefficients as
\begin{equation}
	\Delta \Vec{C}^{[8]}_A=\Vec{C}^{[6]}_X\hat{Z}_{XY,A}\Vec{C}^{[6]}_{Y},
\end{equation}
where $\hat{Z}$ is the divergent renormalization factor and may contain renormalizable couplings. In dimensional regularization, the independence of bare Wilson coefficients from the arbitrary renormalization scale gives the anomalous dimension matrix of eq.~(\ref{eqn4}), which at one-loop and with our conventions takes the following form
\begin{equation}
	\hat{\gamma}\propto 16\pi^2\epsilon \hat{Z}.
\end{equation} 
Note that $\hat{Z}\propto 1/\epsilon$ and the product above is finite as expected. A more detailed derivation of $\hat{\gamma}$ can be found in section \ref{ssec:SMEFTRunning}.  }

Pairs of $\tau\leftrightarrow  l$ operators also contribute to
$\mu\to e$ amplitudes in matching SMEFT onto the low energy EFT  at $m_W$.
In ``integrating out''  the heavy bosons $h$, $Z$ and replacing
the Higgs doublet with its vacuum expectation value,
it is possible to draw diagrams built out of $\tau\leftrightarrow  l$ operators that match onto three or four-legged $\mu\to e$ operators in the low energy EFT. We calculate these matching conditions, which are meant to complete  the tree-level $\order{v^4/\Lambda^4_{\rm NP}}$ matching performed in \cite{Ardu:2021koz}.

Finally, combining two $\tau\leftrightarrow  l$ operators
contributes to the RGEs of Wilson coefficients in the EFT below $m_W$.
We neglect these running contributions because they
carry a suppression factor with respect
to dimension six anomalous dimensions which is
$\lesssim m^2_b/\Lambda_{\rm NP}^2$, given that the bottom quark is
the heaviest dynamical particle in the EFT.
Such suppression is absent in SMEFT,
 where the top quark, the Higgs and gauge bosons are present,  allowing  Higgs legs to be  attached with order one couplings to heavier particles running in loops. 
SMEFT has also the advantage of having two-fermion ``penguin" operators that are efficiently generated in mixing and which match onto vector operators in the low energy EFT. For the above reasons we focus on SMEFT RGEs and matching, while we neglect the running below $m_W$.

 Equation (\ref{eqn4}) has a straightforward solution if the anomalous dimension matrices are  constant, which  occurs when the running of  all-but-one of the SM renormalisable couplings can be neglected.
 We  take  all SM couplings constant  between   $m_W\to \LNP= 4$ TeV,
 in solving eq. (\ref{eqn4}).
 It is augmented by the RGEs of dimension six coefficients:
\begin{equation}
	\frac{d\vec{C}^{[6]}}{dt}=-\vec{C}^{[6]}\tilde{\gamma}
\end{equation}
so the  solution
is
\begin{equation}
	\vec{C}^{[6]}(t)=\vec{C}^{[6]}(0)\exp(-\tilde{\gamma}t)
\end{equation}
\begin{equation}
	\vec{C}^{[8]}(t)=\left[\vec{C}^{[8]}(0)-\int_0^t d\tau \vec{C}^{[6]}(0)\exp(-\tilde{\gamma}\tau)\hat{\gamma}\exp(-\tilde{\gamma}^T\tau)\vec{C}^{[6]}(0)\exp(\gamma \tau)\ \right]\exp(-\gamma t). 
\end{equation}
Expanding the exponential at leading log, the dimension eight coefficients at the electroweak scale take the following form
\begin{equation}
  \vec{C}^{[8]}(m_W)=\vec{C}^{[8]}(\Lambda_{\rm NP})
      {\Big (} 1 - \frac{\gamma}{16\pi^2} \log(\frac{\Lambda_{\rm NP}}{m_W}){\Big )}
      -\vec{C}^{[6]}(\Lambda_{\rm NP})\frac{\hat{\gamma}}{16\pi^2}\vec{C}^{[6]}(\Lambda_{\rm NP})\log(\frac{\Lambda_{\rm NP}}{m_W})+\dots\ . \label{eq:RGEsolution}
\end{equation}
  
\subsection{Operators}\label{sec:Operators}

This subsection lists the operators included in  the SMEFT Lagrangian of eq. (\ref{eq:SMEFTLag}). They are classified  into subgroups ($D_6, 4f_6$...), in order to facilitate the estimates of section \ref{ssec:estimates}.

The SMEFT dimension six operators that are $\tau \to e$ or $\mu \to \tau$ flavour changing are the following, where the indices  $ij$ take the values $e\tau$  or $\tau\mu$ (except for the $4l_6$ operators).
\begin{itemize}
	\item Dipole operators $\equiv D_6$:
	\begin{align}
		\mathcal{O}^{ij}_{eB}&=y_\tau(\bar{\ell}_i H\sigma^{\alpha\beta} e_j)B_{\alpha\beta}\qquad
		\mathcal{O}^{ij}_{eW}=y_\tau(\bar{\ell}_i \tau^a H\sigma^{\alpha\beta} e_j)W^a_{\alpha\beta}
                \label{eq:D6}
	\end{align}
        The Hermitian conjugates with exchanged $i\leftrightarrow j$ match onto the dipole operator with opposite chirality.
	\item Penguin operators $\equiv P_6$:
	\begin{align}
		\mathcal{O}^{ij}_{He}&=i(\bar{e}_i \gamma^\alpha e_j)(H^\dagger\overset\leftrightarrow{D}_\alpha H)\qquad
		\mathcal{O}^{ij}_{H\ell(1)}=i(\bar{\ell}_i \gamma^\alpha \ell_j)(H^\dagger\overset\leftrightarrow{D}_\alpha H) \label{eq:penguinsdim6a}\\
		\mathcal{O}^{ij}_{H\ell(3)}&=i(\bar{\ell}_i \tau^a\gamma^\alpha \ell_j)(H^\dagger\overset{\leftrightarrow}{D_{\alpha} ^a} H) \label{eq:penguinsdim6b}
	\end{align}
		where we have defined 
		\begin{align}
			iH^\dagger\overset\leftrightarrow{D}_\mu H&\equiv iH^\dagger (D_\mu H)-i(D_\mu H^\dagger)H \label{eq:penguinsdefa}\\
			iH^\dagger\overset{\leftrightarrow}{D_{\mu}^a} H&\equiv iH^\dagger \tau^a(D_\mu H)-i(D_\mu H^\dagger)\tau^a H. \label{eq:penguinsdefb}
		\end{align}
	\item Yukawa operators $\equiv Y_6$:
	\begin{align}
		\mathcal{O}^{ij}_{eH}=(\bar{\ell}_i H e_j)(H^\dagger H) \label{eq:yukdim6}
	\end{align}
	and their Hermitian conjugates.
	\item Four lepton operators $\equiv 4l_6$:
	\begin{align}
		\mathcal{O}^{ijkl}_{ee}=(\bar{e}_i\gamma^\alpha e_j)(\bar{e}_k\gamma_\alpha e_l)\qquad \mathcal{O}^{ijkl}_{\ell e}=(\bar{\ell}_i\gamma^\alpha \ell_j)(\bar{e}_k\gamma_\alpha e_l)\\
		\mathcal{O}^{ijkl}_{\ell \ell}=(\bar{\ell}_i\gamma^\alpha \ell_j)(\bar{\ell}_k\gamma_\alpha \ell_l)
	\end{align}
	    { where the pairs $ij, kl, kj, il$ can be $e\tau$ or $\tau\mu$, while the remaining pair is diagonal and can be $\{e,\mu,\tau\}$.}
	\item Two-lepton two-quark operators $\equiv 4f_6$:
	\begin{align}
		\mathcal{O}^{(1)ij nm}_{\ell q}&=(\bar{\ell}_i\gamma^\alpha \ell_j)(\bar{q}_n\gamma_\alpha q_m)\qquad 	\mathcal{O}^{(3)ijnm }_{lq}=(\bar{\ell}_e\tau^a\gamma^\alpha \ell_\mu)(\bar{q}_n\tau^a\gamma_\alpha q_m)\\
		\mathcal{O}^{ijnm}_{\ell u}&=(\bar{\ell}_i\gamma^\alpha \ell_j)(\bar{u}_n\gamma_\mu u_m)\qquad
		\mathcal{O}^{ijnm}_{\ell d}=(\bar{\ell}_i\gamma^\alpha \ell_j)(\bar{d}_n\gamma_\alpha d_m)\\
		\mathcal{O}^{ijnm}_{eq}&=(\bar{e}_i\gamma^\alpha e_j)(\bar{q}_n\gamma_\alpha q_m)\qquad
		\mathcal{O}^{ijnm}_{eu}=(\bar{e}_i\gamma^\alpha e_j)(\bar{u}_n\gamma_\alpha u_m)\\
		\mathcal{O}^{ijnm}_{ed}&=(\bar{e}_i\gamma^\alpha e_j)(\bar{d}_n\gamma_\alpha d_m)\\
		\mathcal{O}^{ij nm}_{\ell edq}&=(\bar{\ell}_i e_j)(\bar{d}_n q_m)\qquad
		\mathcal{O}^{ij nm}_{\ell equ}=(\bar{\ell}_i e_j)\epsilon(\bar{q}_n u_m)
	\end{align}
	with $n,m\in\{1,2,3\}$ running over the three quark families. 
\end{itemize}

At  dimension eight, there are thousands of operators, but  here are listed
only the subset  relevant for our calculations, where relevant means that their contribution could be detectable in the upcoming $\mu \to e$ experimental searches, assuming a NP scale $\Lambda_{\rm NP}\gtrsim 4$ T$e$V.  A
list of such operators  was identified in \cite{Ardu:2021koz}, and is given below.

These include dipole operators $\equiv D_8$
\begin{align} 
    \mathcal{O}^{(1)e\mu}_{ \ell eWH^3}&=y_\mu(\bar{\ell}_e \tau^a H\sigma^{\alpha\beta} e_\mu)W^a_{\alpha\beta}(H^\dagger H) \nonumber \\
	\mathcal{O}^{(2)e\mu}_{\ell eWH^3}&=y_\mu(\bar{\ell}_e H\sigma^{\alpha\beta} e_\mu)W^a_{\alpha\beta}(H^\dagger\tau^a H) \nonumber \\
	\mathcal{O}^{e\mu}_{\ell eBH^3}&=y_\mu(\bar{\ell}_e H\sigma^{\alpha\beta} e_\mu)B_{\alpha\beta}(H^\dagger H) \label{eq:dim8dipoles}
\end{align}
and their Hermitian conjugates with the lepton indices exchanged. Two-lepton two-quark vector $\equiv 4f_8$
\begin{align}
	\mathcal{O}^{(1)e\mu nn}_{\ell^2q^2H^2}&=(\bar{\ell}_e\gamma^\alpha \ell_\mu)(\bar{q}_n\gamma_\alpha q_n)(H^\dagger H)\qquad 	\mathcal{O}^{(2)e\mu nn}_{\ell^2q^2H^2}=(\bar{\ell}_e\tau^a\gamma^\alpha \ell_\mu)(\bar{q}_n\gamma_\alpha q_n)(H^\dagger\tau^a H)\label{eq:dim8vectordoublets1}\\
	\mathcal{O}^{(3)e\mu nn}_{\ell^2q^2H^2}&=(\bar{\ell}_e\tau^a\gamma^\alpha \ell_\mu)(\bar{q}_n \tau^a\gamma_\alpha q_n)(H^\dagger H)\qquad 	\mathcal{O}^{(4)e\mu nn}_{\ell^2q^2H^2}=(\bar{\ell}_e\gamma^\mu \ell_\mu)(\bar{q}_n\tau^a\gamma_\mu q_n)(H^\dagger\tau^a H)\\ 		\mathcal{O}^{(1)e\mu nn}_{\ell^2u^2H^2}&=(\bar{\ell}_e\gamma^\alpha \ell_\mu)(\bar{u}_n\gamma_\mu u_n)(H^\dagger H)\qquad
	\mathcal{O}^{(2)e\mu nn}_{\ell^2u^2H^2}=(\bar{\ell}_e\tau^a\gamma^\alpha \ell_\mu)(\bar{u}_n\gamma_\alpha u_n)(H^\dagger \tau^a H)\\ \mathcal{O}^{(1)e\mu nn}_{\ell^2d^2H^2}&=(\bar{\ell}_e\gamma^\alpha \ell_\mu)(\bar{d}_n\gamma_\alpha d_n)(H^\dagger H)\qquad
	\mathcal{O}^{(2)e\mu nn}_{\ell^2d^2H^2}=(\bar{\ell}_e\tau^a\gamma^\alpha \ell_\mu)(\bar{d}_n\gamma_\alpha d_n)(H^\dagger \tau^a H)\label{eq:dim8vectordoublets2}\\ \mathcal{O}^{(1)e\mu nn}_{e^2q^2H^2}&=(\bar{e}_e\gamma^\alpha e_\mu)(\bar{q}_n\gamma_\alpha q_n)(H^\dagger H)\qquad
	\mathcal{O}^{(2)e\mu nn}_{e^2q^2H^2}=(\bar{e}_e\gamma^\alpha e_\mu)(\bar{q}_n\tau^a\gamma_\alpha q_n)(H^\dagger \tau^a H)\label{eq:dim8vectorsinglets1}\\ \mathcal{O}^{e\mu nn}_{e^2u^2H^2}&=(\bar{e}_e\gamma^\alpha e_\mu)(\bar{u}_n\gamma_\alpha u_n)(H^\dagger H)\qquad
	\mathcal{O}^{e\mu nn}_{e^2d^2H^2}=(\bar{e}_e\gamma^\alpha e_\mu)(\bar{d}_n\gamma_\alpha d_n)(H^\dagger H) \label{eq:dim8vectorsinglets2}
\end{align}
with in most cases  $n=u,d$ belonging to the first generation quarks. There are also penguin operators $\equiv P_8$
\begin{align}
    \mathcal{O}^{(1)e\mu}_{\ell^2H^4D}&=i(\bar{\ell}_e \gamma^\alpha \ell_\mu)(H^\dagger\overset\leftrightarrow{D}_\alpha H)(H^\dagger H)\qquad 
	\mathcal{O}^{(2)e\mu}_{\ell^2H^4D}=i(\bar{\ell}_e \tau^a\gamma^\alpha \ell_\mu)[(H^\dagger\overset{\leftrightarrow}{D_{\alpha} ^a} H)(H^\dagger H)+(H^\dagger\overset\leftrightarrow{D}_{\alpha}  H)(H^\dagger\tau^a H)]\nonumber\\ 
	\mathcal{O}^{e\mu}_{e^2H^4D}&=i(\bar{e}_e \gamma^\alpha e_\mu)(H^\dagger\overset{\leftrightarrow}D_{\alpha}  H)(H^\dagger H). \label{eq:dim8peng}
\end{align}
Furthermore, the following two-fermion two-lepton scalar and tensor operators are also relevant
\begin{align}
	\mathcal{O}^{(1)e\mu nn}_{\ell edqH^2}&=(\bar{\ell}_e e_\mu)(\bar{d}_n q_n)(H^\dagger H)\qquad	\mathcal{O}^{(2)e\mu nn}_{\ell edqH^2}=(\bar{\ell}_e e_\mu)\tau^{a}(\bar{d}_n q_n)(H^\dagger\tau^{a} H) \label{eq:dim8scalardown}\\
	\mathcal{O}^{(1)e\mu nn}_{\ell equH^2}&=(\bar{\ell}_e e_\mu)\epsilon(\bar{q}_n u_n)(H^\dagger H)\qquad
	\mathcal{O}^{(2)e\mu nn}_{\ell equH^2}=(\bar{\ell}_e e_\mu)\tau^{a}\epsilon(\bar{q}_n u_n)(H^\dagger \tau^{a} H) \label{eq:scalarupdim8}\\
	\mathcal{O}^{(3)e\mu nn}_{\ell equH^2}&=(\bar{\ell}_e\sigma^{\alpha\beta} e_\mu)\epsilon(\bar{q}_n \sigma_{\alpha\beta}u_n)(H^\dagger H)\qquad
	\mathcal{O}^{(4)e\mu nn}_{\ell equH^2}=(\bar{\ell}_e\sigma^{\alpha\beta} e_j)\tau^{a}\epsilon(\bar{q}_n \sigma_{\alpha\beta}u_n)(H^\dagger\tau^{a} H)\label{eq:toptensor}\\	
	\mathcal{O}^{(3)e\mu nn}_{\ell edqH^2}&=(\bar{\ell}_e He_\mu)(\bar{q}_n H d_n)\qquad 
	\mathcal{O}^{(4)e\mu nn}_{\ell edqH^2}=(\bar{\ell}_e \sigma^{\alpha\beta}He_\mu)(\bar{q}_n \sigma_{\alpha\beta}H d_n)\\
	\mathcal{O}^{(5)e\mu nn}_{\ell equH^2}&=(\bar{\ell}_eHe_\mu)(\bar{u}_n\tilde{H}^\dagger q_n),
\end{align}
with $n=u,c,t,d,s,b$ running over all quark flavours. Finally the four-lepton operators $\equiv 4l_8$ read
\begin{align}
	\mathcal{O}^{(4)e\mu \tau\tau}_{\ell^2 e^2H^2}&=(\bar{l}_eH\sigma^{\alpha\beta} e_\mu)(\bar{l}_\tau H\sigma_{\alpha\beta} e_\tau)\qquad \mathcal{O}^{(3)e\mu ee}_{\ell^2 e^2H^2}=(\bar{l}_eH e_\mu)(\bar{l}_eH e_e)\\
	\mathcal{O}^{(1)e\mu ee}_{\ell^4H^2}&=(\bar{\ell}_e\gamma^\alpha \ell_\mu)(\bar{\ell}_e\gamma_\alpha \ell_e)(H^\dagger H)\qquad 	\mathcal{O}^{(2)e\mu ee}_{\ell^4 H^2}=(\bar{\ell}_e\gamma^\alpha \ell_\mu)(\bar{\ell}_e\tau^{a}\gamma_\alpha \ell_e)(H^\dagger\tau^{a} H)\\
	\mathcal{O}^{(1)e\mu ee}_{\ell^2 e^2H^2}&=(\bar{\ell}_e\gamma^\alpha \ell_\mu)(\bar{e}_e\gamma_\alpha e_e)(H^\dagger H)\qquad 		\mathcal{O}^{(2)e\mu ee}_{\ell^2 e^2H^2}=(\bar{\ell}_e\tau^{a}\gamma^\alpha \ell_\mu)(\bar{e}_e\gamma_\alpha e_e)(H^\dagger \tau^{a}H)\\
	\mathcal{O}^{e\mu ee}_{e^4H^2}&=(\bar{e}_e\gamma^\alpha e_\mu)(\bar{e}_e\gamma_\alpha e_e)(H^\dagger H).
\end{align}
  Note that in the Lagrangian of eq.~(\ref{eq:SMEFTLag}) we sum over all possible generation indices, and more flavour structures are relevant for low energy LFV interactions. For instance, $\mathcal{O}^{e\mu ee}_{\ell^4 H^2},\mathcal{O}^{eee\mu }_{\ell^4 H^2}$ match onto the same vector operator in the EFT below $m_W$. { Similarly, in the case of $e\mu\tau\tau$ tensor operator, the permutations $\tau\tau e\mu,\tau\mu e \tau, e\tau\tau \mu$ must be considered.  }

\subsection{Equations of Motion}\label{ssec:EoM}

{In this section, we discuss some of the technical subtleties that occur when
  two dimension six operators  mix into dimension eight operators.
	In our calculations of anomalous dimensions we consider two different approaches: we can systematically apply the equations of motions onto the amplitudes of our loop calculations in order to arrive at expressions that are proportional to tree-level amplitudes of the  on-shell,  or ``physical'' operators. Alternatively, we could use a complete set of off-shell operators and project our loop amplitudes onto the on-shell operator basis.
	The situation is slightly complicated by the facts that the dimension six operators will contribute themselves to the equations of motion, and that there are a huge number of dimension eight operators.
	In the following we will show how both approaches are equivalent in our calculation, where we determine the mixing into the subset of dimension eight operators that contribute to LFV at low energy experiments.}

\begin{figure}
	\centering
\begin{tikzpicture}[scale=1.5]
	\begin{feynman}[small]
		\vertex (mu) at (2,0) {\(\ell_\mu\)};
		\vertex (e) at (-1,0) {\(e_\tau\)};
		\vertex (b) at (0,0) [label=90:\(y_{\tau }\)];
		\vertex [style=crossed dot] (a)  at (1,0) [label=90:\( \propto\slashed{q}\)] {};
		\vertex (H2) at (0.5,-1) {\(H\)};
		\vertex (H3) at (1.5,-1) {\(H\)};
		\vertex (H1) at (0,-1) {\(H\)};
		\diagram* [inline=(a.base)]{
			(H3) -- [anti charged scalar] (a),
			(H1) -- [charged scalar] (b),
			(e) -- [fermion] (b)-- [fermion, edge label=\(\ell_\tau\), momentum'=\(q\)] (a) -- [fermion] (mu),
			(H2) -- [charged scalar] (a),
		};
	\end{feynman}
\end{tikzpicture}
\caption{ The diagram shows that the operator $i(\bar{\ell}_\mu \slashed{D} \ell_\tau)(H^\dagger H)$ leads to the same $S-$matrix elements as  $y_\tau(\bar{\ell}_\mu H e_e)(H^\dagger H)$. The non-local momentum dependence of the internal line propagator cancels with the inverse propagator present in the Equation of Motion. }\label{fig:renomEOM}
\end{figure}
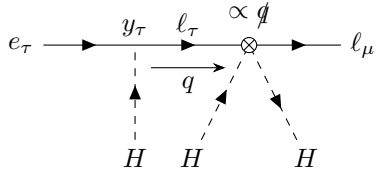

Working with a  on-shell (or physical) operator basis implies the choice of a set of operators that vanish when the Equation of Motions (EOM) are satisfied. Take two operators $\mathcal{O}_1$, $\mathcal{O}_2$ which differ by an operator $\mathcal{O}_{EOM}$ that is EOM vanishing, i.e
\begin{equation}
  \mathcal{O}_1-\mathcal{O}_2=\mathcal{O}_{EOM}\propto \frac{\delta S}{\delta\phi}
  \label{O1-02}
\end{equation}
where $S$ is the action and $\phi$ labels a generic field. $\mathcal{O}_{EOM}$ can be dropped in physical processes because it leads to vanishing $S-$matrix elements, so that the operators $\mathcal{O}_1$, $\mathcal{O}_2$ are physically equivalent and only one of them is retained in the basis.

\begin{figure}
	\centering
	\begin{subfigure}{.8\textwidth}
		\centering
	\begin{tikzpicture}[scale=1.5, baseline={([yshift=-.5ex]current bounding box.center)}]
		\begin{feynman}[small]
			\vertex (mu) at (3,0) {\(\ell_\mu\)};
			\vertex (b) at (2,0);
			\vertex [style=crossed dot] (a)  at (1,0) [label=135:\( C^{\mu \tau }_{Hl}\)] {};
			\vertex (H2) at (0.5,-0.75) {\(H\)};
			\vertex (H3) at (1.5,-0.75) {\(H\)};
			\vertex (in) at (0,0) {\(\ell_\tau\)};
			\diagram* [inline=(a.base)]{
				(H3) -- [anti charged scalar] (a),
				(in) -- [fermion] (a) -- [fermion] (b) -- [fermion] (mu),
				(a) -- [photon, half left] (b),
				(H2) -- [charged scalar] (a),
			};
		\end{feynman}
	\end{tikzpicture}$\propto \frac{C^{(1)\mu \tau }_{Hl}}{16\pi^2 \epsilon}i(\bar{\ell}_\mu \slashed{D} \ell_\tau)(H^\dagger H)$
     \caption{}\label{subfig:offdiagkin}
    \end{subfigure}\\ 
    \begin{subfigure}{.8\textwidth}
    	\centering
	\begin{tikzpicture}[scale=1.5, baseline={([yshift=-.5ex]current bounding box.center)}]
	\begin{feynman}[small]
		\vertex (mu) at (3,0) {\(\ell_\tau\)};
		\vertex (b) at (2,0) [label=-90:\( y_\tau\)];
		\vertex [style=crossed dot] (a)  at (1,0) [label=135:\( C^{\tau\mu }_{Hl}\)] {};
		\vertex (H2) at (1,-0.75) {\(H\)};
		\vertex (in) at (0,0) {\(\mu\)};
		\diagram* [inline=(a.base)]{
			(in) -- [fermion] (a) -- [fermion] (b) -- [fermion] (mu),
			(a) -- [charged scalar, half left] (b),
			(H2) -- [charged scalar] (a),
		};
	\end{feynman}
     \end{tikzpicture}$\propto \frac{C^{(1) \tau \mu }_{Hl}}{16\pi^2 \epsilon}(D^2\bar{\ell}_\tau H\mu)$
    \caption{}\label{subfig:boxop}
    \end{subfigure}
	\caption{One-loop diagrams with the penguin operators of eq.~(\ref{eq:penguinsdim6a}). Matching the divergences off-shell, the redundant operators $i(\bar{\ell}_\mu \slashed{D} \ell_\tau)(H^\dagger H)$, $(D^2\bar{\ell}_\tau H\mu)$ are generated. \label{fig:offdiagkinetic}}
\end{figure}
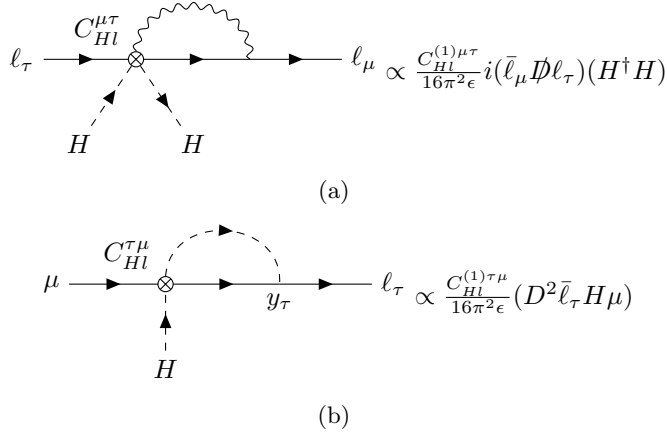

For instance, at dimension six, the  operators 
\begin{equation}
	i(\bar{\ell}_\mu \slashed{D} \ell_\tau)(H^\dagger H)~,~(D^2\bar{\ell}_\tau H\mu)
\end{equation}
 can be generated at one-loop from a penguin operator (see Figure \ref{fig:offdiagkinetic}). The first is relevant here, because it 
is on-shell equivalent to $(\bar{\ell}_\mu H e_\tau)(H^\dagger H)$ by means of the dimension four EOM of the lepton field $i (\slashed{D}\ell_\tau)=y_{\tau}H e_\tau$. (The second operator will be relevant for the $C_{Hl}\times C_{He}$ mixing into dipoles, which is discussed in section \ref{ssec:SMEFTrunningsub}.)

Therefore, we can project an amplitude that is proportional to the left hand side of the previous equation of  motion
\begin{equation}
	i(\bar{\ell}_\mu \slashed{D} \ell_\tau)(H^\dagger H){\to}[i(\bar{\ell}_\mu \slashed{D} \ell_\tau)(H^\dagger H)-y_{\tau}(\bar{\ell}_\mu H e_\tau)(H^\dagger H)] + y_{\tau}(\bar{\ell}_\mu H e_\tau)(H^\dagger H) \label{eq:renomEOM}
\end{equation}
{onto physical and  EOM vanishing -- in brackets -- operators}.
In Figure \ref{fig:renomEOM} we show how the equivalence can be understood diagrammatically: the $\slashed{D}$ operator Feynman rule is proportional to the $\slashed{q}$ momentum of the virtual $\ell_\tau$ line coming out of a renormalizable Yukawa coupling; the momentum dependence cancels with the $\ell_\tau$ propagator, yielding an $S-$matrix element reproduced by the local operator $y_\tau(\bar{\ell}_\mu H e_\tau)(H^\dagger H)$.

Once a reduced physical basis is identified, the theory can be  consistently renormalized among on-shell operators, as redundant counterterms
$A \mathcal{O}_2/\epsilon$ are equivalent to
$A (\mathcal{O}_1-\mathcal{O}_{EOM})/\epsilon$ and EOM vanishing operators mix exclusively among themselves in the RGEs \cite{Simma:1993ky}\footnote{Gauge fixing and ghost terms that appear in the EOM are found to have no physical effects in operator mixing and $S-$matrix elements \cite{Simma:1993ky}.}.

However, in order to consistently renormalize an EFT in a given basis up to dimension eight ($1/\Lambda_{\rm NP}^4$), the dimension six ($1/\Lambda_{\rm NP}^2$) terms in the EOM must be included when removing  redundant operators. Concretely, if a divergent contribution to a redundant dimension six operator,
$\mathcal{O}^{[6]}_{2}/(\Lambda_{\rm NP}^2\epsilon)$
is generated via loops, then it  can be rewritten
\begin{equation}
	\frac{A}{\Lambda_{\rm NP}^2\epsilon}\left(\mathcal{O}^{[6]}_{\rm 1}+\frac{\mathcal{O}^{[8]}}{\Lambda_{\rm NP}^2}-\mathcal{O}_{EOM}\right)
\end{equation}
where ${O}^{[6]}_1$ is equivalent to ${O}^{[6]}_2$ via the renormalizable EOM $\delta S^{d=4}/\delta\phi=0$ of eq. (\ref{O1-02}), and the dimension eight $\mathcal{O}^{[8]}$ is generated by the dimension six corrections $\delta S^{d=6}/\delta\phi$. The dimension eight contribution is proportional to the product of two dimension six operator coefficients, which is the kind of contribution that we are interested in.

As an example of  the impact of dimension six terms in the EOM,
suppose that the  only $\te$ operator at dimension six is  $\mathcal{O}^{ e \tau nm}_{\ell edq}=(\bar{\ell}_\tau e_e)(\bar{d}_n q_m)$,  and that  the
operator $i(\bar{\ell}_\mu \slashed{D} \ell_\tau)(H^\dagger H)$ is generated
via loop corrections. 
Then  eq.~(\ref{eq:renomEOM}), up to dimension 8, becomes
\begin{align}
	i(\bar{\ell}_\mu \slashed{D} \ell_\tau)(H^\dagger H)&=\left[i(\bar{\ell}_\mu \slashed{D} \ell_\tau)(H^\dagger H)-y_\tau(\bar{\ell}_\mu H e_\tau)(H^\dagger H)+\frac{C^{\tau e nm}_{\ell edq}}{\Lambda^2_{\rm NP}}(\bar{\ell}_\mu e_e)(\bar{d}_n q_m)(H^\dagger H)\right]\nonumber \\ 
	&+ y_\tau(\bar{\ell}_\mu H e_\tau)(H^\dagger H)-\frac{C^{\tau e nm}_{ \ell edq}}{\Lambda^2_{\rm NP}}(\bar{\ell}_\mu e_e)(\bar{d}_n q_m)(H^\dagger H) \label{eq:EOMdim8}
\end{align}
where the EOM vanishing operator in square brackets now contains the dimension eight $\mathcal{O}^{(1)iknm}_{\ell edqH^2}=(\bar{\ell}_\mu e_e)(\bar{d}_n q_m)(H^\dagger H)$. 
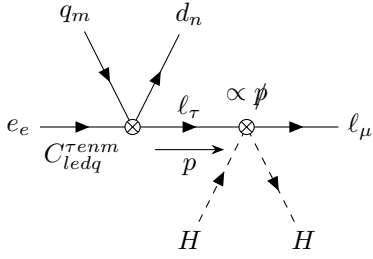
\begin{figure}
	\centering
\begin{tikzpicture}[scale=1.5]
	\begin{feynman}[small]
		\vertex (mu) at (2,0) {\(\ell_\mu\)};
		\vertex (e) at (-1,0) {\(e_e\)};
		\vertex [style=crossed dot] (b) at (0,0) [label=-110:\(C_{ledq}^{\tau e nm}\)] {};
		\vertex [style=crossed dot] (a)  at (1,0) [label=90:\( \propto\slashed{p}\)] {};
		\vertex (H2) at (0.5,-1) {\(H\)};
		\vertex (H3) at (1.5,-1) {\(H\)};
		\vertex (f1) at (-0.5, 1) {\(q_m\)};
		\vertex (f2) at (0.5, 1) {\(d_n\)};
		\diagram* [inline=(a.base)]{
			(H3) -- [anti charged scalar] (a),
			(f1) -- [fermion] (b) -- [fermion] (f2),
			(e) -- [fermion] (b)-- [fermion, edge label=\(\ell_\tau\), momentum'=\(p\)] (a) -- [fermion] (mu),
			(H2) -- [charged scalar] (a),
		};
	\end{feynman}
\end{tikzpicture}
\caption{Correction to the Equation of Motion due to dimension six operators. At $1/\Lambda_{\rm NP}^4$ order, the operator $i(\bar{\ell}_\mu \slashed{D} \ell_\tau)(H^\dagger H)$ is on-shell equivalent to a combination of dimension six and dimension eight operators. The dimension eight contribution can be understood by attaching dimension six interactions to the operator, where the internal line propagator cancels against the vertex Feynman rule. The diagram shows an example with the insertion $\mathcal{O}^{\tau e nm}_{ledq}=(\bar{\ell}_\tau e_e)(\bar{d}_n q_m)$, which reproduces the EOM reduction of eq. (\ref{eq:EOMdim8}).\label{fig:dim6EOM}}
\end{figure}
Similarly to the renormalizable case, the on-shell equivalence is apparent diagrammatically, by  dressing the redundant operator with dimension six contact interactions  as shown in Figure \ref{fig:dim6EOM}.
Once again the inverse propagator that is present in the EOM, and appears in the operator Feynman rule,  cancels  the momentum dependence of the internal line, such that the amplitude is local and equivalent to a dimension eight operator. Its coefficient will be proportional to the product of two dimension six WC. 
For instance, $i(\bar{\ell}_\mu \slashed{D} \ell_\tau)(H^\dagger H)$ is generated in matching off-shell the divergence of the one-loop diagram of Figure \ref{fig:offdiagkinetic} that involves the penguin operators of eq (\ref{eq:penguinsdim6a}).
Eq. (\ref{eq:EOMdim8}) allows to project the divergence onto the on-shell basis, giving a contribution to the renormalisation of  the dimension eight $\me$ operator $\mathcal{O}^{(1)\mu e nm}_{\ell edqH^2}=(\bar{\ell}_\mu e_e)(\bar{d}_n q_m)(H^\dagger H)$  from the product $\mathcal{O}^{(1)}_{H\ell}\times \mathcal{O}_{\ell edq}$.
This contribution from the EOM projection must be included in
calculating  the mixing from (dimension 6)$^2 \to$ dimension 8, together with
 one particle irreducible (1PI) diagrams  $ \propto C^{(1)}_{H\ell}\times C_{\ell edq}$. 
{(Indeed, the anomalous dimension is only gauge invariant if one includes both the IP1 vertex  and  the non-1PI ``wavefunction'' contributions.)}

The EOM contribution can be reproduced by calculating non-1PI divergent diagrams, as shown in figure \ref{fig:dim6EOM}. In working with a subspace of dimension eight operators (as we do here), proceeding diagrammatically can be particularly convenient. Our subspace is phenomenologically selected to contribute to the low energy $\mu\to e$ processes. When using the EOM to project the off-shell divergences, the redundant terms must be written in terms of operators in the full basis (which can include operators outside the subspace) and the EOM vanishing operators that the basis choice implies. In the end, only the interesting operators in the subspace are retained but it required working with the full basis as an intermediate step. On the other hand, in the approach of calculating one-particle -reducible diagrams,  it is often easier to restrict to  diagrams that directly give dimension eight operators of the subspace.
{ In this manuscript, we calculate the  one-particle-reducible diagrams that generate the relevant dimension eight operators. We cross-checked our diagrammatic results by calculating the dimension eight LFV operators obtained from the list of EOM-vanishing operators in \cite{Grzadkowski:2010es}, by using Equations of Motion up to dimension six. }

{ Finally, recall that we work in the   low-energy mass eigenstate basis of the leptons,  where 
the lepton mass matrix is:
\begin{equation}
  m_{e_i}\delta_{ij}=v\left([y_e]_{ij} - C_{eH}^{ij} \right)\label{men}~~.
\end{equation} 
So in the above diagrammatic and EOM-based arguments, the Yukawa  matrix  element  $y_\tau$  is replaced by the  matrix element of the parenthese on the right side of  (\ref{men}), which is also flavour-diagonal\footnote{However, in this basis, the $h$ retains LFV interactions --- see eq. (\ref{eq:LFVhiggs}).}. 
 Therefore we do not include non-1PI diagrams involving a loop on the external leg of  ${\cal O}^{ij}_{eH }$.

}

\subsection{Estimates}
\label{ssec:estimates}

The goal of this section is to better identify the { dimension eight} contributions that are interesting to calculate in the context of $\mu \to e$ LFV, that is,{ those that will be} within the reach of future experiments.
The {  Wilson coefficients} of  the dimension eight operators presented in the previous section were estimated in \cite{Ardu:2021koz} to be within { upcoming } experimental sensitivity if they {   have values $\gtrsim v^4/\Lambda^4$, for}  $\Lambda\gtrsim 4$ T$e$V. We estimate in this section the { additional} loop and small couplings suppression that {  could}  be  encountered in { generating these coefficients} in  running and matching.
{ This will allow to } narrow-down the list of diagrams that should be calculated.

In estimating diagrams built out of $\mu\to \tau  \times \tau\to e $ operators, we take into account the constraints on {
  $\tau \leftrightarrow l$ processes } coming from the bounds reported in the lower part of Table \ref{tab:LFVsearches}. 
Employing the {
  acronyms}  introduced in the previous section for   sets of $\tau$ LFV operators, current and upcoming { one-at-a-time-}limits on their coefficients are written in Table \ref{tab:TauConstraints}. These estimates  assume  that the Branching Ratio sensitivities on $\tau$ decays will improve of an order of magnitude at { BelleII} \cite{belle2t3l}, and use the future sensitivities
  to $h\to \tau^\pm l^\mp$  decays  at the   ILC \cite{ILC}. 
It is understood that the limits do not apply exactly to all operators identified by the collective label, but rather provide the order of magnitude of
the { one-at-a-time-limit, or sensitivity, to} most of them.
In the case where the operators are not (loosely) bounded, we assume
\begin{equation}
  C^{[6]l\tau...}\lsim (v/4\ \text{T}e\text{V})^2\sim 2\times 10^{-3}
\label{minbd}
\end{equation}
corresponding to an $\order{1}$ coefficient { at a New Physics scale of } 4 T$e$V.

\begin{table}[th]
\begin{center}
\begin{tabular}{|l|l|l|l|}
\hline
Operator coefficient & Current sensitivity & Future sensitivity & Process    \\
\hline 
$C^{ l \tau}_{D_6}$ & $\lesssim 7\times 10^{-6}$ & $\lesssim 2\times 10^{-6}$ & $\tau\to  l\gamma$ \\[0.7 mm]
$C^{ l \tau}_{Y_6}$ & $\lesssim10^{-3}$ & $\lesssim 3\times 10^{-4}$ &$h\to  l\tau$\\[0.7 mm]
   $C^{ l \tau}_{P_6}$ & $\lesssim 4\times 10^{-4}$ & $\lesssim  10^{-4}$ & $\tau\to \bar{ l} l l$ \\[0.7 mm]
            $C^{ l \tau l l}_{4l_6}$ & $\lesssim 3\times 10^{-4}$ & $\lesssim  10^{-4}$ & $\tau\to \bar{ l} l l$ \\[0.7 mm]
            $C^{ l\tau qq}_{4f_6}$ & $\lesssim 3\times 10^{-4}$ & $\lesssim  10^{-4}$ &$\tau\to  l \pi(\eta)$ \\
            \hline
		\end{tabular}
	\end{center}
	\caption{Sensitivities to  $\tau\leftrightarrow  l$  dimension six operator coefficients, normalized as in eq. (\ref{eq:SMEFTLag}). 
           Current limits come from the Branching ratio bounds of Table \ref{tab:LFVsearches}, while  the third column assumes that the experimental  sensitivity  to $\tl$  decays will improve by an order of magnitude.  }\label{tab:TauConstraints}
\end{table}
\begin{table}[th]
	\begin{center}
		\begin{tabular}{|l|l|l|l|}
			\hline
			Operator coefficient & Current sensitivity & Future sensitivity & Process    \\ 
			\hline
$C^{e\mu}_{D_8}$ & $\lesssim 10^{-8}$ & $\lesssim 1.5\times 10^{-9}$ & $\mu\to e\gamma$ \\ [0.7 mm]
			$C^{e\mu tt}_{4f_8,T}$ & $\lesssim 3\times10^{-11}$ & $\lesssim 5\times10^{-12}$ & $\mu\to e\gamma$\\ [0.7 mm]
			$C^{e\mu \tau\tau}_{4l_8,T},C^{e\mu cc}_{4f_8,T}$ & $\lesssim 10^{-8}$ & $\lesssim 1.5\times 10^{-9}$ & $\mu\to e\gamma$\\ [0.7 mm]
			$C^{e\mu bb}_{4f_8,T}$ & $\lesssim 8\times 10^{-9}$ & $\lesssim  10^{-9}$ & $\mu\to e\gamma$\\ [0.7 mm]
			$C^{e\mu}_{P_8}$ & $\lesssim  10^{-7}$ & $\lesssim  10^{-9}$ & $\mu A\to e A$ \\ [0.7 mm]
			$C^{e\mu e e}_{4l_8}$ & $\lesssim 8\times 10^{-7}$ & $\lesssim  8\times10^{-9}$ & $\mu\to \bar{e}ee$ \\ [0.7 mm]
			$C^{e\mu uu, e\mu dd}_{4f_8,S}$ & $\lesssim 10^{-8}$ & $\lesssim  10^{-10}$ & $\mu A\to e A$ \\[0.7 mm]
			\hline
		\end{tabular}
	\end{center}
	\caption{Sensitivities to  $\mu\to e$  dimension eight operator coefficients, normalized as in eq. (\ref{eq:SMEFTLag}). Current and future limits correspond to  the experimental sensitivities of Table \ref{tab:LFVsearches}. $T,S$ label the Lorentz structure of the operator for tensor and scalar respectively. For instance, $C^{e\mu tt}_{4f_8,T}$ is the coefficient of the dimension eight tensor in eq.~(\ref{eq:toptensor}) with top quarks. \label{tab:mutoeConstraints}}
\end{table}

Diagrams that can generate  the dimension eight $\mu \leftrightarrow e$
operators of section \ref{sec:Operators}, in matching or in running,  are drawn with a pair of $\tau\leftrightarrow  l$ operators. The contribution to the coefficients are estimated as
\begin{equation}
	\Delta C^{[8]e\mu}\simeq C_1^{[6]e\tau}C_2^{[6]\tau \mu}\left(\frac{1}{16\pi^2}\right)^n \times \left\{y^k\ g^l\ \lambda^m\dots\right\} \times \log
\end{equation}
where $n$ is the number of loops, SM couplings are factored out into the curly brackets, and the $\log(4 ~{\rm TeV}/m_W)$ factor is present in running,  while  absent in matching. In running,  we restrict the number of loops to $n=1$, while  up to two loop diagrams contribute in ``tree-level''(in the low-energy EFT) matching.
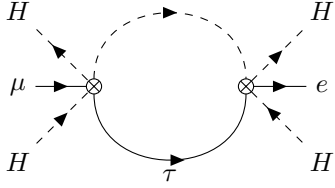
\begin{figure}
	\centering
\begin{tikzpicture}
	\begin{feynman}[small]
		\vertex (H1) at (-1,1) {\(H\)};
		\vertex (mu) at (-1,0) {\(\mu\)};
		\vertex (e) at (3,0) {\(e\)};
		\vertex[style=crossed dot] (a)  at (0,0) {};
		\vertex[style=crossed dot] (b) at (2,0) {};
		\vertex (H2) at (-1,-1) {\(H\)};
		\vertex (H3) at (3,1) {\(H\)};
		\vertex (H4) at (3,-1) {\(H\)};
		\diagram* [inline=(a.base)]{
			(H1) -- [anti charged scalar] (a),
			(mu) -- [fermion] (a),
			(b) -- [fermion] (e),
			(H2) -- [charged scalar] (a),
			(a) -- [charged scalar, half left] (b) -- [anti fermion, half left, edge label=\(\tau\)] (a),
			(b) -- [charged scalar] (H3),
			(b) -- [anti charged scalar] (H4),
		};
	\end{feynman}
\end{tikzpicture}
\caption{Mixing to the dimension eight $\mu\to e$ penguin operator from double insertion of dimension six Yukawas $Y_6\times Y_6\to P_8$. \label{fig:yuks2penguin}}
\end{figure}

{ An example of a  diagram contributing to the RGEs is}  shown in the diagram of Figure \ref{fig:yuks2penguin}, 
where two Yukawa operators $\mathcal{O}^{e\tau}_{eH}\times\mathcal{O}^{\tau\mu}_{eH}\sim Y_6\times Y_6$ mix into dimension eight $\mu\to e$ penguin operators $\mathcal{O}^{e\mu}_{e^2H^2D}, \mathcal\mathcal{O}^{e\mu}_{l^2H^2D}\sim P_8$ by exchanging the $\tau$ and closing the loop with a Higgs line.
The estimated contribution to the penguin coefficients is then
\begin{equation}
	\Delta C_{P_8}\sim (C_{Y_6})^2\frac{\log(4\ \text{T}e\text{V}/m_W)}{16\pi^2}\sim 3\times 10^{-9}.
\end{equation}
Future $\mu A\to e A$ experiments will be sensitive to penguin coefficients larger than $\sim 10^{-9}$, hence our estimate lies within experimental reach and $Y_6\times Y_6 \to P_8$ mixing is calculated in section \ref{sec:Calculations}.

As another example, $\tau\leftrightarrow  l$ dipoles $D_6$ are defined with a built-in $\tau$ Yukawa suppression {--- see eq. (\ref{eq:D6})--- so
  $y_\tau \sim 10^{-2}$ multiplies  any dipole insertion.}
For instance, if $D_6\times \mathcal{O}_6$ mix into a dimension eight operator $\mathcal{O}_8$, its coefficient is estimated to be
\begin{equation}
  \Delta C_8\sim y_\tau C_{D_6} C_6 \frac{\log(4\ \text{T}e\text{V}/m_W)}{16\pi^2}\lesssim  10^{-12},
  \label{eq:D*O=8}
\end{equation}
where we took $C_6\lsim v^2/\Lambda^2$, for  $\Lambda\sim 4$ T$e$V.  Equation (\ref{eq:D*O=8}) is smaller than any future $\mu \to e$ sensitivity to operator coefficients,  so  we disregard mixing that involves $\tau$ dipoles  in our calculations.

The results of our estimates are summarized in Tables \ref{tab:RunningEstimate} and \ref{tab:MatchingEstimate}, referring respectively to RGEs and matching contributions. There, we report the   potentially detectable dimension eight operators generated by a given pair of dimension six operators.
\begin{table}[h]
	\begin{center}
		\begin{tabular}{c|c|c|c|c}
			& $P_6$ & $Y_6$ & $4l_6$ & $4f_6$\\
			\hline
			$P_6$ & $D_8\equiv 0$ & $D_8\equiv0$ &  $\cross$ & $4f_8$\\
			\hline
			$Y_6$ & $D_8\equiv0$ & $P_8$ & $\cross$ & $\cross$\\
			\hline
			$4l_6$ & $\cross$& $\cross$& $\cross$& $\cross$\\
			\hline
			$4f_6$ & $4f_8$ &  $\cross$& $\cross$& $4f_8$\\
		\end{tabular}
		\caption{We present the dimension eight operators that we estimate to be generated within experimental sensitivity through $\text{(dimension six)}^2$ mixing in the RGEs. The $\cross$ means that the contributions is too small or that there is no one-loop diagram that can generate the desired dimension eight operators with the given pair. $P_6\times P_6\to D_8$, $Y_6\times P_6\to D_8$ mixing diagrams exist and appear to be interesting, however we find that the anomalous dimension vanishes {(see section \ref{ssec:SMEFTrunningsub}).} }\label{tab:RunningEstimate}
	\end{center}
\end{table}

\begin{table}[h]
	\begin{center}
		\begin{tabular}{c|c|c|c|c}
			& $P_6$ & $Y_6$ & $4l_6$ & $4f_6$\\
			\hline
			$P_6$ & $\cross$ & $D_8$ &  $\cross$ & $\cross$\\
			\hline
			$Y_6$ & $\cross$ & $D_8,4l_8$ & $\cross$ & $\cross$\\
			\hline
			$4l_6$ & $\cross$& $\cross$& $\cross$& $\cross$\\
			\hline
			$4f_6$ & $\cross$ &  $\cross$& $\cross$& $\cross$\\
		\end{tabular}
		\caption{We present the dimension eight operators that we estimate to be generated within experimental sensitivity through $\text{(dimension six)}^2$ in matching. The $\cross$ means that the contributions is too small or that there is no tree-level matching that can generate the desired dimension eight operators with the given pair.}\label{tab:MatchingEstimate}
	\end{center}
\end{table}

\section{Calculation}\label{sec:Calculations}

The contributions that were estimated in the previous section to be within experimental sensitivity are  calculated here. Section \ref{ssec:SMEFTRunning}
determines the divergences of the relevant one-loop diagrams and relates them to the anomalous dimensions of the dimension eight Wilson coefficients in SMEFT,
and in Section \ref{ssec:Matching}, pairs of $\tau \leftrightarrow  l$ dimension six operators are tree-level-matched at $m_W$ onto the low energy $\mu \to e$ EFT.

\subsection{SMEFT Running}\label{ssec:SMEFTRunning}

In this section, we outline the calculation of the anomalous dimension matrix $\hat{\gamma}_{XY,A}$, that mixes the dimension six $\tau\leftrightarrow l$ operators $\mathcal{O}^{[6]}_{X},\mathcal{O}^{[6]}_{Y}$ into the $\mu\to e$ dimension eight $\mathcal{O}_A^{[8]}$.
We work in dimensional regularization in $4-2\epsilon$ dimensions and renormalize in  the $\overline{\rm MS}$ scheme, where we label the renormalization scale with $M$ (rather than the usual $\mu$).
Double insertions of dimension six operators renormalize dimension eight coefficients  as
\begin{equation}
	\Delta \Vec{C}^{[8]}_A=\Vec{C}^{[6]}_X\hat{Z}_{XY,A}\Vec{C}^{[6]}_{Y},
\end{equation}
where the Wilson coefficients of dimension eight and six are respectively aligned in the row vectors $\vec{C}^{[8]}$, $\vec{C}^{[6]}$,
dimension eight and six  operator labels are  respectively capitals from the beginning  and end of the alphabet,   and  flavour indices are suppressed.
The bare dimension eight coefficients can be written as
\begin{align}
	\Vec{C}^{[8]}_{A,bare}=M^{a_A\epsilon}(\Vec{C}^{[8]}_BZ_{BA}+\Vec{C}^{[6]}_X\hat{Z}_{XY,A}\Vec{C}^{[6]}_{Y})
\end{align}
where we have factored out the sliding scale power $M^{a_A\epsilon}$ to assure that the renormalized WC stay dimensionless in $d=4-2\epsilon$ dimensions.
The RGEs can be obtained  from the independence of the
bare Lagrangian  from  the arbitrary renormalization scale $M$
\begin{equation}
	(16\pi^2)\frac{d\vec{C}^{[8]}_{A,bare}}{d\log M}=0, \label{eq:renscaleindep}
\end{equation}
which implies the following differential equation for the renormalized Wilson coefficients
\begin{align}
	(16\pi^2
	)\frac{d\vec{C}^{[8]}_{A}}{d\log M}=(16\pi^2
	)\bigg[-a_A\epsilon(\vec{C}^{[8]}_A+\vec{C}^{[6]}_{X}\vec{C}^{[6]}_{Y}\hat{Z}_{XY,B} Z^{-1}_{BA})-\vec{C}^{[8]}_B\frac{d{Z}_{BC}}{d\log M}Z^{-1}_{CA}+\nonumber\\
	-\frac{d\vec{C}^{[6]}_{X}}{d\log M}\hat{Z}_{XY,B}\vec{C}^{[6]}_{Y}Z^{-1}_{BA}	-\vec{C}^{[6]}_{X}\hat{Z}_{XY,B}\frac{d\vec{C}^{[6]}_{Y}}{d\log M}Z^{-1}_{BA}-\vec{C}^{[6]}_{X}\frac{d\hat{Z}_{XY,B}}{d\log M}\vec{C}^{[6]}_{Y}Z^{-1}_{BA}\bigg]. \label{eq:RGEs1}
\end{align}
The RGEs of dimension six Wilson coefficients are the following 
\begin{equation}
	(16\pi^2
	)\frac{d\vec{C}^{[6]}_{X}}{d\log M}=
	-(16\pi^2
	)a_X\epsilon\vec{C}^{[6]}_{X}+\vec{C}^{[6]}_{Y}\tilde{\gamma}_{YX}+\dots \label{eq:dimsixRGE}
\end{equation}
where $a_X\epsilon$ is the mass dimension of the  bare coefficient of
$\mathcal{O}_X$ and $\tilde{\gamma}$ is the anomalous dimension matrix for dimension six operators.
In the limit $\epsilon\to 0$,
the term proportional to $\epsilon$ is irrelevant for the dimension six renormalization,
while it plays a crucial role in $\text{(dimension6)}^2$ to dimension eight mixing. Upon substitution, eq.~(\ref{eq:RGEs1}) becomes
\begin{align}
	(16\pi^2
	)\frac{d\vec{C}^{[8]}_{A}}{d\log M}=(16\pi^2
	)\bigg[-a_A\epsilon\vec{C}^{[8]}_A+\vec{C}^{[8]}_B\gamma_{BA}-(a_A-a_X-a_Y)\epsilon(\vec{C}^{[6]}_{X}\vec{C}^{[6]}_{Y}\hat{Z}_{XY,B} Z^{-1}_{BA})-\vec{C}^{[6]}_{X}\frac{d\hat{Z}_{XY,B}}{d\log M}\vec{C}^{[6]}_{Y}Z^{-1}_{BA}\bigg]\nonumber\\
	-\vec{C}^{[6]}_{W}\tilde{\gamma}_{WX}\hat{Z}_{XY,B}\vec{C}^{[6]}_{Y}Z^{-1}_{BA}	-\vec{C}^{[6]}_{X}\hat{Z}_{XY,B}\vec{C}^{[6]}_{W}\tilde{\gamma}_{WY}Z^{-1}_{BA} \nonumber \label{eq:RGEs2}
\end{align}
having defined $\gamma_{BA}\equiv-(16\pi^2
)\frac{d{Z}_{BC}}{d\log M}Z^{-1}_{CA}$, which is the anomalous dimension matrix of dimension eight operators. At one-loop we can replace $Z$ with the identity and neglect the second line of the above equation, since $\tilde{\gamma}$ and $\hat{Z}$ both appear at one loop at leading order. The product $\epsilon \hat{Z}$ is finite, and the RGEs in $d=4$ dimensions read
\begin{align}
	(16\pi^2)\frac{d\vec{C}^{[8]}_{A}}{d\log M}&=\vec{C}^{[8]}_B\gamma_{BA}-(16\pi^2)(a_A-a_X-a_Y)\vec{C}^{[6]}_{X}\vec{C}^{[6]}_{Y}\epsilon\hat{Z}_{XY,A} -(16\pi^2)\vec{C}^{[6]}_{X}\frac{d\hat{Z}_{XY,A}}{d\log M}\vec{C}^{[6]}_{Y}\\
	&\equiv \vec{C}^{[8]}_B\gamma_{BA}+\vec{C}^{[6]}_{X}\hat{\gamma}_{XY,A}\vec{C}^{[6]}_{Y}
	\label{eq:RGEs3}
\end{align}
The one-loop $\hat{\gamma}$ anomalous dimension matrix that mixes two dimension six operators into dimension eight is finally
\begin{equation}
	\hat{\gamma}_{XY,A}=(16\pi^2)\left[(a_X+a_Y-a_A)\epsilon\hat{Z}_{XY,A}-\frac{d\hat{Z}_{XY,A}}{d\log M}\right].
\end{equation}
The second term contribute to the mixing when renormalizable couplings appear in $\hat{Z}$, which carry an implicit dependence on the renormalization scale $M$. The beta functions of renormalized SM couplings for $\epsilon>0$ take the form $\beta_\epsilon (\{g,g',y\})=-\epsilon \{g,g',y\}+\beta(\{g,g',y\})$ and at one-loop 
\begin{equation} 
	-\frac{d\hat{Z}_{XY,A}}{d\log M}=-\frac{d\hat{Z}_{XY,A}}{d\{g,g',y\}}\beta_\epsilon (\{g,g',y\})= \epsilon \frac{d\hat{Z}_{XY,A}}{d\{g,g',y\}} \times \{g,g',y\} + \text{higher loops.}
\end{equation}

\subsubsection{$\mu\to \tau \times \tau \to e$ in SMEFT}\label{ssec:SMEFTrunningsub}

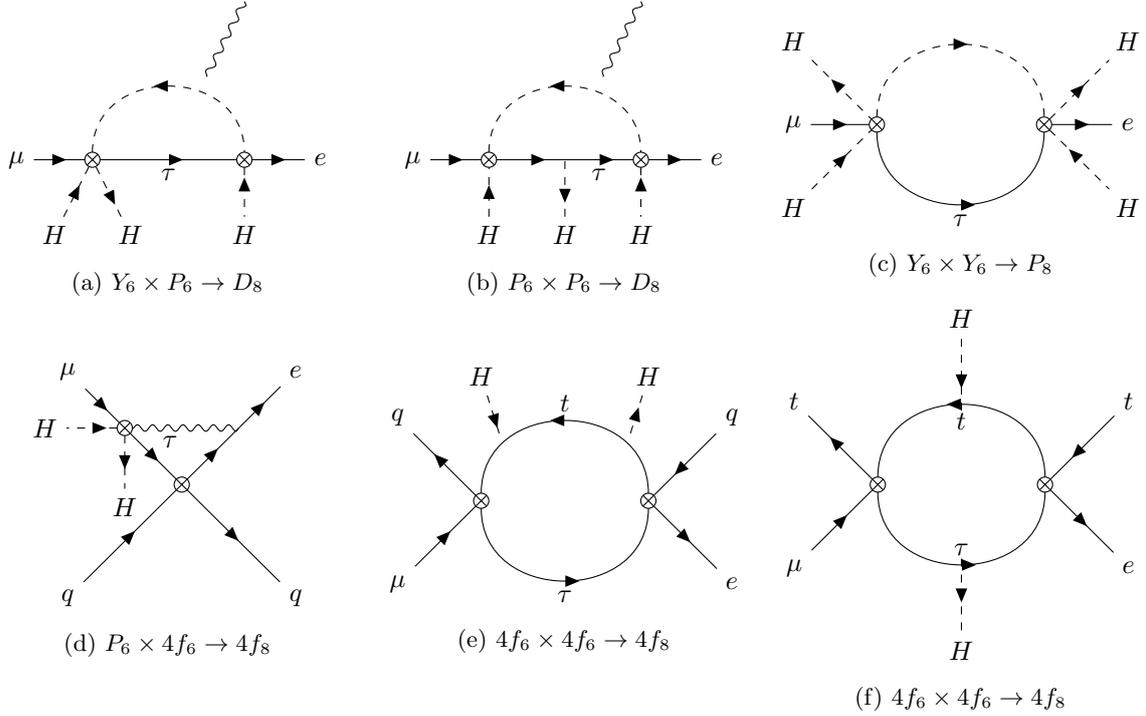
\begin{figure}[t]
	\centering
	\begin{subfigure}{.3\textwidth}
		\centering
		\begin{tikzpicture}
			\begin{feynman}[small]
				\vertex (mu) at (-1,0) {\(\mu\)};
				\vertex (e) at (3,0) {\(e\)};
				\vertex [style=crossed dot] (a)  at (0,0) {};
				\vertex [style=crossed dot] (b) at (2,0) {};
				\vertex (boson1) at (1.5, 1.1);
				\vertex (boson2) at (2, 2.1);
				\vertex (H2) at (-0.5,-1) {\(H\)};
				\vertex (H3) at (0.5,-1) {\(H\)};
				\vertex (H4) at (2,-1) {\(H\)};
				\diagram* [inline=(a.base)]{
					(H3) -- [anti charged scalar] (a),
					(boson1) -- [photon] (boson2),
					(mu) -- [fermion] (a),
					(b) -- [fermion] (e),
					(H2) -- [charged scalar] (a),
					(a) -- [anti charged scalar, half left] (b) -- [anti fermion, edge label=\(\tau\)] (a) ,
					(b) -- [anti charged scalar] (H4),
				};
			\end{feynman}
		\end{tikzpicture}
		\caption{$Y_6\times P_6\to D_8$}\label{subfig:1}
	\end{subfigure}
	\begin{subfigure}{.3\textwidth}
		\centering
		\begin{tikzpicture}
			\begin{feynman}[small]
				\vertex (mu) at (-1,0) {\(\mu\)};
				\vertex (e) at (3,0) {\(e\)};
				\vertex [style=crossed dot] (a)  at (0,0) {};
				\vertex [style=crossed dot] (b) at (2,0) {};
				\vertex (boson1) at (1.5, 1.1);
				\vertex (boson2) at (2, 2.1);
				\vertex (H2) at (0,-1) {\(H\)};
				\vertex (c) at (1,0);
				\vertex (H3) at (1,-1) {\(H\)};
				\vertex (H4) at (2,-1) {\(H\)};
				\diagram* [inline=(a.base)]{
					(H3) -- [anti charged scalar] (c),
					(boson1) -- [photon] (boson2),
					(mu) -- [fermion] (a),
					(b) -- [fermion] (e),
					(H2) -- [charged scalar] (a),
					(a) -- [anti charged scalar, half left] (b) -- [anti fermion, edge label=\(\tau\)] (c) -- [anti fermion] (a) ,
					(b) -- [anti charged scalar] (H4),
				};
			\end{feynman}
		\end{tikzpicture}
		\caption{$P_6\times P_6\to D_8$}\label{subfig:2}
	\end{subfigure}
	\begin{subfigure}{.3\textwidth}
		\centering
		\begin{tikzpicture}[scale=1.1]
			\begin{feynman}[small]
				\vertex (H1) at (-1,1) {\(H\)};
				\vertex (mu) at (-1,0) {\(\mu\)};
				\vertex (e) at (3,0) {\(e\)};
				\vertex [style=crossed dot] (a)  at (0,0) {};
				\vertex [style=crossed dot] (b) at (2,0) {};
				\vertex (H2) at (-1,-1) {\(H\)};
				\vertex (H3) at (3,1) {\(H\)};
				\vertex (H4) at (3,-1) {\(H\)};
				\diagram* [inline=(a.base)]{
					(H1) -- [anti charged scalar] (a),
					(mu) -- [fermion] (a),
					(b) -- [fermion] (e),
					(H2) -- [charged scalar] (a),
					(a) -- [charged scalar, half left] (b) -- [anti fermion, half left, edge label=\(\tau\)] (a),
					(b) -- [charged scalar] (H3),
					(b) -- [anti charged scalar] (H4),
				};
			\end{feynman}
		\end{tikzpicture}
		\caption{$Y_6\times Y_6\to P_8$}\label{subfig:3}
	\end{subfigure}\\
	\begin{subfigure}{.3\textwidth}
		\centering
		\begin{tikzpicture}[scale=1.5]
			\begin{feynman}[small]
				\vertex (a) at (-1,1) {\(\mu\)};
				\vertex  [style=crossed dot] (b) at (0,0) {};
				\vertex [style=crossed dot] (boson1) at (-0.5, 0.5) {};
				\vertex (boson2) at (0.5, 0.5);
				\vertex (H1) at (-0.5, -0.2) {\(H\)};
				\vertex (H2) at (-1.2, 0.5) {\(H\)};
				\vertex  (c) at (1,1) {\(e\)} ;
				\vertex (d) at (-1,-1) {\(q\)};
				\vertex (e) at (1,-1){\(q\)};
				
				\diagram* {
					(a)  --  [fermion] (boson1) -- [fermion, edge label=\(\tau\)] (b)  -- [fermion] (boson2) -- [fermion] (c),
					(boson1) -- [photon] (boson2),
					(boson1) -- [charged scalar] (H1),
					(boson1) -- [anti charged scalar] (H2),
					(d) --  [fermion] (b) -- [fermion] (e),
				};
			\end{feynman}
		\end{tikzpicture}
		\caption{$P_6\times 4f_6\to 4f_8$}\label{subfig:4}
	\end{subfigure}
	\begin{subfigure}{.3\textwidth}
		\centering
		\begin{tikzpicture}[scale=1.1]
			\begin{feynman}[small]
				\vertex (fin) at (-1,1) {\(q\)};
				\vertex [style=crossed dot] (a) [label=0:] at (0,0) {};
				\vertex [style=crossed dot] (b) [label=180:]at (2,0) {};
				\vertex (mu) at (-1,-1) {\(\mu\)};
				\vertex  (c) at (0.25, 0.70) {};
				\vertex (H1)  at (0,1.5) {\(H\)};
				\vertex (d) at (1.75, 0.70) {};
				\vertex (H2) at (2, 1.5) {\(H\)};
				\vertex (fout) at (3,1) {\(q\)};
				\vertex (e) at (3,-1) {\(e\)};
				\diagram* {
					(fin) -- [anti fermion] (a),
					(mu) -- [fermion] (a),
					(a) -- [anti fermion, half left, edge label=\(t\)] (b) -- [anti fermion, half left, edge label=\(\tau\)] (a),
					(b) -- [fermion] (e),
					(b) -- [anti fermion] (fout),
					(c) -- [anti charged scalar] (H1),
					(d) -- [charged scalar] (H2),
				};
			\end{feynman}
		\end{tikzpicture}
		\caption{$4f_6\times 4f_6\to 4f_8$}\label{subfig:5}
	\end{subfigure}
	\begin{subfigure}{.3\textwidth}
	\centering
	\begin{tikzpicture}[scale=1.1]
		\begin{feynman}[small]
			\vertex (fin) at (-1,1) {\(t\)};
			\vertex [style=crossed dot] (a) [label=0:] at (0,0) {};
			\vertex [style=crossed dot] (b) [label=180:]at (2,0) {};
			\vertex (mu) at (-1,-1) {\(\mu\)};
			\vertex  (c) at (1, 0.78) {};
			\vertex (H1)  at (1,2) {\(H\)};
			\vertex (d) at (1, -0.78) {};
			\vertex (H2) at (1, -2) {\(H\)};
			\vertex (fout) at (3,1) {\(t\)};
			\vertex (e) at (3,-1) {\(e\)};
			\diagram* {
				(fin) -- [anti fermion] (a),
				(mu) -- [fermion] (a),
				(a) -- [anti fermion, half left, edge label'=\(t\)] (b) -- [anti fermion, half left, edge label'=\(\tau\)] (a),
				(b) -- [fermion] (e),
				(b) -- [anti fermion] (fout),
				(c) -- [anti charged scalar] (H1),
				(d) -- [charged scalar] (H2),
			};
		\end{feynman}
	\end{tikzpicture}
	\caption{$4f_6\times 4f_6\to 4f_8$}\label{subfig:6}
\end{subfigure}
	\caption{Classes of divergent diagrams that give observable contributions to $\mu\to e$ processes, as identified in Table \ref{tab:RunningEstimate}.}\label{fig:diagramsRunning}
\end{figure}

We calculate the divergent part of one-loop diagrams with the product of  $\mu\to \tau \times \tau\to e$ operator insertions, which, according to the estimates summarized in Table \ref{tab:RunningEstimate}, give potentially detectable contributions to $\mu\to e$ observables in the dimension eight running.
We work in SMEFT and unbroken SU(2), where all SM particles are taken massless, including the Higgs doublet.
The diagrams have been drawn by hand and were also generated with a code based on FeynArts \cite{Hahn:2000kx} and FeynRules \cite{Alloul:2013bka}. In most cases\footnote{The exception  is the $\mu e \tau \tau$ tensors,
 but the leading contribution to these is  from tree-level matching onto the low energy EFT, which is discussed  the next section.},
   the dimension eight operators to which $\mu \to e $ observables are sensitive do not contain $\tau$ external legs, so we here consider diagrams with a virtual $\tau$ line connecting two dimension six SMEFT operators.
   We are interested in one-particle-irreducible divergent diagrams (which restrict the number of internal propagators) that can generate the dimension eight operators of section \ref{ssec:EFT} (which constrain the external legs),
   and also  in some one-particle-reducible divergent diagrams that reproduce the contribution of the dimension six correction in the EOM, as discussed in section \ref{ssec:EoM}.
   Yukawa couplings smaller than $y_\tau\sim 10^{-2}$ are neglected, because they lead to $\mu\to e$ coefficients below experimental sensitivity, assuming dimension six WC $C^{[6]}\lesssim v^2/\Lambda_{\rm NP}^2$ and $\Lambda_{\rm NP}=4$ T$e$V.
   However, the estimates of section \ref{ssec:estimates} select diagrams that only involve top Yukawas $y_t$ and single insertions of $y_\tau$, while the bottom and charm Yukawas $y_b$,$y_c$ do not appear.  

 In Figure \ref{fig:diagramsRunning} we show the ``classes" of diagrams listed in Table \ref{tab:RunningEstimate}, that were estimated to be within $\mu\to e$ experimental sensitivity. Each class is described below. The divergences were calculated both by hand and with an in-house developed Mathematica program, making use of the Feynman Rules listed in Appendix \ref{appendix:FeynRules}.

\begin{itemize}

\item \textbf{Figure \ref{subfig:1}: $Y_6\times P_6\to D_8$}\\
  The penguin operators of eq.s~(\ref{eq:penguinsdim6a})-(\ref{eq:penguinsdim6b}) can be combined  with the Yukawa operators of eq. (\ref{eq:yukdim6}).
 The chirality  flips on  the lepton line, so attaching a gauge boson potentially generates the $\mu\to e$ dipoles of eq.~(\ref{eq:dim8dipoles}).
  The gauge bosons can be inserted on the internal Higgs and lepton lines or can come out of penguin operators, while the three external Higgs can be permuted in several ways among the dimension six vertices.
  Also,  in the diagram depicted, the Yukawa operator is $\mu\to \tau$ and the penguin is $\tau\to e$,  but the two vertices can be exchanged:
  for instance, in the case of external left-handed electrons, the possible operator combinations are:
  $\mathcal{O}^{\tau \mu}_{He}\times \mathcal{O}^{e\tau}_{eH}$,
  $\mathcal{O}^{\tau \mu}_{eH}\times \mathcal{O}^{(1)e\tau}_{Hl}$,
  $\mathcal{O}^{\tau \mu}_{ eH}\times \mathcal{O}^{(3)e\tau}_{Hl}$.
  We find that these anomalous dimensions vanish.
  This is consistent with  the dimension six version of this calculation,
  where { neither penguin operators dressed with renormalizable Yukawa couplings,
  nor  $\mathcal{O}_{eH}$ dressed with a gauge loop,  mix} into the  dimension six dipoles \cite{Manohar2}. Note that in broken SU(2) and unitary gauge, dimension six penguins and Yukawas give Feynman rules that look like SM renormalisable interactions. By analogy with the SM, we expect them to not generate divergent non-renormalisable dipoles. 
\item \textbf{Figure \ref{subfig:2}: $P_6\times P_6\to D_8$}

  The diagrams feature double insertions of penguin operators - see eq.s~(\ref{eq:penguinsdim6a})-(\ref{eq:penguinsdim6b}).
 The two vertices couple to vector currents of leptons, so to mix into the $\mu\to e$ dipoles, the chirality flip is achieved by attaching a Higgs to the $\tau$ virtual line.
 The contribution is estimated to lie within experimental sensitivity, because the generated $\mu\to e$ dipole coefficient is enhanced by the ratio $y_\tau/y_\mu$ due to the Yukawa couplings in the dipole operator  definitions in eq.~(\ref{eq:dim8dipoles}).
 The gauge bosons can be attached to the Higgs and $\tau$ in the loop, or can belong to one of the penguin vertices.
 Furthermore, all possible permutations of the external Higgses are taken into account.
 The operator pairs are $\mathcal{O}_{He} \times \mathcal{O}^{(1),(3)}_{H \ell}$, where the  $\tau\to e$ LFV can be mediated by either right-handed or left-handed penguins, depending on the chirality of the external legs.  As the previous case, the mixing into the $\mu \to e$ dipole is found to vanish.

  In addition to the 1PI diagrams of Figure \ref{subfig:2}, dimension six terms in the EOM contribute to the mixing.
  Loop diagrams where the Higgs leg of a penguin operator closes into the $\tau$ line via a Yukawa interaction renormalize the redundant operator $(D^2\bar{\ell}_\tau) H e_i$ { (see  Figure \ref{subfig:boxop})}.
  When the divergence is projected onto the on-shell basis, the penguin correction to the EOM gives additional $P_6\times P_6\to D_8$ mixing. However, the combination of SMEFT $\mu\to e$ dipoles that is generated is orthogonal to the $\gamma$ dipole and does not contribute to low energy $\mu\to e$ observables.
   This is also apparent in  considering non-1PI diagrams (see section \ref{ssec:EoM}) where a penguin operator is inserted in the $\tau$ line of $ D^2\bar{\ell}_\tau H e_i$; the amplitude is local and reproduces the EOM result when the external gauge boson belongs to the penguin vertex. In broken SU(2), penguins give flavour changing (and correct the flavour diagonal) couplings with the $Z$, but leave QED interactions untarnished.

	\item \textbf{Figure \ref{subfig:3}: $Y_6\times Y_6\to P_8$}\\
	  In this class of diagrams the loop is closed with  Higgs exchange between two Yukawa operators. The superficial degree of divergence is 1, and the divergence is linear in momentum. With four external Higgses, it mixes into the dimension eight $\mu\to e$ penguin operators of eq. (\ref{eq:dim8peng}). For right-handed leptons the inserted operators are $\mathcal{O}^{\tau\mu}_{eH}\times \mathcal{O}^{*\tau e}_{eH}$, while $\mathcal{O}^{*\mu\tau}_{eH}\times \mathcal{O}^{e\tau}_{eH}$ gives mixing into left-handed penguins.

\item \textbf{Figure \ref{subfig:4}: $4f_6\times P_6\to 4f_8$}\\
 Two-lepton two-quark $\tau\to l$ operators can mix into $\mu\to e$ dimension eight four fermion operators by inserting a penguin in the tau line and closing the loop with a gauge boson.
 Only two-lepton two quark operators are considered because they contribute to $\mu\to e$ conversion (while tensors with heavy quarks contribute to $\mu\to e\gamma$), which is the process with the best upcoming sensitivity to operator coefficients.
 The gauge boson is attached to the other fermion lines in every possible way, and the diagram shows just one example.
 As discussed in section \ref{ssec:EoM}, we also include dimension six corrections to the EOM or, equivalently, non-1PI diagrams where the loop of Figure \ref{fig:offdiagkinetic} dresses one of the lepton lines. These diagrams are analogous to fermion wave function renormalization and are pure-gauge, i.e $\propto \xi$ in the $R_\xi$ gauge; 
 to avoid calculating wave function-like diagrams, the calculation is done for $\xi=0$, commonly known as Landau gauge.
 In Table \ref{tab:4f6penguins} we summarize the $\mu\to e$ dimension eight operators generated by the product of  $\tau\to l$ penguins with four fermion operators. 

      \item \textbf{Figure \ref{subfig:5}-\ref{subfig:6}: $4f_6\times 4f_6\to 4f_8$}
In the last two diagrams, pairs of two-lepton two-quark dimension six operators are connected through a fermion loop, where two Higgs legs are inserted. With the exception of dimension eight tensor with tops, $\mu\to e$ observables are sensitive to the resulting dimension eight coefficients only if the Higgs are attached to a top internal line.
In the case of tensors with tops, the better sensitivity allows for the topology of Figure \ref{subfig:6}, where a $\tau$ Yukawa is present. In Table \ref{tab:4f6squared} we list the dimension eight operators that are generated for every pair of dimension six four fermion operators.
\end{itemize}
\begin{table}[th]
	\centering
	\begin{tabular}{c|ccccccccc}
		& $C^{(1)}_{\ell q}$ & $C^{(3)}_{\ell q}$ & $C_{\ell u}$ & $C_{eq}$ & $C_{eu}$ & $C_{\ell edq}, C^*_{\ell edq}$ & $C^{(1)}_{\ell equ},C^{*(1)}_{\ell equ}$ & $C^{(3)}_{\ell equ},C^{*(3)}_{\ell equ}$\\  [1 mm]
		\hline
		$C^{(1)}_{\ell q}$  & $C^{(1),(4)}_{\ell^2q^2H^2}$ & $C^{(2),(3),(5)}_{\ell^2q^2H^2}$ & $\cross$ & $\cross$ & $C^{(3),(4)}_{\ell equH^2}$&  \makecell{$C^{(1),(2)}_{\ell edqH^2}$\\  $C^{*(1),(2)}_{\ell edqH^2}$} & \makecell{$C^{(1),(2),(3),(4)}_{\ell equH^2}$\\ $C^{*(1),(2),(3),(4)}_{\ell equH^2}$} & \makecell{$C^{(1),(2),(3),(4)}_{\ell equH^2}$\\ $C^{*(1),(2),(3),(4)}_{\ell equH^2}$}\\ [1 mm]
		\hline
		$C^{(3)}_{\ell q}$ & & $C^{(1),(2)}_{\ell^2q^2H^2}$ & $\cross$ & $\cross$ & $C^{(3),(4)}_{\ell equH^2}$ & \makecell{$C^{(1),(2)}_{\ell edqH^2}$\\ $C^{*(1),(2)}_{\ell edqH^2}$} & \makecell{$C^{(1),(2),(3),(4)}_{\ell equH^2}$\\ $C^{*(1),(2),(3),(4)}_{\ell equH^2}$} & \makecell{$C^{(1),(2),(3),(4)}_{\ell equH^2}$\\ $C^{*(1),(2),(3),(4)}_{\ell equH^2}$}\\ [1 mm]
		\hline
		$C_{\ell u}$ & & & $C^{(1)}_{\ell^2u^2H^2}$ &  $C^{(3),(4)}_{\ell equH^2}$ & $\cross$ & $\cross$ & \makecell{$C^{(1),(3)}_{\ell equH^2}$\\ $C^{*(1),(3)}_{\ell equH^2}$} & \makecell{$C^{(1),(3)}_{\ell equH^2}$\\ $C^{*(1),(3)}_{\ell equH^2}$}\\ [1 mm]
		\hline
		$C_{eq}$ & & & & $C^{(1),(2)}_{e^2q^2H^2}$ & $\cross$ &  \makecell{$C^{(1),(2)}_{\ell edqH^2}$\\ $C^{*(1),(2)}_{\ell edqH^2}$} & \makecell{$C^{(1),(2),(3),(4)}_{\ell equH^2}$\\ $C^{*(1),(2),(3),(4)}_{\ell equH^2}$} & \makecell{$C^{(1),(2),(3),(4)}_{\ell equH^2}$\\ $C^{*(1),(2),(3),(4)}_{\ell equH^2}$}\\ [1 mm]
		\hline
		$C_{eu}$ & & & & & $C_{e^2u^2H^2}$ & $\cross$ & \makecell{$C^{(1),(3)}_{\ell equH^2}$\\ $C^{*(1),(3)}_{\ell equH^2}$} & \makecell{$C^{(1),(3)}_{\ell equH^2}$\\
			$C^{*(1),(3)}_{\ell equH^2}$}\\ [1 mm]
		\hline
		$C_{\ell edq}, C^*_{\ell edq}$ & & & & & & \makecell{$C_{e^2d^2H^2}$\\ $C^{(1),(2)}_{\ell^2d^2H^2}$} & $\cross$ & $\cross$\\ [1 mm]
		$C^{(1)}_{\ell equ},C^{*(1)}_{\ell equ}$ & & & & & & & \makecell{$C_{e^2u^2H^2}$\\ $C^{(1)}_{e^2q^2H^2}$\\ $C^{(1),(2)}_{\ell^2u^2H^2}$ \\  $C^{(1),(3)}_{\ell^2q^2H^2}$} & \makecell{$C_{e^2u^2H^2}$\\ $C^{(1)}_{e^2q^2H^2}$\\ $C^{(1),(2)}_{\ell^2u^2H^2}$ \\  $C^{(1),(3)}_{\ell^2q^2H^2}$} \\ [1 mm]
		\hline
		$C^{(3)}_{\ell equ},C^{*(3)}_{\ell equ}$ & & & & & & & & \makecell{$C_{e^2u^2H^2}$\\ $C^{(1)}_{e^2q^2H^2}$\\ $C^{(1),(2)}_{\ell^2u^2H^2}$ \\  $C^{(1),(3)}_{\ell^2q^2H^2}$}\\ [1 mm]
\end{tabular}
\caption{Dimension eight operators generated through the diagrams of Figure \ref{subfig:5} and \ref{subfig:6} with pairs of two-lepton two-quark operators, $4f_6\times 4f_6$. Most of dimension eight coefficients are proportional to $y_t^2$, with the exception of $\mathcal{O}_{\ell u}\times \mathcal{O}_{eq}$, $\mathcal{O}_{\ell q}\times \mathcal{O}_{eu}$ mixing into the tensors $\mathcal{O}^{(3),(4)}_{\ell equH^2}$, where the Yukawa couplings $y_\tau y_t$ multiply the coefficient.
	\label{tab:4f6squared}}
\end{table}
\begin{table}[th]
	\centering
	\begin{tabular}{c|cccccccc}
		
		& $C^{(1)}_{\ell q}$ & $C^{(3)}_{\ell q}$ & $C_{\ell u}$ & $C_{\ell d}$ &  $C_{\ell edq}, C^*_{\ell edq}$ & $C^{(1)}_{\ell equ},C^{*(1)}_{\ell equ}$ & $C^{(3)}_{\ell equ},C^{*(3)}_{\ell equ}$\\
		\hline
		
	    $C_{H\ell(1)}$ & $C^{(1),(4)}_{\ell^2q^2H^2}$& $C^{(2),(3),(5)}_{\ell^2q^2H^2}$& $C^{(1)}_{\ell^2u^2H^2}$ & $C^{(1)}_{\ell^2d^2H^2}$  &  \makecell{$C^{(1)}_{\ell edqH^2}$\\ $C^{*(1)}_{\ell edqH^2}$} & \makecell{$C^{(1),(3),(4)}_{\ell equH^2}$\\ $C^{*(1),(3),(4)}_{\ell equH^2}$}&  \makecell{$C^{(1),(2),(3),(4)}_{\ell equH^2}$\\ $C^{*(1),(2),(3),(4)}_{\ell equH^2}$} \\
	    \hline
	    $C_{H\ell(3)}$ & $C^{(2),(3)}_{\ell^2q^2H^2}$& $C^{(1),(4),(5)}_{\ell^2q^2H^2}$& $C^{(2)}_{\ell^2u^2H^2}$ & $C^{(2)}_{\ell^2d^2H^2}$  &  \makecell{$C^{(2)}_{\ell edqH^2}$\\ $C^{*(2)}_{\ell edqH^2}$} & \makecell{$C^{(2),(3),(4)}_{\ell equH^2}$\\ $C^{*(2),(3),(4)}_{\ell equH^2}$}&  \makecell{$C^{(1),(2),(3),(4)}_{\ell equH^2}$\\ $C^{*(1),(2),(3),(4)}_{\ell equH^2}$} \\
\end{tabular}\\ \vspace{1 cm}
\begin{tabular}{c|ccccccc}
		 & $C_{eq}$ & $C_{eu}$ & $C_{ed}$ &  $C_{\ell edq}, C^*_{\ell edq}$ & $C^{(1)}_{\ell equ},C^{*(1)}_{\ell equ}$ & $C^{(3)}_{\ell equ},C^{*(3)}_{\ell equ}$\\
		\hline
	    $C_{He}$& $C^{(1),(2)}_{e^2q^2H^2}$& $C_{e^2u^2H^2}$& $C_{e^2d^2H^2}$ &  \makecell{$C^{(1),(2)}_{\ell edqH^2}$\\ $C^{*(1),(2)}_{\ell edqH^2}$} & \makecell{$C^{(1),(2),(3),(4)}_{\ell equH^2}$\\ $C^{*(1),(2),(3),(4)}_{\ell equH^2}$}&  \makecell{$C^{(1),(2),(3),(4)}_{\ell equH^2}$\\ $C^{*(1),(2),(3),(4)}_{\ell equH^2}$} \\
    \end{tabular}
	\caption{Dimension eight operators generated via the diagrams of Figure \ref{subfig:4} with pairs of two-lepton two-quark $4f_6$ and penguins $P_6$ 
	\label{tab:4f6penguins}}
\end{table}
The complete  anomalous dimensions for the above classes of diagrams can be found in Appendix \ref{appendix:AnomalousDimensions}.

We discuss the example of a pair of  dimension six $\tl$ Yukawa operators mixing into the $\mu\to e$ dimension eight penguins, depicted in  the representative diagram of Figure \ref{subfig:3}.  The counterterms that renormalize the divergences are the following
\begin{align}
	\left(C^{\tau \mu}_{eH}\hat{Z}C^{*\tau e}_{eH}\right)^{e\mu}_{e^2H^4D}&=-\frac{C^{\tau \mu}_{eH}C^{*\tau e}_{eH}}{32\pi^2\epsilon}\qquad
	\left(C^{\tau \mu}_{eH}\hat{Z}C^{*\tau e}_{eH}\right)^{e\mu}_{ve^2H^4D}=-\frac{3C^{\tau \mu}_{eH}C^{*\tau e}_{eH}}{32\pi^2\epsilon}\nonumber\\
	\left(C^{* \mu\tau}_{eH}\hat{Z}C^{e\tau }_{eH}\right)^{(1)e\mu}_{\ell^2H^4D}&=\frac{C^{* \mu\tau}_{eH}C^{e\tau }_{eH}}{64\pi^2\epsilon}\qquad
	\left(C^{* \mu\tau}_{eH}\hat{Z}C^{e\tau }_{eH}\right)^{(1)e\mu}_{v\ell^2H^4D}=-\frac{C^{* \mu\tau}_{eH}C^{e\tau }_{eH}}{64\pi^2\epsilon}\nonumber \\
	\left(C^{* \mu\tau}_{eH}\hat{Z}C^{e\tau }_{eH}\right)^{(2)e\mu}_{\ell^2H^4D}&=\frac{C^{* \mu\tau}_{eH}C^{e\tau }_{eH}}{128\pi^2\epsilon}\qquad
	\left(C^{* \mu\tau}_{eH}\hat{Z}C^{e\tau }_{eH}\right)^{(2)e\mu}_{v\ell^2H^4D}=-\frac{C^{* \mu\tau}_{eH}C^{e\tau }_{eH}}{16\pi^2\epsilon}\nonumber \\
	\left(C^{* \mu\tau}_{eH}\hat{Z}C^{e\tau }_{eH}\right)^{(4)e\mu}_{\ell^2H^4D}&=\frac{C^{* \mu\tau}_{eH}C^{e\tau }_{eH}}{128\pi^2\epsilon} \label{eq:yuk2Pengcount}
\end{align}
where the subscript in the brackets label the corresponding dimension eight operators. The operator\\ $\mathcal{O}^{(4)e\mu}_{\ell^2H^4 D}=\epsilon^{IJK}(\bar{\ell}_e \tau^I\gamma^\alpha \ell_\mu)(H^\dagger\tau^J H)D_{\alpha}(H^\dagger\tau^K H)$ is not in the list of section \ref{ssec:EFT} because it does not contribute to low energy $\mu\to e$ observables, although it appears as a counterterm.
Furthermore, the following redundant operators are radiatively generated in our off-shell calculation
\begin{equation}
	\mathcal{O}^{e\mu}_{ve^2H^4D}=i(\bar{e}\overset{\leftrightarrow}{\slashed{D}}\mu)(H^\dagger H)^2\equiv i(\bar{e}\slashed{D}\mu)(H^\dagger H)^2-i(\slashed{D}\bar{e}\mu)(H^\dagger H)^2\label{eq:offshelloperator1}
\end{equation}
\begin{equation}
	\mathcal{O}^{(1)e\mu}_{v\ell^2H^4D}=i(\bar{\ell}_e\overset{\leftrightarrow}{\slashed{D}}\ell_\mu)(H^\dagger H)^2\label{eq:offshelloperator2}
\end{equation}
\begin{equation}
	\mathcal{O}^{(2)e\mu}_{v\ell^2H^4D}=i(\bar{\ell}_{eI}\overset{\leftrightarrow}{\slashed{D}} \ell_{\mu J})(H_I H^\dagger_J)(H^\dagger H)\label{eq:offshelloperator3}.
\end{equation}
These are related to the physical/on-shell basis as follows 
\begin{equation}
	\mathcal{O}^{e\mu}_{ve^2H^4D}=\mathcal{O}^{e\mu}_v+[y^*_e]_{i\mu}\mathcal{O}^{*i e}_{leH^5}+[y_e]_{ie}\mathcal{O}^{i \mu}_{leH^5}
\end{equation}
\begin{equation}
	\mathcal{O}^{(1)e\mu}_{v\ell^2H^4D}=\mathcal{O}^{(1)e\mu}_v+[y_e]_{\mu i}\mathcal{O}^{e i}_{leH^5}+[y^*_e]_{ei}\mathcal{O}^{*\mu i}_{leH^5}
\end{equation}
\begin{equation}
	\mathcal{O}^{(2)e\mu}_{v\ell^2H^4D}=\mathcal{O}^{(2)e\mu}_v+[y_e]_{\mu i}\mathcal{O}^{e i}_{leH^5}+[y^*_e]_{ei}\mathcal{O}^{*\mu i}_{leH^5}
\end{equation}
where $\mathcal{O}^{ij}_{\ell eH^5}=(\bar{\ell}_i H e_j)(H^\dagger H)^2$,
and each of $\mathcal{O}^{e\mu}_v$, $\mathcal{O}^{(1)e\mu}_v$, $\mathcal{O}^{(2)e\mu}_v$  vanishes,
when the renormalizable EOM on singlet and doublet leptons  $i(\slashed{D}\mu)-[y^*_e]_{i\mu}(H^\dagger \ell_i)=0$ , $i(\slashed{D}\ell_\mu)-[y_e]_{\mu i}(H e_i)=0$ are satisfied. The off-shell counterterms are on-shell equivalent to  $[y^*_e]_{i\mu}\mathcal{O}^{*i e}_{\ell eH^5}+[y_e]_{ie}\mathcal{O}^{i \mu}_{\ell eH^5}$, which is beyond $\mu\to e$ experimental reach. The resulting RGEs are obtained from eq.~(\ref{eq:yuk2Pengcount}) and (\ref{eq:RGEs3}), and read 
\begin{align}
	16\pi^2 \dot{C}^{e \mu }_{e^2H^4D}=-C^{\tau\mu}_{eH}C^{*\tau e}_{eH}
\end{align}
\begin{align}
	16\pi^2 \dot{C}^{(1)e \mu }_{\ell^2H^4D}=\frac{1}{2}C^{*\mu\tau }_{eH}C^{e\tau}_{eH}\qquad 	16\pi^2 \dot{C}^{(2)e \mu }_{\ell^2H^4D}=\frac{1}{4}C^{*\mu\tau }_{eH}C^{e\tau}_{eH}
\end{align}
where the dot on the dimension eight coefficients corresponds to $d/d\log M$.

\subsection{Matching SMEFT onto the low energy EFT}\label{ssec:Matching}

\begin{figure}[th]
	\centering
	\begin{subfigure}{.3\textwidth}
		\centering
\begin{tikzpicture}[scale=0.8]
	\begin{feynman}[small]
		\vertex (a) at (-1,1){\(\mu\)};
		\vertex [style=crossed dot] (b) at (0,0) {};
		\vertex (c) at (-1,-1) {\(\tau\)};
		\vertex [style=crossed dot] (d) at (2,0) {};
		\vertex (e) at (3,1) {\(e \)};
		\vertex (f) at (3,-1) {\(\tau\)};

		\diagram* {
			(a)  --  [fermion] (b)  -- [fermion] (c),
			(b) --  [photon, edge label=\(Z\)] (d),
			(d) -- [fermion] (e),
			(d) -- [anti fermion] (f),
		};
	\end{feynman}
\end{tikzpicture}
		\caption{}\label{subfig:matching1}
	\end{subfigure}
	\begin{subfigure}{.3\textwidth}
	\centering
\begin{tikzpicture}[scale=0.8]
	\begin{feynman}[small]
		\vertex (a) at (-1,1){\(\mu\)};
		\vertex [style=crossed dot] (b) at (0,0) {};
		\vertex (c) at (-1,-1) {\(\tau\)};
		\vertex [style=crossed dot] (d) at (2,0) {};
		\vertex (e) at (3,1) {\(e \)};
		\vertex (f) at (3,-1) {\(\tau\)};

		\diagram* {
			(a)  --  [fermion] (b)  -- [fermion] (c),
			(b) --  [scalar, edge label=\(h\)] (d),
			(d) -- [fermion] (e),
			(d) -- [anti fermion] (f),
		};
	\end{feynman}
\end{tikzpicture}
	\caption{}\label{subfig:matching2}
\end{subfigure}
\begin{subfigure}{.3\textwidth}
	\centering
\begin{tikzpicture}[scale=1.3]
	\begin{feynman}[small]
		\vertex (mu) at (-1,0) {\(\mu\)};
		\vertex (e) at (3,0) {\(e\)};
		\vertex[style=crossed dot] (a)  at (0,0) {};
		\vertex[style=crossed dot] (b) at (2,0) {};
		\vertex (c) at (1,0) ;
		\vertex (loop1) at (1,0.5);
		\vertex (loop2) at (1,1.6);
		\vertex (photon) at (0.7,0.6);
		\vertex (scalar) at (1.3,0.6);
		\vertex (photon2) at (1, 2.6) {\(\gamma\)};
		\diagram* [inline=(a.base)]{
			(mu) -- [fermion] (a),
			(b) -- [fermion] (e),
			(a) -- [fermion, edge label'=\(\tau\)] (b),
			(loop1) -- [fermion, insertion=0.75, half left] (loop2) -- [fermion, half left, edge label'=\(t\)] (loop1),
			(a) -- [photon, edge label=\(Z\)] (photon),
			(b) -- [scalar, edge label'=\(h\)] (scalar),
			(loop2) -- [photon] (photon2),
		};
	\end{feynman}
\end{tikzpicture}
	\caption{}\label{subfig:matchingBarzeet}
\end{subfigure}\qquad
\begin{subfigure}{.3\textwidth}
	\centering
	\begin{tikzpicture}[scale=1.3]
		\begin{feynman}[small]
			\vertex (mu) at (-1,0) {\(\mu\)};
			\vertex (e) at (3,0) {\(e\)};
			\vertex[style=crossed dot] (a)  at (0,0) {};
			\vertex[style=crossed dot] (b) at (2,0) {};
			\vertex (c) at (1,0) ;
			\vertex (loop1) at (1,0.5);
			\vertex (loop2) at (1,1.6);
			\vertex (photon) at (0.7,0.6);
			\vertex (scalar) at (1.3,0.6);
			\vertex (photon2) at (1, 2.6) {\(\gamma\)};
			\diagram* [inline=(a.base)]{
				(mu) -- [fermion] (a),
				(b) -- [fermion] (e),
				(a) -- [fermion, edge label'=\(\tau\)] (b),
				(loop1) -- [photon, half left] (loop2) -- [photon, half left, edge label'=\(W\)] (loop1),
				(a) -- [photon, edge label=\(Z\)] (photon),
				(b) -- [scalar, edge label'=\(h\)] (scalar),
				(loop2) -- [photon] (photon2),
			};
		\end{feynman}
	\end{tikzpicture}
	\caption{}\label{subfig:matchingBarzeeW}
\end{subfigure}
\caption{Diagrams matching  pairs of dimension six $\tau\to l$ SMEFT operators
  onto low energy $\mu\to e$ operators. \label{fig:matchingdiagrams}}
\end{figure}
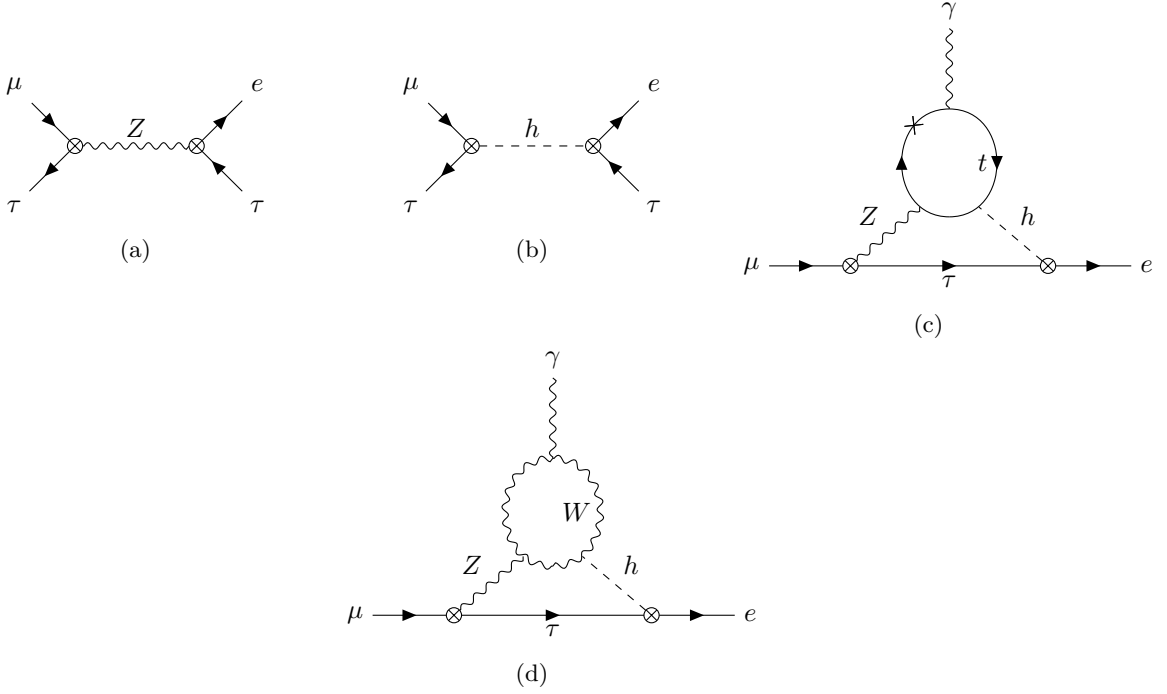

In Table \ref{tab:MatchingEstimate} of Section \ref{sec:Operators}, we identified the relevant matching contributions to low energy $\me$ interactions from the double insertion of $\mu\to \tau$ $ \times$ $\tau\to e$  dimension six SMEFT operators.
{ At the matching scale $m_W$, the electroweak symmetry is spontaneously broken by the Higgs  VEV,  and the $h,Z, W$ and $t$ are removed from the low energy EFT.}
{ The matching is performed by identifying the matrix elements
  of  a $\mu\to e$ process  calculated  in the theories above and below the matching scale,  with the electroweak symmetry  broken in both theories.}
As a result,  products of $\tl$  SMEFT operators can match onto $\mu\to e$ three and four point functions.
The interesting diagrams are illustrated in Figure \ref{fig:matchingdiagrams}.
When the Higgs doublet acquires a VEV,
Yukawa operators contribute to the mass matrix
\begin{equation}
	m_{e_i} \delta_{ij}=v\left([y_e]_{ij}-C^{ij}_{eH}\right)
\label{me}
\end{equation} 
and the $h$ couplings
\begin{equation}
	-\frac{h}{\sqrt{2}}\bar{e}_iP_Re_j\left([y_e]_{ij}-3C^{ij}_{eH}\right)+\text{h.c}=-\frac{h}{\sqrt{2}}\bar{e}_iP_Re_j\left(\frac{	m_{e_i} \delta_{ij}}{v}-2C^{ij}_{eH}\right)+\text{h.c} \label{eq:LFVhiggs}
\end{equation}
of charged leptons with a different prefactor,
such that  $h$ acquires LFV couplings in the lepton mass eigenstate basis.

The two-loop Barr-Zee diagrams (Figure \ref{subfig:matchingBarzeet}-\ref{subfig:matchingBarzeeW}) match  to the dipole at tree level in the low energy EFT.  The lepton line is connected via  $Z$ and $h$ exchange to a top or $W$ loop,
where the $Z$ and $h$  respectively couple to the lepton line  via a penguin and an off-diagonal Yukawa operator.
{ Such diagrams can be significant\cite{Bjorken:1977br} (despite the two-loop suppression), because they  are not suppressed by  small Yukawa couplings.
  We estimate these diagrams from the results of \cite{Chang:1993kw}, who calculated the Barr-Zee diagrams  in the two Higgs Doublet Model
  (2HDM) with LFV couplings, where they provide the leading contribution
  to $\mu\to e \gamma$ (because the diagrams are not suppressed by $y_\mu$).
  In the 2HDM results of \cite{Chang:1993kw}, the $Z$-diagrams  are suppressed (relative to $\gamma$ diagrams) because
  the $C$-even dipole moment only couples the $Z$ to the vector current of leptons,
  so there  is a suppression  of  $(1-{ 4}\sin^2 \theta_W) \lsim 0.03 $.
  However in our case, the $Z$-lepton vertex  is a penguin operator  with a flavour-changing coefficient  that we wish to constrain, and so  does not suffer from such SM factors. } The estimated contributions to the dipole coefficients are \cite{Chang:1993kw} :
\begin{align}
	C_{D,R}&\simeq \frac{9e\alpha_e}{64\pi^3}\frac{v}{m_\mu}\left[C^{\tau \mu}_{He}C^{e\tau}_{eH}+\left(C^{(1)e\tau }_{H\ell}+C^{(3)e\tau }_{H\ell}\right)C^{\tau\mu}_{eH}\right]\\
	C_{D,L}&\simeq \frac{9e\alpha_e}{64\pi^3}\frac{v}{m_\mu}\left[C^{*\tau e}_{eH}\left(C^{(1)\tau \mu}_{H\ell}+C^{(3)\tau \mu}_{H\ell}\right)+C^{*\mu \tau}_{eH}C^{e\tau}_{He}\right]
\end{align}A dipole is also generated at one-loop with a pair of penguin operators, which look like the flavor changing version of the electroweak correction to $(g-2)_\mu$ with a $Z$ exchange. However, assuming the future limits on penguin coefficients shown in Table \ref{tab:TauConstraints}, the contribution is below $\mu\to e\gamma$ upcoming experimental sensitivity.

Four lepton $e\mu \tau\tau$ operators get matching contribution from tree-level diagrams with a $Z,h$ exchange between penguin vertices or LFV Higgs boson couplings, as illustrated in the diagrams of Figure \ref{subfig:matching1} and \ref{subfig:matching2}. SMEFT $\tau-$LFV penguins and Yukawa corrections are matched at $m_W$ onto low energy four lepton operator coefficients as follows
\begin{align}
	C^{e\mu\tau\tau}_{T,RR}&=-\frac{1}{4}C^{e\tau}_{eH}C^{\tau \mu}_{eH}\frac{v^2}{m_h^2} \label{eq:tensormatchingR}\\
	C^{e\mu\tau\tau}_{T,LL}&=-\frac{1}{4}C^{*\tau e}_{eH}C^{* \mu\tau}_{eH}\frac{v^2}{m_h^2} \label{eq:tensormatchingL}\\
	C^{e\mu\tau\tau}_{S,RR}&=-C^{e\tau}_{eH}C^{\tau \mu}_{eH}\frac{v^2}{m_h^2}\\
	C^{e\mu\tau\tau}_{S,LL}&=-C^{*\tau e}_{eH}C^{* \mu\tau}_{eH}\frac{v^2}{m_h^2}\\
	C^{e\mu\tau\tau}_{S,RL}&=\frac{2g^2}{\cos^2\theta_W} \left(C^{\tau\mu}_{He}C^{(1)e\tau}_{H\ell}+C^{\tau\mu}_{He}C^{(3)e\tau}_{H\ell}\right)\frac{v^2}{M_Z^2}\\
	C^{e\mu\tau\tau}_{S,LR}&=\frac{2g^2}{\cos^2\theta_W} \left(C^{e\tau}_{He}C^{(1)\tau\mu}_{Hl}+C^{e\tau}_{He}C^{(3)\tau\mu}_{H\ell}\right)\frac{v^2}{M_Z^2}\\
	C^{e\mu\tau\tau}_{V,LL}&=-\frac{g^2}{\cos^2\theta_W} \left(C^{(1)e\tau}_{H\ell}C^{(1)\tau\mu}_{H\ell}+C^{(3)e\tau}_{H\ell}C^{(3)\tau\mu}_{H\ell}+C^{(1)e\tau}_{H\ell}C^{(3)\tau\mu}_{H\ell}+C^{(1)\tau\mu}_{H\ell}C^{(3)e\tau}_{H\ell}\right)\frac{v^2}{M_Z^2}\\
	C^{e\mu\tau\tau}_{V,RR}&=-\frac{g^2}{\cos^2\theta_W} C^{e\tau}_{He}C^{\tau\mu}_{He}\frac{v^2}{M_Z^2}
\end{align}
where the low energy EFT basis is in the notation of \cite{Davidson:2020hkf}.
We report for completeness the matching conditions for $e\mu\tau\tau$ vector coefficients, although $\mu\to e$ observables are not sensitive to them. 

\section{Phenomenological implications}\label{sec:Pheno}

This section 
gives limits  on pairs of 
$\tau\leftrightarrow l$  coefficients
from their contribution to $\mu\to e$ processes,
and  we discuss some examples where the upcoming sensitivity of $\mu\to e$ observables  is complementary to  the future direct limits from   $\tau\rightarrow l$  processes. Section \ref{ssec:Fishtop} considers   $\me$ amplitudes generated by the fish diagrams of Figure \ref{subfig:5}-\ref{subfig:6},
and compares with the limits arising from $B\to \tau$ LFV decays (summarised  in Appendix \ref{appendix:Bdecays}).
An example of $\me$  from  matching out the Higgs  is given in
Section \ref{ssec:HiggsLFV},  where we compare the sensitivity of $\mu\to e$
processes  to $h\to \tau^\pm l^\mp$ decays. Appendix \ref{appendix:Sensitivities} gives results for the cases where the $\me$ sensitivity is marginal or uninteresting.

{ The limits we quote apply to
	pairs of $\tau\leftrightarrow l$
	coefficients at a New Physics scale $\LNP = $ 4 TeV.
	We assume   that dimension six
	$\tau\leftrightarrow l$ operators are  generated at  $\Lambda_{\rm NP}=4$ T$e$V and
	contribute to $\mu\to e$ observables in two ways: first,  as discussed in section \ref{ssec:SMEFTRunning},  via  
	Renormalisation Group mixing into   dimension eight $\mu\to e$
	operators in SMEFT between $\Lambda_{\rm NP}$ and $m_W$, and second via  the matching  at $m_W$  of combined dimension six $\tau\leftrightarrow l$ operators onto
$\mu\to e$ operators as calculated in section \ref{ssec:Matching}.
The running is  described with  the  solution   of  the RGEs  given in eq.~(\ref{eq:RGEsolution}), then the  dimension eight $\mu\to e$ operators
are matched onto  the low energy EFT as given in \cite{Ardu:2021koz}.
The  sensitivity of
current $\me$ experiments to coefficients at $m_W$ is tabulated in \cite{Davidson:2020hkf}; we extrapolate
	these limits  to the future experimental reaches given in table
	\ref{tab:LFVsearches}, in order to
	determine the experimental sensitivities of $\mu\to e$ processes to the product of $\tau\to l$ operator coefficients.}
{ In most cases,  we just rescale the sensitivities of \cite{Davidson:2020hkf}.
But for  the limits  from $\mu A \to e A$ on vector operators with quarks,
we recalculate the sensitivities on an Aluminium target, as will be used by
upcoming experiments. The current bounds  are from  Gold targets,
which have  more neutrons than protons, whereas  Aluminium  contains  equal numbers of protons and neutrons ($u$ and $d$ quarks).
So Gold has comparable sensitivity to $(\bar{e}\g \mu)(\bar{u}\g u +\bar{d}\g d)$  and $(\bar{e}\g \mu)(\bar{u}\g u -\bar{d}\g d)$, whereas the sensitivity of Aluminium to  $(\bar{e}\g \mu)(\bar{u}\g u -\bar{d}\g d)$ is suppressed by a loop. }

\begin{figure}
	\centering
	\includegraphics[scale=0.6]{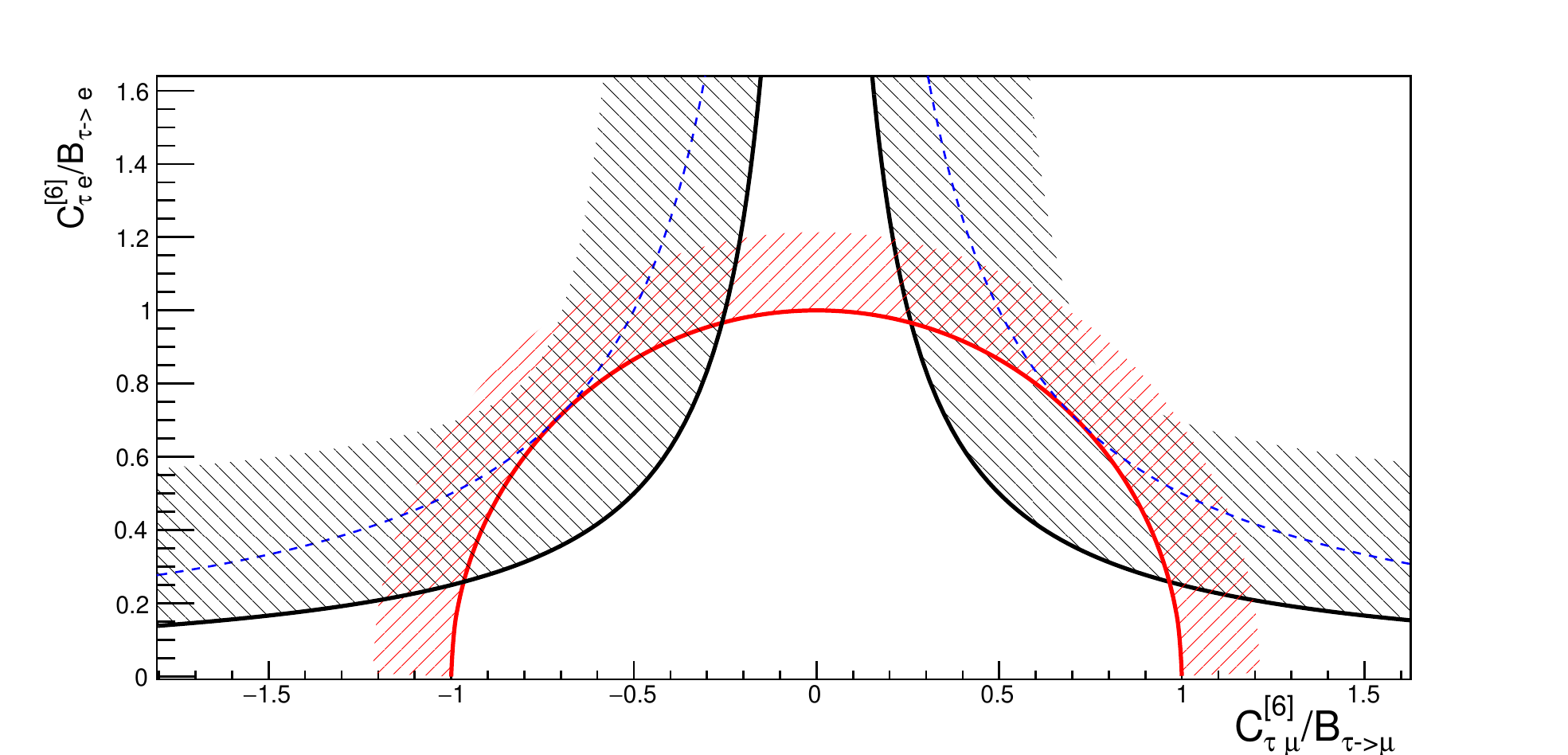}
	\caption{The plot shows the parameter space probed by direct $\tau\leftrightarrow l$ searches and by $\mu\to e$ observables, in the $C_{\tau\mu}-C_{\tau e}$ plane. The direct searches can probe the region outside the ellipse of eq.~(\ref{eq:ellipse}) (which correspond to the red circle when the Wilson coefficients are normalized by the sensitivities $B_{\tl}$ of the $\tl$ processes), while $\mu\to e$ is sensitive to the area above the hyperbolae, as defined in eq.~(\ref{eq:hyperbole}). The blue dashed hyperbolae correspond to the boundary condition $B_{\me}/(B_{\te}B_{\tm})=1/2$, while the black ones satisfies $B_{\me}/(B_{\te}B_{\tm})<1/2$. In this second case $\mu\to e$ searches are able to probe parameter space missed by $\tl$ observables.}\label{fig:circlevshyperbola}
\end{figure}

Note that  we distinguish {\it sensitivities}  from  {\it constraints}
or {\it bounds}. But we use {\it limits} to mean either.
A constraint  identifies the region of parameter space where the coefficients must sit, while  a sensitivity  represents the smallest absolute  value that
can be experimentally detected.
The notion of sensitivity is particularly useful when the number of parameters is larger than the number of observables,
so that  exclusion bounds on single coefficients cannot be inferred.
A coefficient smaller than the sensitivity escapes experimental detection but larger values can also escape detection if accidental cancellations occur.
In practise, in this manuscript  we obtain sensitivities,
because we consider one non-zero pair of $\tau \leftrightarrow l$ operators
at a time and compute the contribution to $\mu\to e$ observables.

{  Our results  are interesting, because they  show that upcoming $\me$ experiments could be sensitive to  $\tl$ coefficients beyond the reach of $\tl$ searches.
	We obtain }experimental sensitivities $B_{\me}$  to  the  product  of coefficients
\begin{equation}
	\abs{C^{[6]\tau \mu}C^{[6]e\tau}}\lesssim B_{\me}.\label{eq:hyperbole}
\end{equation}
The same coefficients $C^{[6]\tau \mu}$, $C^{[6]e\tau}$ might contribute to constrained $\tau \leftrightarrow l$ processes and be respectively subjected to the sensitivity ``limits" $B_{\tau\to \mu}$, $B_{\tau\to e}$ imposed by direct $\tau$LFV
searches. In the $C^{[6]\tau \mu}-C^{[6]e\tau}$ plane, this identifies an ellipse
\begin{equation}
	\frac{\abs{C^{[6]\tau \mu}}^2}{B^2_{\tm}}+\frac{\abs{C^{[6] e \tau }}^2}{B^2_{\te}}\lesssim1 \label{eq:ellipse}
\end{equation}
that encloses the coefficient space to which $\tau\leftrightarrow l$ observables are not sensitive. On the other hand, $\mu\to e$ searches can detect coefficients in the region bounded by the hyperbola in eq.~(\ref{eq:hyperbole}). If the following inequality is satisfied
\begin{equation}
	B_{\me}< \frac{B_{\te}B_{\tm}}{2} \label{eq:hyperbolevsellipse}
\end{equation}
the hyperbola  enters the ellipse  and $\mu\to e$ processes are able to probe a region of parameter space that eludes the direct $\tau \leftrightarrow l$ searches. This is illustrated in Figure \ref{fig:circlevshyperbola}. In the subsequent sections we discuss examples where eq.~(\ref{eq:hyperbolevsellipse}) is satisfied considering the upcoming experimental sensitivities on $\mu\to e$ and $\tau\to l$ processes.

\subsection{ Fish diagrams with internal top quarks}\label{ssec:Fishtop}

{
	In this section, we discuss  some  examples where  the sensitivity of $\mu\to e$ conversion to  some $\tl$ coefficients   is complementary to  $B$ decays.
	The ``fish" diagrams that mix four fermion $\tau\leftrightarrow l$ interactions into dimension eight $\mu\to e$ operators  are illustrated in 
	Figure \ref{subfig:5}-\ref{subfig:6} of section \ref{ssec:SMEFTRunning}.
	In these diagrams,}  one or two Higgs are attached to a heavy top internal line, so  the $\tau\leftrightarrow l$ operators that our calculation can probe contain one quark doublet or up-type singlet in the third generation. In the former case, the operator can contribute to the LFV decays of the $B$ mesons with a $\tau \ (\nu_\tau)$ in the final state.  The following subsections are organized by the different $\mu\to e$ interactions that the $\tl$ operators mix into.

\subsubsection{$\mu\to e$ scalars}

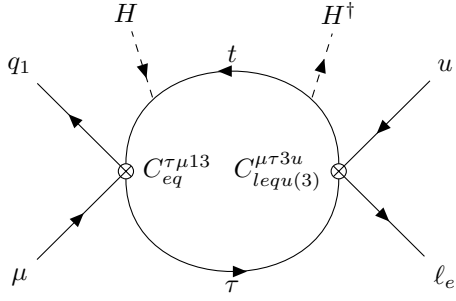
\begin{figure}
	\centering
	\begin{tikzpicture}[scale=1.4]
		\begin{feynman}[small]
			\vertex (fin) at (-1,1) {\(q_1\)};
			\vertex [style=crossed dot] (a) [label=0:\(C^{\tau\mu 13}_{eq}\)] at (0,0) {};
			\vertex [style=crossed dot] (b) [label=180:\(C^{\mu\tau 3u}_{lequ(3)}\)]at (2,0) {};
			\vertex (mu) at (-1,-1) {\(\mu\)};
			\vertex (c) at (0.25, 0.70);
			\vertex (H1)  at (0,1.5) {\(H\)};
			\vertex (d) at (1.75, 0.70);
			\vertex (H2) at (2, 1.5) {\(H^\dagger\)};
			\vertex (fout) at (3,1) {\(u\)};
			\vertex (e) at (3,-1) {\(\ell_e\)};
			\diagram* {
				(fin) -- [anti fermion] (a),
				(mu) -- [fermion] (a),
				(a) -- [anti fermion, half left, edge label=\(t\)] (b) -- [anti fermion, half left, edge label=\(\tau\)] (a),
				(b) -- [fermion] (e),
				(b) -- [anti fermion] (fout),
				(c) -- [anti charged scalar] (H1),
				(d) -- [charged scalar] (H2),
			};
		\end{feynman}
	\end{tikzpicture}
	\caption{The operators $C^{e \tau 31}_{eq}, C^{(3)e\tau 31}_{\ell equ}$ are inserted in the left diagram and mix into the dimension eight $\mu\to e$ scalar/tensor operators $\mathcal{O}^{(1),(2),(3),(4)}_{\ell equH^2}$ of eq.s~(\ref{eq:scalarupdim8})-(\ref{eq:toptensor}).
\label{fig:fishBexample}}
\end{figure}

Consider, for example, the operators
$\mathcal{O}^{\tau\mu 13}_{eq}=(\bar{\tau} \gamma \mu)(\bar{q}_1 \gamma q_3)$ and
$\mathcal{O}^{(3)e\tau 31}_{\ell equ}=(\bar{\ell}_e \sigma \tau)(\bar{q}_3\sigma u)$, which  mix into the $\mu\to e$ scalar and tensor dimension eight operators  $\mathcal{O}^{(1),(2),(3),(4)}_{\ell equH^2}$ of eq.s~(\ref{eq:scalarupdim8})-(\ref{eq:toptensor}) (with up quarks) via the diagram
of Figure \ref{fig:fishBexample}.
\begin{figure}[th]
	\centering
	\begin{subfigure}{\textwidth}
		\centering
		\includegraphics[scale=0.45]{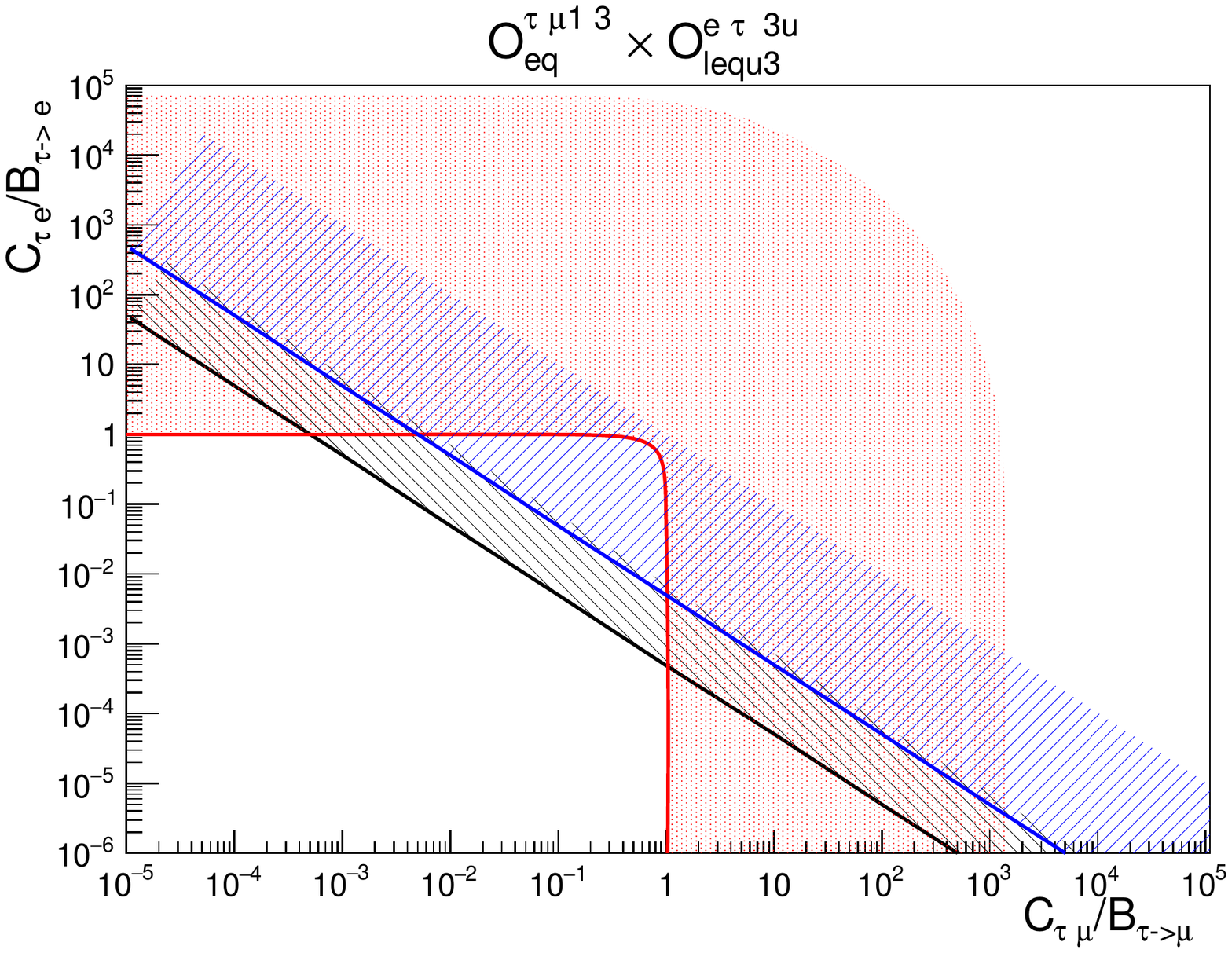}
		\caption{ Parameter space probed by $\mue$ conversion (straight lines) and LFV $B$ decays (box), in the $C^{\tau\mu 13}_{eq}-C^{(3)e\tau 3u}_{\ell equ}$ plane. The blue line correspond to the current experimental reach, while in the black one we assume $Br(\mu Al\to e Al)\sim 10^{-16}$. In both cases, the $\mu\to e$ hyperbole enter the ellipse beyond the reach of $B$ decays.\label{fig:examplelog}}
	\end{subfigure}\\
	\begin{subfigure}{\textwidth}
		\centering
		\includegraphics[scale=0.45]{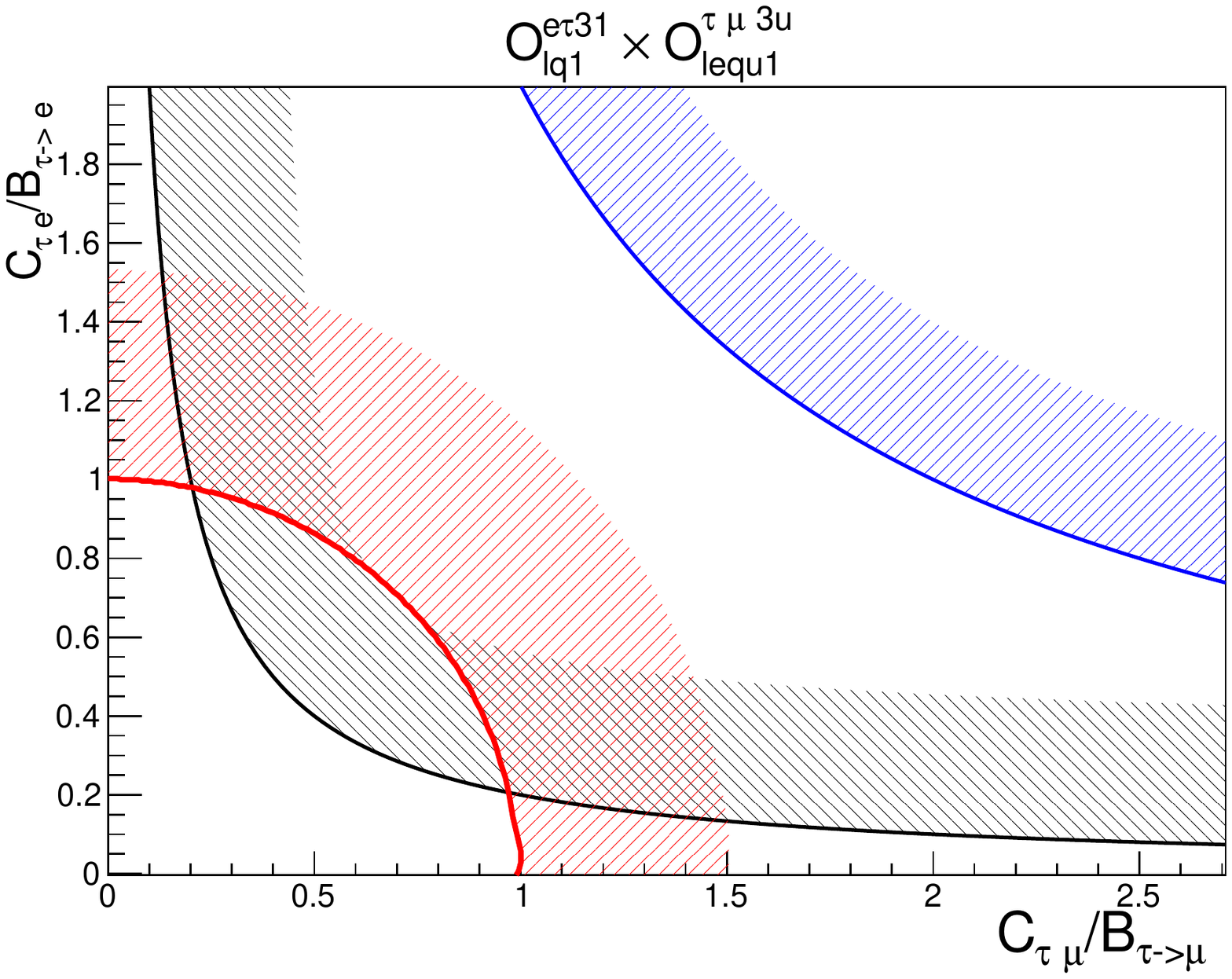}
		\caption{	 Similar to Figure \ref{fig:examplelog}, in the $C^{(1)e\tau 13}_{\ell q}-C^{(1)\tau\mu 3u}_{\ell equ}$ plane. For this pair of operators, $\mu\to e$ will have a better sensitivity to the coefficient product than $B$ decays with the upcoming experimental improvement.\label{fig:examplelin}}
	\end{subfigure}
	\caption{
	    \label{fig:hyperbolavsellipseexample}}
\end{figure}
These match  at $m_W$ onto scalar and tensor operators in the low energy EFT, with the following coefficients\footnote{{ This simple solution does not include the QCD running of tensors and scalars from $\LNP \to m_W$.  Since QCD does not renormalize vector coefficients, this QCD running is  analogous to the rescaling of QED tensor$\leftrightarrow$scalar mixing below $m_W$ \cite{Davidson:2020hkf},and can be estimated to be a  $\lesssim 10\%$ effect. It is therefore neglected.}}
\begin{equation}
	C^{uu}_{S,RR}(m_W)=\frac{3}{2\pi^2}\frac{m^2_t}{v^2}C^{ \tau\mu 13}_{eq}(\Lambda_{\rm NP})\times C^{(3)e\tau 3u}_{\ell equ}(\Lambda_{\rm NP})\log(\frac{m_W}{\Lambda_{\rm NP}})
\end{equation}
\begin{equation}
	C^{uu}_{T,RR}(m_W)=\frac{3}{8\pi^2}\frac{m^2_t}{v^2}C^{ \tau\mu 13}_{eq}(\Lambda_{\rm NP})\times C^{(3)e\tau 3u}_{\ell equ}(\Lambda_{\rm NP})\log(\frac{m_W}{\Lambda_{\rm NP}}) \label{eq:exampletensor}
\end{equation}    
where $m_t\sim v$ is the top quark mass.

Scalar operators  with up quarks contribute at tree-level to $\mu\to e$ conversion in nuclei (see eg  \cite{Kitano:2002mt}), where a muon is stopped in a target, captured by a nucleus,  and converts into an electron in the presence of LFV interaction with nucleons.
Scalar interactions of first generation quarks match onto nucleon operators with large matching coefficients, and  the rate for spin-independent conversion is enhanced by the atomic number of the target, giving a good  current sensitivity to scalar  coefficients   $C^{uu}_S\lsim 10^{-8}$ \cite{Davidson:2020hkf}.
Including the impressive improvement in sensitivity promised by upcoming experiments,
$Br(\mu Au\to e Au)\lsim 10^{-12} \to $ $Br(\mu Al\to e Al)\sim 10^{-16}$, $\mu\to e$ conversion will be able to probe scalar coefficients as small as $C^{uu}_S\sim 10^{-10}$.

Tensors with light-quarks contribute to the spin-independent rate via their QED mixing into scalars, which introduces a $\sim 1/10$ suppression. For this reason, the tensor of eq.~(\ref{eq:exampletensor}) contribute to the $\mu\to e$ conversion rate as $C_T\sim C_S/40$ and is therefore neglected.
So the upcoming $\mu\to e$ conversion experiments can set the following limit (sensitivity) on the product of the coefficients at $\LNP= 4$ T$e$V
\begin{equation}
	C^{ \tau\mu 13}_{eq}\times C^{(3)e\tau 31}_{\ell equ}\lesssim 1.5\times 10^{-10} \label{eq:limitmutoeconv}
\end{equation}
The two $\tl$  operators  could also induce  the  leptonic  decays of  $B$ mesons
$B^0_d\to \mu^\pm\tau^\mp $ and  $ B^+\to \overline{\tau} \nu$.
The current  95\%C.L. experimental  constraints  on these processes
lead to the following  limits  on the coefficients
\begin{align}
	Br(B^0_d\to \mu^\pm\tau^\mp)<1.4\times 10^{-5}\qquad &\to\qquad C^{ \tau \mu 13}_{eq}\lesssim 1.4\times 10^{-3} \nonumber \\
	Br(B^+\to \overline{\tau} \nu)=1.6\times 10^{-4}\qquad &\to\qquad C^{(3)e \tau 31}_{lequ}\lesssim 2.2\times 10^{-3}. \label{eq:limitsBdecays}
\end{align}
These limits  were obtained  with the  public code Flavio \cite{Straub:2018kue}
 and analytically, and  are discussed in more detail in 
Appendix \ref{appendix:Bdecays}, which    reviews  the sensitivity of
B decays    to interesting operator coefficients.

In order to compare future $B$ decay sensitivities to the future reach of
$\mu\to e$ conversion, we suppose that  Belle II could improve
the  sensitivities to  $B$ decays  by  an order of magnitude, so the limits of eq.~(\ref{eq:limitsBdecays}) on the Wilson coefficients will get $\sim \sqrt{10}$ better. Comparing the product of the upcoming $B$  sensitivities with the limit in eq.~(\ref{eq:limitmutoeconv}) that arise from future $\mu\to e$ conversion gives (the $(f)$ superscript stands for ``future") 
\begin{equation}
	B^{(f)}_{\me}=(B^{(f)}_{\te}B^{(f)}_{\tm})\times (5\times 10^{-4}) \label{eq:exampleBdecay}
\end{equation}
which satisfies the condition of eq.~(\ref{eq:hyperbolevsellipse}). We fall in the scenario depicted in Figure \ref{fig:examplelog}, where $\mu\to e$ probes a region inside the ellipse, beyond the reach of $B\to \tau$ direct searches.
 Notice
  { that the hyperbola of the current $\mu\to e$ conversion results already enters the ellipse 
    of the  $B\to \tau$ LFV decays
    (with the current sensitivities $B^{(c)}_{\me}/(B^{(c)}_{\te}B^{(c)}_{\tm})\sim 5\times 10^{-3}$). This is because tensors contribute to the $B$ decays rate via the one-loop QED mixing to scalars, while the $\text{(dimension six)}^2\to$ (dimension eight) mixing benefits from a large anomalous dimension. 

    The pair of $\tl$ dimension six operators $C^{(1)e\tau 13}_{\ell q} C^{ (1)\tau \mu 3u}_{\ell equ }$ similarly mixes into the dimension eight $\mu\to e$ scalars with a singlet $u$.
In this case,
$B$ decays are currently  more sensitive than $\mu\to e$ processes
to the product of the coefficients
($B^{(c)}_{\me}/(B^{(c)}_{\te}B^{(c)}_{\tm})\sim 2$). However, in the next generation of experiments,
the sensitivity ratio will  be reduced by one order of magnitude
to $B^{(f)}_{\me}/(B^{(f)}_{\te}B^{(f)}_{\tm})\sim 0.2$,
allowing the $\mu\to e$ conversion hyperbola to enter the
ellipse of the direct $\tl$ searches (see Figure \ref{fig:examplelin}). }

  In Tables \ref{tab:Bdecaysscalarup} and
 \ref{tab:Bdecaysscalardown},
  we compare the sensitivities of $\tl$ and $\mu\to e$ processes to the product of several operators that mix into scalars with first generation quarks, via diagrams similar to Figure \ref{fig:fishBexample}. { Note that the pairs in the table feature an electron doublet and a singlet muon, but opposite chiralities are also possible. For instance, $C^{\tau \mu 13}_{e q}C^{(1)e\tau 3u}_{\ell equ}$ mix into $C^{(1),(2)e\mu 1u}_{\ell e quH^2}$ while $C^{e\tau  31}_{e q}C^{*(1)\mu\tau 3u}_{\ell equ}$ contributes to the RGEs of $C^{*(1),(2)\mu e 1u}_{\ell e quH^2}$. Although the anomalous dimensions are the same (and so are the $\mue$ sensitivities), the dimension six operator that was $\te$ is now $\tm$ and vice-versa, which might lead to slightly different direct limits on the $\tl$ interactions. In the above-example, the branching ratios sensitivities of the $B^0_d$ decay into $\tau e, \tau \mu$ differ by a factor $\sim 3$, and as a result the limits on the vector coefficients $C^{e\tau 13}_{eq},C^{\tau \mu 31}_{eq}$ is $\sim \sqrt{3}$ different. We do not present the tables for the pairs with exchanged $\me$, as the marginally different numbers do not modify our conclusions.}

\begin{table}[!h]
	\begin{center}
		\begin{tabular}{|l|c|c|}
			\hline
			coefficients &     $B^{(c)}_{\te} B^{(c)}_{\tm}$  & $B^{(f)}_{\mue}$  \\
			\hline
			$C^{e\tau tu}_{\ell u} C^{(1)\tau \mu 1t}_{\ell equ }$ & ---$ \times $---
			&$2 \times 10^{-9}  $ \\
			$C^{e\tau tu}_{\ell u} C^{(3)\tau \mu 1t}_{\ell equ }$ &---$ \times $---
			&$1.5 \times 10^{-10}  $ \\
			$C^{ \tau\mu 13}_{eq} C^{(1)e\tau 3u}_{\ell equ}$ &$  1.5 \times 10^{-3}(c) \times 4.3 \times 10^{-4}(c) $
			&$2 \times 10^{-9}  $  \\
			$C^{ \tau\mu 13}_{eq} C^{(3)e\tau 3u}_{\ell equ}$ &$  1.5 \times 10^{-3}(c) \times 2.4 \times 10^{-3}(c) $ %
			&$1.5 \times 10^{-10}  $  \\
			$C^{\tau \mu tu}_{eu} C^{(1)e \tau 1t}_{\ell equ }$ & ---$   \times $ ---
			&-$2 \times 10^{-9}  $  \\
			$C^{\tau \mu tu}_{eu} C^{(3)e \tau 1t}_{\ell equ }$ & ---$   \times $ ---
			&$1.5 \times 10^{-10}  $  \\
			$C^{(1)e\tau 13}_{\ell q} C^{ (1)\tau \mu 3u}_{\ell equ }$ & $2.3 \times 10^{-3}(c)   \times 4.3 \times 10^{-5}(c) $ 
			&$2 \times 10^{-9}  $  \\
			$C^{(3)e\tau 13}_{\ell q} C^{ (1)\tau \mu 3u}_{\ell equ }$ & $2.3 \times 10^{-3}(c)   \times 4.3 \times 10^{-5}(c) $ 
			&$2 \times 10^{-9}  $  \\
			$C^{(1)e\tau 13}_{\ell q} C^{(3) \tau \mu 3u}_{\ell equ }$ & $2.3 \times 10^{-3}(c)   \times 1.8 \times 10^{-4}(c) $ 
			&$1.5 \times 10^{-10}  $  \\
			$C^{(3)e\tau 13}_{\ell q} C^{(3) \tau \mu 3u}_{\ell equ }$ & $2.3 \times 10^{-3}(c)   \times 1.8 \times 10^{-4}(c) $ 
			&$1.5 \times 10^{-10}  $  \\
			\hline
		\end{tabular}
		\caption{The product of current (c) direct  limits  $B^{(c)}_{\te} B^{(c)}_{\tm}$  on pairs of  coefficients
			that  mix to  a $\mue$ dimension eight scalar operator with a singlet $u$ quark (see eq.~(\ref{eq:scalarupdim8})),
			upon which applies the limit $B^{(f)}_{\mue}$ arising from future $\mec$ ($Br(\mu Al\to e Al)\sim 10^{-16}$).
			The ``limits'' are on coefficients at $\LNP\sim 4$ TeV. { Details on the limits that apply to operators with permuted indices are given in the text below eq.~(\ref{eq:exampleBdecay})}. To compare $B^{(f)}_{\mue}$ with the future sensitivity of direct $\tl$ searches, the product $B^{(c)}_{\te} B^{(c)}_{\tm}$ should be divided by 10: $B^{(f)}_{\te} B^{(f)}_{\tm}\sim B^{(c)}_{\te} B^{(c)}_{\tm}/10$.
			\label{tab:Bdecaysscalarup}} 
	\end{center}
\end{table}

\begin{table}[!h]
	\begin{center}
		\begin{tabular}{|l|c|c|}
			\hline
			coefficients &     $B^{(c)}_{\te} B^{(c)}_{\tm}$  & $B^{(f)}_{\mue}$  \\
			\hline
			$C^{(3)e\tau 31}_{\ell q} C^{\tau \mu d3}_{\ell edq }$ & $ 2.3 \times 10^{-3} (c)\times  2.2 \times 10^{-4} (c)$
			&$1 \times 10^{-9}  $ \\
			$C^{(3)\tau \mu 1 3}_{\ell q} (C^{\tau e d3}_{\ell edq })^*$ & $ 1.5 \times 10^{-3}(c) \times 3.4 \times 10^{-4} (c)$
			&$1 \times 10^{-9}  $ \\
			\hline
		\end{tabular}
		\caption{Similar to table \ref{tab:Bdecaysscalarup}, for 
			dimension eight  scalar  $\mue$ operators involving a singlet $d$  quark (see eq.~(\ref{eq:dim8scalardown})). The  limit $B^{(f)}_{\mue}$ arises from $\mec$ ($Br(\mu Al\to e Al)\sim 10^{-16}$).
			\label{tab:Bdecaysscalardown}} 
	\end{center}
\end{table}

\subsubsection{$\mu\to e$ tensors with heavy quarks}
\label{ssec:Banom}

\begin{figure}
	\centering
	\includegraphics[scale=0.6]{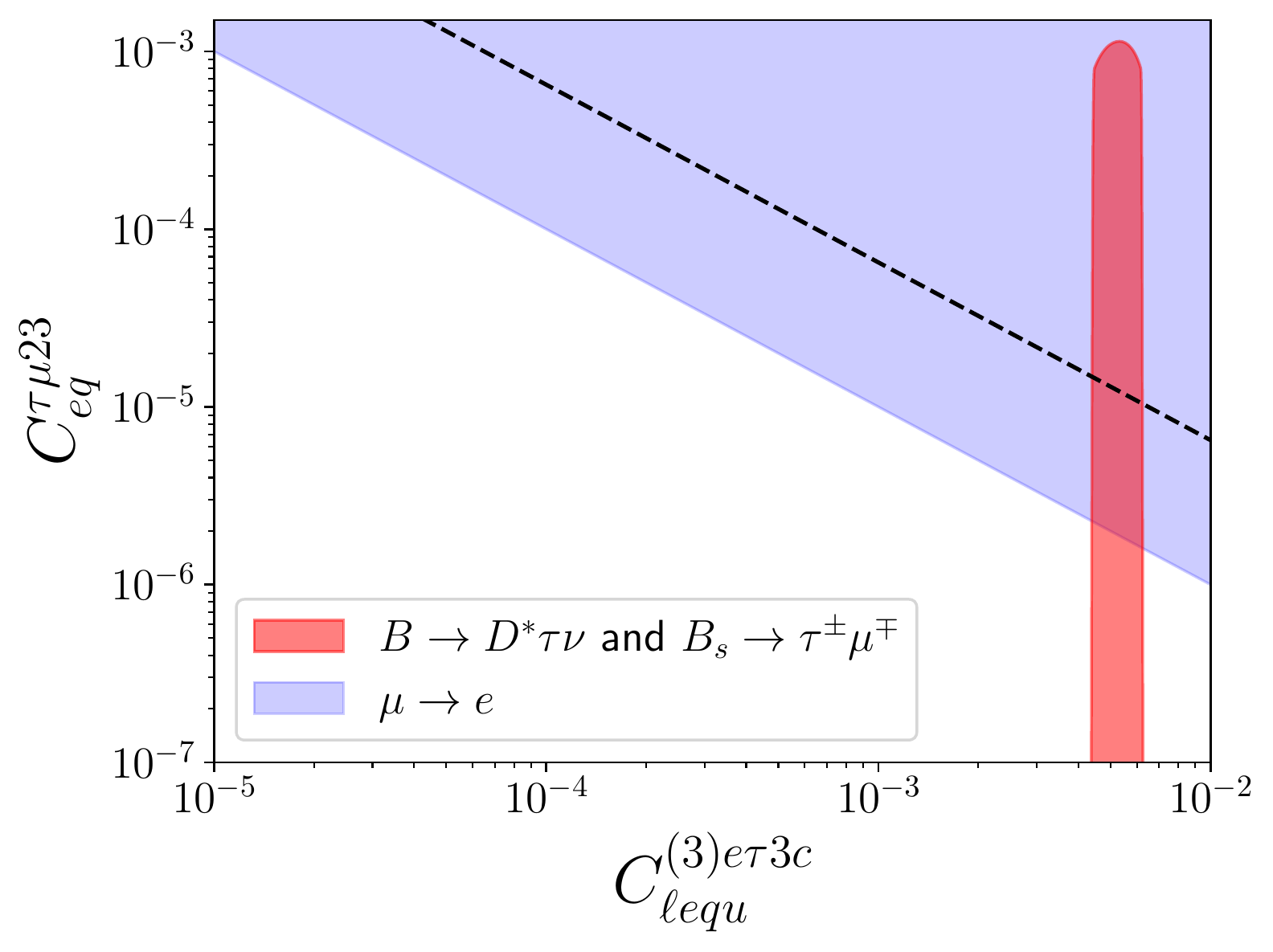}
	\caption{The plot shows the parameter space probed by  $B$ LFV decays and by future $\meg$, in the $C^{(3)e\tau 3c}_{\ell e qu}-C^{\tau\mu 23}_{eq}$ plane. The ellipse is centered to the best-fit value of $C^{(3)e\tau 3c}_{\ell e qu}$ that can explain the $R_{D^*}$ anomaly (see text for details). Non-observation of $\meg$ can give a limit on $C^{\tau\mu 23}_{eq}$ (assuming only this pair to be non-zero). The dashed line correspond to the current MEG upper bound $Br(\meg)<4\times 10^{-13}$.  \label{fig:fitBanom}}
\end{figure}

{  The fish diagrams that generated scalar and tensor $\mu\to e$ operators on $u$ quarks, arise also  with external $c$ quarks.
	Although the sensitivity of  $\mu\to e$ conversion to charm scalars   is insufficient for our purposes,  $\mu\to e \gamma$ has interesting sensitivity to
	the   charm tensors, because their mixing to  the dipole is enhanced  $\propto m_c/m_\mu$.}
The pairs of $\tl$  operators  that mix to $\me$  tensors with external charms,
and the  sensitivities of $B$ decays and  $Br(\mu\to e \gamma)< 10^{-14}$ 
are summarized in Table \ref{tab:Bdecaystensorcharm}. 

 Leptonic and semi-leptonic $B$ decays have recently attracted  attention
  due to
  several anomalies with respect to SM expectations, see e.g. Ref. \cite{Shi:2019gxi}.
  Our LFV operators  could  potentially address the anomalies in ``charged current" $b$ transitions (such as $B^+\to \tau^+ \nu$),
however the observed rates are often  below the SM expectations,
so cannot be explained by lepton-flavour-changing interactions that necessarily increase the rates(because they cannot interfere destructively with the SM).
An exception is the SM expectation for $R^{\rm SM}_{D^*\tau/l}\equiv Br(B\to D^* \tau \bar{\nu})/Br(B\to D^* l \bar{\nu})\sim 0.24$ \cite{Straub:2018kue}
which  is smaller than the observed value $R^{exp}_{D^*\tau/l}\sim 0.3$ \cite{HFLAV:2019otj}.
We can fit the difference  by enhancing the branching fraction
in the numerator with the tensor operator $C^{(3)l\tau 3c}_{\ell e qu}$.
The latter can be paired with the vector $C^{\tau\mu 23 }_{eq}$ to mix
into a dimension eight tensor with external charms,
to which $Br(\meg)\sim 10^{-14}$ has
the sensitivity $B_{\mue}$ reported in Table \ref{tab:Bdecaystensorcharm}.
In the $C^{(3)e\tau 3c }_{\ell e qu}-C^{\tau\mu 23 }_{eq}$ plane,
the ellipse is now shifted to the right and centered on
the best-fit value of $C^{(3)e\tau 3c }_{\ell e qu}$
(see Figure \ref{fig:fitBanom}).
In the simplified scenario where the discrepancy $\abs{R^{\rm SM}_{D^*\tau/e}-R^{\rm exp}_{D^*\tau/e}}$ is fully explained by the presence of the $\te$ tensor, non-observation $\meg$ signal in future experiments would make it unlikely for the coefficients to occupy the portion of the red ellipse overlaping the blue region. 

 Table  \ref{tab:toptensors}  summarises the case   of  $\me$ operator with external top quarks.
 The mixing of tensors with  a top bilinear  into the dipole is  enhanced by the ratio $m_t/m_\mu$,
  so the upcoming $\mu\to e\gamma$ experiments can probe  dimension 6 coefficients $C^{[6] e\mu tt}_T\gtrsim 5\times 10^{-12}$.
 We suppose that  the  SMEFT mixing of dimension eight tensors into the dimension eight dipoles is comparable to the dimension six mixing \cite{Manohar2}. This impressive sensitivity explains why the diagrams of Figure \ref{subfig:6} with external top legs are interesting regardless of the $y_\tau$ Yukawa suppression.
	
The SMEFT $\tl$ operators that are inserted in those diagrams contain a flavour diagonal quark pair in the third generation. Vectors with tops contribute to the rate of $\tau\to 3l$ via one-loop penguin diagrams, while the dimension six tensors contribute to $\tau\to l\gamma$ via the above-discussed mixing into the $\tl$ dipole. Tensors are not considered in our tables, because $\tau\to l\gamma$ has already an excellent sensitivity to the operator coefficients. In Table \ref{tab:toptensors} the direct ``limits" on the product of $\tl$ dimension six vectors arising from  $\tau\to 3l$ searches are compared with the sensitivity of $Br(\mu\to e\gamma)< 10^{-14}$.  

\begin{table}[!h]
	\begin{center}
		\begin{tabular}{|l|c|c|}
			\hline
			coefficients &     $B^{(c)}_{\te} B^{(c)}_{\tm}$  & $B^{(f)}_{\mue}$  \\
			\hline
			$C^{e\tau tc}_{\ell u} C^{(1)\tau \mu 2t}_{\ell equ }$ & ---$ \times $---
			&$1.2 \times 10^{-7}  $ \\
			$C^{e\tau tc}_{\ell u} C^{(3)\tau \mu 2t}_{\ell equ }$ &---$ \times $---
			&$1 \times 10^{-8}  $ \\
			$C^{ \tau\mu 23}_{eq} C^{(1)e\tau 3c}_{\ell equ}$ &$ 2.3\times 10^{-3}(c) \times 1.0\times 10^{-2}(c) $
			&$1.2\times 10^{-7}  $  \\
			$C^{ \tau\mu 23}_{eq} C^{(3)e\tau 3c}_{\ell equ}$ &$2.3\times 10^{-3}(c) \times 5.0\times 10^{-3}(c) $
			&$1  \times 10^{-8}  $  \\
			$C^{\tau \mu tc}_{eu} C^{(1)e \tau 2t}_{\ell equ }$ & ---$   \times $ ---
			&$1.2 \times 10^{-7}  $  \\
			$C^{\tau \mu tc}_{eu} C^{(3)e \tau 2t}_{\ell equ }$ & ---$   \times $ ---
			&$1 \times 10^{-8}  $  \\
			$C^{(1)e\tau 23}_{\ell q} C^{(1) \tau \mu 3c}_{\ell equ }$ &$2.3\times 10^{-3}(c)   \times 9.0\times 10^{-3}(c) $
			&$1.2 \times 10^{-7}  $  \\
			$C^{(3)e\tau 23}_{\ell q} C^{(1) \tau \mu 3c}_{\ell equ }$ & $2.3\times 10^{-3}(c)   \times 9.0\times 10^{-3}(c) $ 
			&$1  \times 10^{-7}  $  \\
			$C^{(1)e\tau 23}_{\ell q} C^{(3) \tau \mu 3c}_{\ell equ }$ &$2.3\times 10^{-3}(c)   \times 6.4\times 10^{-3}(c) $ 
			&$1.2 \times 10^{-7}  $  \\
			$C^{(3)e\tau 23}_{\ell q } C^{(3) \tau \mu 3c}_{\ell equ }$ & $2.3\times 10^{-3}(c)   \times  6.4\times 10^{-3}(c) $
			&$1 \times 10^{-9}  $  \\
			\hline
		\end{tabular}
		\caption{ Similar to table \ref{tab:Bdecaysscalarup}, 
			for  $\mue$ dimension eight tensor operators (see eq.~(\ref{eq:toptensor})) with a $c$ quark bilinear.
			The sensitivity $B^{(f)}_{\mue}$ arises from $\meg$ with a branching ratio  $Br(\meg)\sim 10^{-14}$.
			The ``limits'' are on coefficients at $\LNP\sim 4$ TeV.
			\label{tab:Bdecaystensorcharm}} 
	\end{center}
\end{table}
\begin{table}[!h]
	\begin{center}
		\begin{tabular}{|l|c|c|}
			\hline
			coefficients &  $B^{(c)}_{\te} B^{(c)}_{\tm}$  & $B^{(f)}_{\mue}$  \\
			\hline
			$C^{e\tau tt}_{\ell u} C^{\tau \mu 33}_{eq }$ &$ 1.0 \times 10^{-2} (c)\times 2.0 \times 10^{-2} (c)$
			&$1.0 \times 10^{-6}  $  \\
			$C^{(3)e\tau 33}_{\ell q} C^{\tau \mu tt}_{eu }$ &$4.5 \times 10^{-3} (c)\times1.0 \times 10^{-2} (c) $
			&$1.0 \times 10^{-6}   $  \\
			$C^{(1)e\tau 33}_{\ell q}C^{\tau \mu tt}_{eu }$ &$4.0 \times 10^{-2}(c)\times1.0 \times 10^{-2} (c) $
			&$-1.0 \times 10^{-6}   $  \\
			\hline
		\end{tabular}
		\caption{ Similar to table 8, with the product of (current) direct limits  $B^{(c)}_{\te} B^{(c)}_{\tm}$  on pairs of $\tl$ coefficients
			that  mix to  a $\mue$ dimension eight tensor operator (see eq.~(\ref{eq:toptensor})) with two top quarks,
			upon which applies the limit $B_{\mue}$.
			All the  limits apply  to the  coefficients at $\LNP\sim 4$ TeV.
			The  limit $B^{(f)}_{\mue}$ arises  from $\meg$ ($Br(\meg)< 10^{-14}$), due to the large
			mixing of the top-tensor to the dipole, while the limits $B^{(c)}_{\tl}$ are from the current upper limits on $Br(\tau\to 3l)$ given in Table \ref{tab:LFVsearches}. Future limits $B^{(f)}_{\te}B^{(f)}_{\tm}$ are $\sim B^{(c)}_{\te}B^{(c)}_{\tm}/10$.   
			\label{tab:toptensors}} 
	\end{center}
\end{table}
\clearpage

\subsubsection{$\mu\to e$ vectors}
The remaining fish diagrams give mixing of two dimension six $\tl$ SMEFT operators into dimension eight $\mu\to e$ vectors with first generation quarks.
The sensitivities of $\mu\to e$ conversion and $B$ decays on the product of the operator coefficients are summarized in Table \ref{tab:BdecaysvectR} for lepton singlets and in Table \ref{tab:BdecaysvectL} for lepton doublets. (The $\mu\to e$ conversion estimates assume an  Aluminium  target --- see  the beginning of Section \ref{sec:Pheno}.) 
\begin{table}[!h]
	\begin{center}
		\begin{tabular}{|l|c|c|}
			\hline
			coefficients &     $B^{(c)}_{\te} B^{(c)}_{\tm}$  & $B^{(f)}_{\mue}$  \\
			\hline
			$C^{e\tau 31}_{eq} C^{\tau \mu 13}_{eq }$ & $ 2.3 \times 10^{-3}(c)\times 1.5 \times 10^{-3}(c)  $
			&$2.5 \times 10^{-9}  $ \\
			$C^{e\tau 13}_{eq} C^{\tau \mu 31}_{eq }$ & $ 2.3 \times 10^{-3}(c)\times 1.5 \times 10^{-3}(c)  $
			&$1 \times 10^{-8}  $ \\
			&&\\
			$C^{e\tau tu}_{eu} C^{\tau \mu ut}_{eu }$ & ---$ \times  $---
			&$2.5 \times 10^{-9}  $ \\
			$C^{e\tau ut}_{eu} C^{\tau \mu tu}_{eu }$ & ---$ \times  $---
			&$2.5 \times 10^{-9}  $ \\
			&&\\
			$(C^{\tau e d 3}_{\ell edq})^* C^{\tau \mu d3}_{\ell edq }$ & $  3.4 \times 10^{-4}(c) \times 2.2 \times 10^{-4}(c) $
			&$4 \times 10^{-8}  $ \\
			&&\\
			$(C^{(1)\tau e 1 t}_{\ell equ})^* C^{(1)\tau \mu 1t}_{\ell equ }$ & ---$ \times  $---
			&$2 \times 10^{-8}  $ \\
			$(C^{(1)\tau e  3u}_{\ell equ})^* C^{(1)\tau \mu 3u}_{\ell equ }$ & $5.8 \times 10^{-5}(c) \times 4.3 \times 10^{-5}(c) $
			&$4 \times 10^{-8}  $ \\
			&&\\
			$(C^{(3)\tau e 1 t}_{\ell equ })^* C^{(3)\tau \mu 1t}_{\ell equ }$ & ---$ \times  $---
			&$ 1 \times 10^{-10}  $ \\
			$(C^{(3)\tau e 3u}_{\ell equ})^* C^{(3)\tau \mu 3u}_{\ell equ }$ & $2.4 \times 10^{-4}(c) \times 2.4 \times 10^{-4}(c)  $
			&$2.5 \times 10^{-10}  $ \\
			&&\\
			$(C^{(1)\tau e 1 t}_{\ell equ})^* C^{(3)\tau \mu 1t}_{\ell equ }$ & ---$ \times  $---
			&$2 \times 10^{-9}  $ \\
			$(C^{(3)\tau e 1 t}_{\ell equ})^* C^{(1)\tau \mu 1t}_{\ell equ }$ & ---$ \times  $---
			&$2 \times 10^{-9}  $ \\
			$(C^{(1)\tau e 3u}_{\ell equ})^* C^{(3)\tau \mu 3u}_{\ell equ }$ & $5.8 \times 10^{-5}(c) \times2.4 \times 10^{-4}(c)  $
			&$4 \times 10^{-9}  $ \\
			$(C^{(3)\tau e 3u}_{\ell equ})^* C^{(1)\tau \mu 3u}_{\ell equ }$ & $ 2.4 \times 10^{-4}(c)\times 4.3 \times 10^{-5}(c) $
			&$4 \times 10^{-9}  $ \\
			\hline
		\end{tabular}
		\caption{Similar to tables \ref{tab:Bdecaysscalarup}, for
			dimension eight  $\mue$ vector  operators with  SU(2) singlet leptons (see eq.s~(\ref{eq:dim8vectorsinglets1})-(\ref{eq:dim8vectorsinglets2})). 
			\label{tab:BdecaysvectR}.} 
	\end{center}
\end{table}

\begin{table}[!h]
	\begin{center}
		\begin{tabular}{|l|c|c|}
			\hline
			coefficients &     $B^{(c)}_{\te} B^{(c)}_{\tm}$  & $B^{(f)}_{\mue}$  \\
			\hline
			$C^{(1)e\tau 31}_{\ell q} C^{(1)\tau \mu 13}_{\ell q }$ & $2.3 \times 10^{-3}(c) \times 1.5 \times 10^{-3}(c)  $
			&$1 \times 10^{-8}  $ \\
			$C^{(1)e\tau 13}_{\ell q} C^{(1)\tau \mu 31}_{\ell q }$ & $2.3 \times 10^{-3}(c) \times 1.5 \times 10^{-3}(c)  $
			&$2.5 \times 10^{-9}  $ \\
			&&\\
			$C^{(3)e\tau 31}_{\ell q} C^{(3)\tau \mu 13}_{\ell q }$ & $2.3 \times 10^{-3}(c) \times 1.5 \times 10^{-3}(c)  $
			&$2 \times 10^{-9}  $ \\
			$C^{(3)e\tau 13}_{\ell q} C^{(3)\tau \mu 31}_{\ell q }$ & $2.3 \times 10^{-3}(c) \times 1.5 \times 10^{-3}(c)  $
			&$2.5 \times 10^{-9}  $ \\
			&&\\
			$C^{(3)e\tau 13}_{\ell q} C^{(1)\tau \mu 31}_{\ell q }$ & $2.3 \times 10^{-3}(c) \times 1.5 \times 10^{-3}(c)  $
			&$2.5 \times 10^{-9}  $ \\
			$C^{(1)e\tau 13}_{\ell q} C^{(3)\tau \mu 31}_{\ell q }$ & $2.3 \times 10^{-3}(c) \times 1.5 \times 10^{-3}(c)  $ 
			&$2.5 \times 10^{-9}  $ \\
			$C^{(3)e\tau 31}_{\ell q} C^{(1)\tau \mu 13}_{\ell q }$ & $2.3 \times 10^{-3}(c) \times 1.5 \times 10^{-3}(c)  $
			&$1 \times 10^{-8}  $ \\
			$C^{(1)e\tau 31}_{\ell q} C^{(3)\tau \mu 13}_{\ell q }$ & $2.3 \times 10^{-3}(c) \times 1.5 \times 10^{-3}(c)  $ 
			&$1 \times 10^{-8}  $ \\
			&&\\
			$C^{e\tau ut}_{\ell u} C^{\tau \mu tu}_{\ell u }$ & ---$ \times  $---
			&$1 \times 10^{-8}  $ \\
			$C^{e\tau tu}_{\ell u} C^{\tau \mu ut}_{\ell u }$ & ---$ \times  $---
			&$2.5\times 10^{-9}  $ \\
			&&\\%
			$(C^{(1)e\tau  3u}_{\ell equ})^* C^{(1)\mu \tau  3u}_{\ell equ }$ &  $4.5 \times 10^{-4}(c) \times 4.5 \times 10^{-4}(c)  $
			&$4 \times 10^{-8}  $ \\
			$(C^{(1)e \tau  1 t}_{\ell equ})^* C^{(1)\mu \tau 1t}_{\ell equ }$ & ---$ \times  $---
			&$4 \times 10^{-8}  $ \\
			&&\\
			$(C^{(3)e\tau  3u}_{\ell equ})^* C^{(3)\mu \tau  3u}_{\ell equ }$ &  $1.8 \times 10^{-3}(c) \times 1.8 \times 10^{-3}(c)  $
			&$1.25 \times 10^{-10}  $ \\
			$(C^{(3)e\tau  1 t}_{\ell equ})^* C^{(3)\mu \tau  1t}_{\ell equ }$ & ---$ \times  $---
			&$1.25 \times 10^{-10}  $ \\
			&&\\
			$(C^{(1)e\tau  1 t}_{\ell equ})^* C^{(3) \mu\tau  1t}_{\ell equ }$ & ---$ \times  $---
			&$3 \times 10^{-9}  $ \\
			$(C^{(1)e\tau  3u}_{\ell equ})^* C^{(3)\mu\tau 3u}_{\ell equ }$ & $4.5 \times 10^{-5}(c) \times  1.8 \times 10^{-3}(c)$
			&$1.6 \times 10^{-9}  $ \\
			$(C^{(3)e\tau  1 t}_{\ell equ})^* C^{(1) \mu\tau  1t}_{\ell equ }$ &---$ \times  $--- 
			&$3 \times 10^{-9}  $ \\
			$(C^{(3)e\tau  3u}_{\ell equ})^* C^{(1)\mu\tau  3u}_{\ell equ }$ & $  1.8 \times 10^{-3}(c)\times 4.5 \times 10^{-5}(c) $
			&$1.6 \times 10^{-9}  $ \\
			\hline
		\end{tabular}
		\caption{  
			Similar to tables \ref{tab:Bdecaysscalarup},  to generate  $\mue$  vector operators with a doublet  lepton bilinear (see eq.s~(\ref{eq:dim8vectordoublets1})-(\ref{eq:dim8vectordoublets2})).
			\label{tab:BdecaysvectL}} 
	\end{center}
\end{table}
\clearpage

\subsection{Higgs LFV couplings}\label{ssec:HiggsLFV}
In this section we discuss the sensitivities of $\mu\to e$ observables  to dimension six  Yukawa operators ${\cal O}_{eH}$ (eq.~(\ref{eq:yukdim6})), and  compare them   with  the upcoming direct limits imposed by $h\to \tau l$ decays.
Pairs of Yukawa $\tl$ operators contribute to various $\mu\to e$ interactions at dimension eight.
They mix into penguins via the divergent diagrams of Figure \ref{subfig:3}, which match onto the vector operators involved at tree-level in the $\mu\to e$ conversion and $\mu\to \bar{e}ee$ rates.
In addition, dimension six Yukawas are inserted in the diagrams of Figures \ref{subfig:matching2}-\ref{subfig:matching3}, that give matching contributions to the  $\mu e\tau\tau$ tensor and dipole respectively. The matching conditions are written in eq.s (\ref{eq:matchdipoleR}) and (\ref{eq:tensormatchingR})-(\ref{eq:tensormatchingL}).
$\meg$ is marginally more sensitive to the $\mu e\tau\tau$ tensor than on the dipole; this is due to the large tensor-to-dipole mixing and the built-in $y_\mu$ Yukawa suppression in the dipole definition, which lead to the already discussed enhancement $m_\tau/m_\mu$.
As a result, $\meg$ is the  most sensitive process, and an
upcoming experimental reach of   $Br(\meg)\lesssim 10^{-14}$
gives :
\begin{equation}
	\abs{C^{e\tau}_{eH} C^{\tau\mu}_{eH}},\abs{C^{\tau e}_{eH} C^{\mu\tau}_{eH}}\lesssim 6 \times 10^{-9}. \label{eq:HiggsdecayLFV}
\end{equation} 
In the charged lepton mass-eigenstate basis,  the dimension six Yukawas
induce flavour-changing interactions of 125 G$e$V-Higgs (see eq.~\ref{eq:LFVhiggs}), so  $h\to \tau l$ decays probe the off-diagonal coefficients $C^{\tau l, l\tau}_{eH}$.
The most stringent upper limits on the rates are currently set by CMS \cite{cms2021}, and ILC is expected to improve them by one order of magnitude \cite{ILC}. The projected sensitivities to the branching ratios lead to the bounds
\begin{align}
	Br(h\to \tau e)<2.3\times 10^{-4}\qquad &\to \qquad \sqrt{\abs{C_{eH}^{e\tau}}^2+\abs{C_{eH}^{\tau e}}^2}<3.2\times 10^{-4}\nonumber\\
	Br(h\to \tau \mu)<2.4\times 10^{-4}\qquad &\to \qquad \sqrt{\abs{C_{eH}^{\tau\mu}}^2+\abs{C_{eH}^{\mu\tau }}^2}<3\times 10^{-4} \label{eq:sensHiggsdecay}.
\end{align}
The product of the direct limits is larger than $2\times$ the sensitivity of eq.~(89), so that $\mu\to e$ probe a region of parameter space that is beyond the reach of future LFV Higgs decays (see Figure \ref{fig:circlevshyperbola}). 

\clearpage

\section{Summary}
The $\mu\to e$ experiments under construction are expected to improve the current branching ratio sensitivities by several orders of magnitude. In some cases the improvement is such that the upcoming experiments will be able to probe contributions to $\mu\to e$ observables that are the result of combined $\mu\to \tau$ and $\tau\to e$ interactions in loop diagrams, beyond the reach of direct $\tl$ searches (where $l \in \{e,\mu\}$).
However, the relationship between $\tl$ and $\me$ observables {is generically
	model-dependent}, as we discussed in Section \ref{ssec:models}. The goal of this paper is to retain the model-independent contributions to $\mu\to e$ processes from $\tl$ lepton flavour change, although these may be subdominant. To do so, we assume that the New Physics responsible for $\tl$ LFV is heavy ($\Lambda_{\rm NP}\gtrsim 4$ T$e$V) and we parameterise it with $\tl$ dimension six operators in the  ``on-shell'' operator basis of SMEFT.
We briefly introduce our EFT formalism in section  \ref{ssec:EFT}.

We insert $\mu\to \tau$ and $\tau\to e$ dimension six interactions $\order{1/\LNP^2}$ in diagrams that generates $\mu\to e$ amplitudes at dimension eight $\order{1/\LNP^4}$. We only compute the contributions that are phenomenologically relevant, i.e within the reach of future experiments. Firstly, we focus on a subspace of dimension eight operators to which $\mu\to e$ observables are sensitive, as given in \cite{Ardu:2021koz} and presented in Section \ref{sec:Operators}. Secondly, in Section \ref{ssec:estimates} we draw and estimate diagrams with two $\tl$ dimension six interactions generating the above-mentioned dimension eight operators, and we disregard the contributions smaller than the upcoming experimental sensitivity. 

Log-enhanced corrections to $\mu\to e$ dimension eight coefficients are the result of the $\text{(dimension 6)}^2\to \text{(dimension 8)}$ mixing which appear in the Renormalization Group evolution, that we review in Section \ref{ssec:EFT}. Calculating this mixing present some technical challenges.
The ``on-shell" operator bases   we use at dimension six and eight are reduced using the Equation of Motion { (EOM)}, i.e do not contain operators that are related by applying the classical EOM on some field.
In order to include the dimension 8 contributions that arise from using the EOM up to dimension 6 in reducing  to the on-shell basis at  dimension 6,
we include some not-1PI diagrams in our calculations.
This is more carefully discussed in Section \ref{ssec:EoM}.

In Section \ref{sec:Calculations} we describe the calculation of the interesting contributions to $\mu\to e$ processes from $\tl$ interactions, depicted in the diagrams of Figure \ref{fig:diagramsRunning} and Figure \ref{fig:matchingdiagrams}.
Pairs of $\tau\leftrightarrow l$ operators are assumed to be generated at a New Physics scale $\Lambda_{\rm NP}=4$ T$e$V and mix into dimension eight $\mu\to e$ interactions when evolved down to the experimental scale of $\mu\to e$ observables. Between $\Lambda_{\rm NP}$ and $m_W$, the running is perfomed in SMEFT as described in section \ref{ssec:SMEFTRunning} and employing the RGEs solution of eq.~(\ref{eq:RGEsolution}). The complete  list of the
$\text{(dimension 6)}^2\to \text{(dimension 8)}$ anomalous dimensions that we obtained  is given in Appendix \ref{appendix:AnomalousDimensions}. 

The dimension eight SMEFT operators that are generated in running are matched onto low energy interactions at $m_W$ as described in \cite{Ardu:2021koz}.
We also include the contribution from  pairs of $\tau\leftrightarrow l$ operators  which generate $\me$ operators at tree level in matching, 
as discussed  in section \ref{ssec:Matching}.  
Between $m_W$ and the experimental scale $\Lambda_{exp}$, the running of low energy  Wilson Coefficients is taken from \cite{Davidson:2020hkf}, while we find that $\mu\to \tau \times \tau\to e$ RGEs mixing is negligible in the EFT below $m_W$, as discussed at the end of section \ref{ssec:EFT}. 

We thus determined the  sensitivity of $\mu\to e$ processes to  products of $\tl$ operator coefficients. Sensitivities represent the largest absolute value that is experimentally detectable and are obtained by considering one non-zero pair of $\tl$ operators at a time. They give a hyperbola in the $C^{[6]\tau \mu}$-$C^{[6]e\tau}$ plane of the dimension six coefficients (see Figure \ref{fig:circlevshyperbola}),
outside which $\mu\to e$ observables can probe. In the same plane, direct $\tl$ searches are sensitive to the region outside an ellipse.
In Section \ref{sec:Pheno} we discuss  two examples where the hyperbola passes inside the ellipse: 
Section \ref{ssec:Fishtop}  shows that the  contributions of  fish diagrams (see Figure \ref{subfig:5}-\ref{subfig:6}) to $\mu\to e$ observables allow to probe  products of $\tl$  coefficients involving third generation quarks. These same interactions contribute to  the rate of LFV $B\to \tau (\nu_\tau)$ meson decays, which can directly probe the size of the Wilson Coefficients (The ``limits" arising from the upper bounds on $B\to \tau (\nu_\tau)$+\dots are summarized in Appendix \ref{appendix:Bdecays}). In most cases, we find that upcoming $\mu\to e$ experiments   are sensitive to coefficients beyond the reach of future $B\to \tau (\nu_\tau)$+\dots searches. 
In Section \ref{ssec:HiggsLFV}, we study the  sensitivity of upcoming  $\mu\to e$ searches  to products of LFV Higgs couplings, which overcomes the projected reach of the ILC to $h\to \tau^\pm l^\mp$. 

In this paper, we computed in SMEFT the contributions to $\mu\to e$ observables arising from $(\mu \to \tau)\times (\tau \to e)$ interactions. This required calculating a subset of the RGEs for  dimension eight operators, so far missing in the literature. As a result, we obtained limits on  products of $\tl$ SMEFT coefficients assuming non-observation of $\mue$ in future experiments. This can give model-independent relations among $\me$, $\te$ and $\tm$ LFV: in the event of a detected $\tm$  signal, the non-observation of $\me$  would suggest that some $\te$ interactions are unlikely(if they occur, additional $\me$ interactions are required to  obtain  a cancellation in the $\me$ amplitude). 
This could provide theoretical guidance on where to search, or not, for $\te$.

We find that $\mu\to e$  processes have a good sensitivity to products of  $\tl$  operators that involve  $b$ quarks.
These mediate leptonic flavour changing $B$ decays, which are a promising avenue for New Physics in light of the recent anomalies. In most cases the
anomalous rates are below the SM expectations, requiring destructive interference with the SM that cannot be addressed by our LFV operators. An exception is the $R_{D^*}$ anomaly, where the experimental value is larger than the SM prediction and, as discussed in Section \ref{ssec:Banom}
(see Figure \ref{fig:fitBanom}), can be fitted by increasing the rate of $B\to D^* \tau \nu$ with $\te$ operators. This is an example of the above-discussed relations that we can extrapolate from our calculation; the non-observation of $\mue$ processes can identify values where $\tm$ is unlikely to be seen.

$~$\\

\subsection*{Acknowledgements}
The work of MG has been supported by STFC under the Consolidated Grant ST/T000988/1. MA is supported by a doctoral fellowship from the IN2P3.

\appendix
\section{Feynman Rules}\label{appendix:FeynRules}
In this section we list the Feynman Rules for the interactions involved in the diagrams of section \ref{ssec:SMEFTRunning}. Capital letters $I,J,L,K\dots$ are used to label SU(2) indices, while lower-case letters $i,j,l,k$ are generation indices. $\tau^a$ are the Pauli matrices and $\epsilon_{12}=-\epsilon_{21}=1, \epsilon_{11}=\epsilon_{22}=0$ is the anti-symmetric SU(2) tensor. The Feynman rules are obtained calculating by hand the $iM$ amplitude of the tree-level processes.
\begin{figure}[!h]
	\centering
	\begin{subfigure}{.3\textwidth}
		\centering
		\begin{tikzpicture}[scale=1.2, baseline={([yshift=-.5ex]current bounding box.center)}]
			\begin{feynman}[small] 
				\vertex (a) at (-1,-1){\(\ell_{iJ}\)};
				\vertex (b) at (0,0) ;
				\vertex (c) at (1,-1) {\(\ell_{iI}\)};
				\vertex (d) at (0,1) {\(B\)};
				
				\diagram*[inline=b]{
					(a)  --  [fermion] (b)  -- [fermion] (c),
					(b) --  [photon] (d),
				};
			\end{feynman}
		\end{tikzpicture}$-ig'Y(\ell) \delta_{IJ}\gamma^\alpha P_L$ 
	\end{subfigure}
	\begin{subfigure}{.3\textwidth}
		\centering
		\begin{tikzpicture}[scale=1.2, baseline={([yshift=-.5ex]current bounding box.center)}]
			\begin{feynman}[small]
				\vertex (a) at (-1,-1){\(\ell_{iJ}\)};
				\vertex (b) at (0,0) ;
				\vertex (c) at (1,-1) {\(\ell_{iI}\)};
				\vertex (d) at (0,1) {\(W^a\)};
				
				\diagram* {
					(a)  --  [fermion] (b)  -- [fermion] (c),
					(b) --  [photon] (d),
				};
			\end{feynman}
		\end{tikzpicture}$-ig\frac{\tau^a_{IJ}}{2}\gamma^\alpha P_L$
	\end{subfigure}
	\begin{subfigure}{.3\textwidth}
		\centering
		\begin{tikzpicture}[scale=1.2, baseline={([yshift=-.5ex]current bounding box.center)}]
			\begin{feynman}[small]
				\vertex (a) at (-1,-1){\(e_{i}\)};
				\vertex (b) at (0,0) ;
				\vertex (c) at (1,-1) {\(e_{i}\)};
				\vertex (d) at (0,1) {\(B\)};
				
				\diagram* {
					(a)  --  [fermion] (b)  -- [fermion] (c),
					(b) --  [photon] (d),
				};
			\end{feynman}
		\end{tikzpicture}$-ig'Y(e)\gamma^\alpha P_R$ 
	\end{subfigure}\\
	\begin{subfigure}{.3\textwidth}
		\centering
		\begin{tikzpicture}[scale=1.2, baseline={([yshift=-.5ex]current bounding box.center)}]
			\begin{feynman}[small]
				\vertex (a) at (-1,-1){\(q_{jJ}\)};
				\vertex (b) at (0,0) ;
				\vertex (c) at (1,-1) {\(q_{iI}\)};
				\vertex (d) at (0,1) {\(B\)};
				
				\diagram* {
					(a)  --  [fermion] (b)  -- [fermion] (c),
					(b) --  [photon] (d),
				};
			\end{feynman}
		\end{tikzpicture}$-ig'Y(q) \delta_{IJ}\gamma^\alpha P_L$
	\end{subfigure}
	\begin{subfigure}{.3\textwidth}
		\centering
		\begin{tikzpicture}[scale=1.2, baseline={([yshift=-.5ex]current bounding box.center)}]
			\begin{feynman}[small]
				\vertex (a) at (-1,-1){\(q_{iJ}\)};
				\vertex (b) at (0,0) ;
				\vertex (c) at (1,-1) {\(q_{iI}\)};
				\vertex (d) at (0,1) {\(W^a\)};
				
				\diagram* {
					(a)  --  [fermion] (b)  -- [fermion] (c),
					(b) --  [photon] (d),
				};
			\end{feynman}
		\end{tikzpicture}$-ig\frac{\tau^a_{IJ}}{2}\gamma^\alpha P_L$
	\end{subfigure}
	\begin{subfigure}{.3\textwidth}
		\centering
		\begin{tikzpicture}[scale=1.2, baseline={([yshift=-.5ex]current bounding box.center)}]
			\begin{feynman}[small]
				\vertex (a) at (-1,-1){\(u_{i}\)};
				\vertex (b) at (0,0) ;
				\vertex (c) at (1,-1) {\(u_{i}\)};
				\vertex (d) at (0,1) {\(B\)};
				
				\diagram* {
					(a)  --  [fermion] (b)  -- [fermion] (c),
					(b) --  [photon] (d),
				};
			\end{feynman}
		\end{tikzpicture}$-ig'Y(u)\gamma^\alpha P_R$
	\end{subfigure}\\
	\begin{subfigure}{.3\textwidth}
		\centering
		\begin{tikzpicture}[scale=1.2, baseline={([yshift=-.5ex]current bounding box.center)}]
			\begin{feynman}[small]
				\vertex (a) at (-1,-1){\(d_{i}\)};
				\vertex (b) at (0,0) ;
				\vertex (c) at (1,-1) {\(d_{i}\)};
				\vertex (d) at (0,1) {\(B\)};
				
				\diagram* {
					(a)  --  [fermion] (b)  -- [fermion] (c),
					(b) --  [photon] (d),
				};
			\end{feynman}
		\end{tikzpicture}$-ig'Y(d)\gamma^\alpha P_R$
	\end{subfigure}\qquad
	\begin{subfigure}{.3\textwidth}
		\centering
		\begin{tikzpicture}[scale=1.2, baseline={([yshift=-.5ex]current bounding box.center)}]
			\begin{feynman}[small]
				\vertex (a) at (-1,-1){\(e_{i}\)};
				\vertex (b) at (0,0) ;
				\vertex (c) at (1,-1) {\(\ell_{iI}\)};
				\vertex (d) at (0,1) {\(H_J\)};
				
				\diagram* {
					(a)  --  [fermion] (b)  -- [fermion] (c),
					(b) --  [anti charged scalar] (d),
				};
			\end{feynman}
		\end{tikzpicture}$-iy_{e_i}\delta_{IJ}P_R$ 
	\end{subfigure}
	\begin{subfigure}{.3\textwidth}
		\centering
		\begin{tikzpicture}[scale=1.2, baseline={([yshift=-.5ex]current bounding box.center)}]
			\begin{feynman}[small]
				\vertex (a) at (-1,-1){\(u_{i}\)};
				\vertex (b) at (0,0) ;
				\vertex (c) at (1,-1) {\(q_{iI}\)};
				\vertex (d) at (0,1) {\(H_J\)};
				
				\diagram* {
					(a)  --  [fermion] (b)  -- [fermion] (c),
					(b) --  [charged scalar] (d),
				};
			\end{feynman}
		\end{tikzpicture}$-iy_{u_i}\epsilon_{IJ}P_R$
	\end{subfigure}\\
	\begin{subfigure}{.45\textwidth}
		\centering
		\begin{tikzpicture}[scale=1.2, baseline={([yshift=-.5ex]current bounding box.center)}]
			\begin{feynman}[small]
				\vertex (a) at (-1,-1){\(H_{J}\)};
				\vertex (b) at (0,0) ;
				\vertex (c) at (1,-1) {\(H_{I}\)};
				\vertex (d) at (0,1) {\(B\)};
				
				\diagram* {
					(a)  --  [charged scalar] (b)  -- [charged scalar] (c),
					(b) --  [photon] (d),
				};
			\end{feynman}
		\end{tikzpicture}$-ig'Y(H)\delta_{IJ}(p_{H_I}+p_{H_J})_\alpha$ 
	\end{subfigure}
	\begin{subfigure}{.45\textwidth}
		\centering
		\begin{tikzpicture}[scale=1.2, baseline={([yshift=-.5ex]current bounding box.center)}]
			\begin{feynman}[small]
				\vertex (a) at (-1,-1){\(H_{J}\)};
				\vertex (b) at (0,0) ;
				\vertex (c) at (1,-1) {\(H_{I}\)};
				\vertex (d) at (0,1) {\(W^a\)};
				
				\diagram* {
					(a)  --  [charged scalar] (b)  -- [charged scalar] (c),
					(b) --  [photon] (d),
				};
			\end{feynman}
		\end{tikzpicture}$-ig\frac{\tau^a_{IJ}}{2}(p_{H_I}+p_{H_J})_\alpha$ 
	\end{subfigure}\\
	\caption{Feynman rules for the dimension four interaction. The Higgs momenta follow the hypercharge arrow.}\label{fig:FeynRulesdim4}
\end{figure}
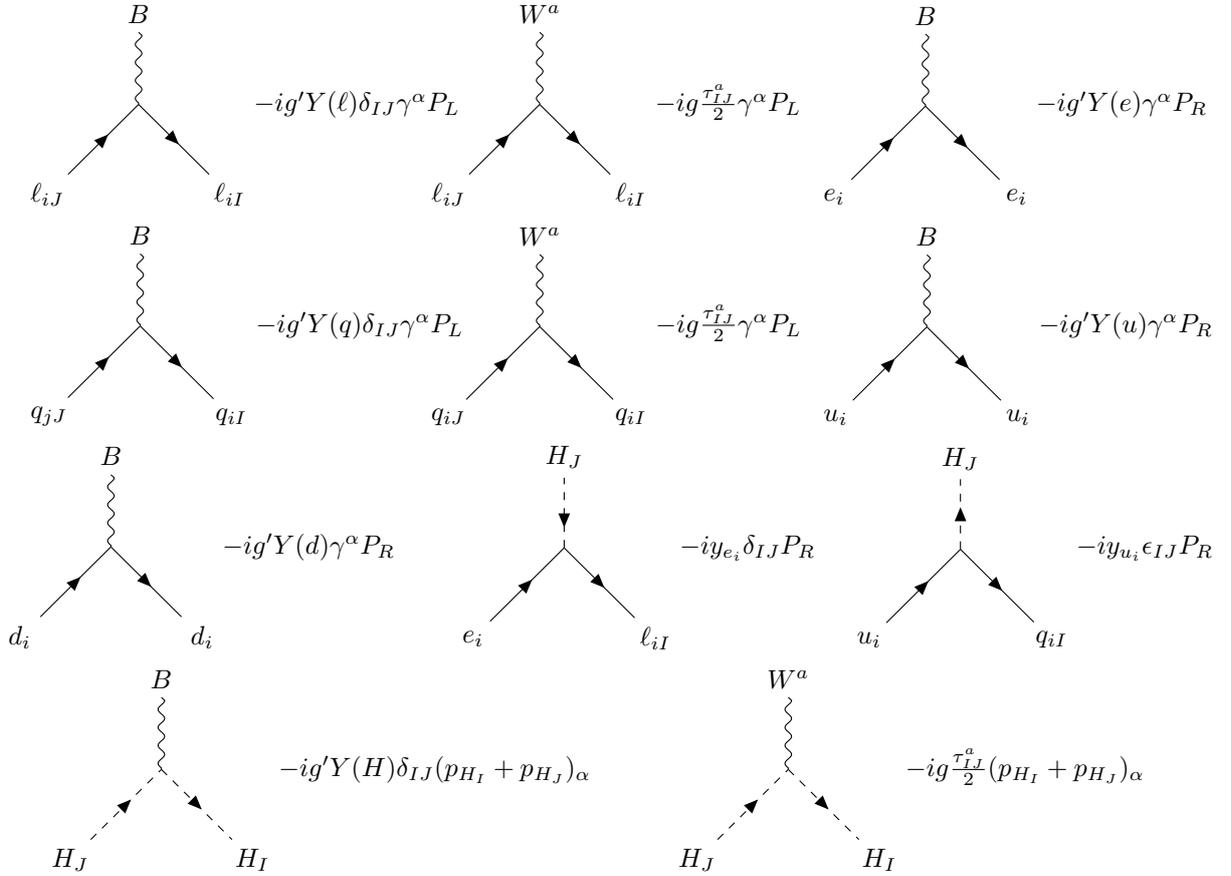

\begin{figure}[!h]
	\centering
	\begin{subfigure}{.4\textwidth}
		\centering
		\begin{tikzpicture}[scale=1.2, baseline={([yshift=-.5ex]current bounding box.center)}]
			\begin{feynman}[small]
				\vertex (a) at (-1,1){\(e_{j}\)};
				\vertex[style=crossed dot] (b) at (0,0) {};
				\vertex (c) at (1,1) {\(e_{i}\)};
				\vertex (d) at (-1,-1) {\(q_{kK}\)};
				\vertex (e) at (1,-1){\(q_{lL}\)};
				
				\diagram* {
					(a)  --  [fermion] (b)  -- [fermion] (c),
					(d) --  [fermion] (b) -- [fermion] (e),
				};
			\end{feynman}
		\end{tikzpicture}$iC^{ijlk}_{eq}\delta_{LK}\gamma^\alpha P_R \otimes \gamma_\alpha P_L$
	\end{subfigure}
	\begin{subfigure}{.4\textwidth}
		\centering
		\begin{tikzpicture}[scale=1.2, baseline={([yshift=-.5ex]current bounding box.center)}]
			\begin{feynman}[small]
				\vertex (a) at (-1,1){\(e_j\)};
				\vertex[style=crossed dot] (b) at (0,0) {};
				\vertex (c) at (1,1) {\(e_i\)};
				\vertex (d) at (-1,-1) {\(u_{k}\)};
				\vertex (e) at (1,-1){\(u_{l}\)};
				
				\diagram* {
					(a)  --  [fermion] (b)  -- [fermion] (c),
					(d) --  [fermion] (b) -- [fermion] (e),
				};
			\end{feynman}
		\end{tikzpicture}$iC^{ijlk}_{eu}\gamma^\alpha P_R \otimes \gamma_\alpha P_R $
	\end{subfigure}\\
	\begin{subfigure}{.4\textwidth}
		\centering
		\begin{tikzpicture}[scale=1.2, baseline={([yshift=-.5ex]current bounding box.center)}]
			\begin{feynman}[small]
				\vertex (a) at (-1,1){\(e_{j}\)};
				\vertex[style=crossed dot] (b) at (0,0) {};
				\vertex (c) at (1,1) {\(e_{i}\)};
				\vertex (d) at (-1,-1) {\(d_{k}\)};
				\vertex (e) at (1,-1){\(d_{l}\)};
				
				\diagram* {
					(a)  --  [fermion] (b)  -- [fermion] (c),
					(d) --  [fermion] (b) -- [fermion] (e),
				};
			\end{feynman}
		\end{tikzpicture}$iC^{ijlk}_{ed}\gamma^\alpha P_R \otimes \gamma_\alpha P_R$
	\end{subfigure}
	\begin{subfigure}{.4\textwidth}
		\centering
		\begin{tikzpicture}[scale=1.2, baseline={([yshift=-.5ex]current bounding box.center)}]
			\begin{feynman}[small]
				\vertex (a) at (-1,1){\(\ell_{jJ}\)};
				\vertex[style=crossed dot] (b) at (0,0) {};
				\vertex (c) at (1,1) {\(\ell_{iI}\)};
				\vertex (d) at (-1,-1) {\(u_{k}\)};
				\vertex (e) at (1,-1){\(u_{l}\)};
				
				\diagram* {
					(a)  --  [fermion] (b)  -- [fermion] (c),
					(d) --  [fermion] (b) -- [fermion] (e),
				};
			\end{feynman}
		\end{tikzpicture}$iC^{ijlk}_{\ell u}\delta_{IJ}\gamma^\alpha P_L \otimes \gamma_\alpha P_R$
	\end{subfigure}\\
	\begin{subfigure}{.4\textwidth}
		\centering
		\begin{tikzpicture}[scale=1.2, baseline={([yshift=-.5ex]current bounding box.center)}]
			\begin{feynman}[small]
				\vertex (a) at (-1,1){\(\ell_{jJ}\)};
				\vertex[style=crossed dot] (b) at (0,0) {};
				\vertex (c) at (1,1) {\(\ell_{iI}\)};
				\vertex (d) at (-1,-1) {\(d_{k}\)};
				\vertex (e) at (1,-1){\(d_{l}\)};
				
				\diagram* {
					(a)  --  [fermion] (b)  -- [fermion] (c),
					(d) --  [fermion] (b) -- [fermion] (e),
				};
			\end{feynman}
		\end{tikzpicture}$iC^{ijlk}_{\ell d}\delta_{IJ}\gamma^\alpha P_L \otimes \gamma_\alpha P_R$
	\end{subfigure}
	\begin{subfigure}{.4\textwidth}
		\centering
		\begin{tikzpicture}[scale=1.2, baseline={([yshift=-.5ex]current bounding box.center)}]
			\begin{feynman}[small]
				\vertex (a) at (-1,1){\(\ell_{jI}\)};
				\vertex[style=crossed dot] (b) at (0,0) {};
				\vertex (c) at (1,1) {\(\ell_{iI}\)};
				\vertex (d) at (-1,-1) {\(q_{kK}\)};
				\vertex (e) at (1,-1){\(q_{lL}\)};
				
				\diagram* {
					(a)  --  [fermion] (b)  -- [fermion] (c),
					(d) --  [fermion] (b) -- [fermion] (e),
				};
			\end{feynman}
		\end{tikzpicture}$\begin{array}{c}     i(C^{(1)ijlk}_{\ell q}\delta_{IJ}\delta_{KL}+\nonumber\\+C^{(3)ijlk}_{\ell q}\tau^a_{IJ}\tau^a_{LK})\gamma^\alpha P_L \otimes \gamma_\alpha P_L \nonumber \end{array} $
	\end{subfigure} \\
	\begin{subfigure}{.4\textwidth}
		\centering
		\begin{tikzpicture}[scale=1.2, baseline={([yshift=-.5ex]current bounding box.center)}]
			\begin{feynman}[small]
				\vertex (a) at (-1,1){\(e_{j}\)};
				\vertex[style=crossed dot] (b) at (0,0) {};
				\vertex (c) at (1,1) {\(\ell_{iI}\)};
				\vertex (d) at (-1,-1) {\(q_{kK}\)};
				\vertex (e) at (1,-1){\(d_{l}\)};
				
				\diagram* {
					(a)  --  [fermion] (b)  -- [fermion] (c),
					(d) --  [fermion] (b) -- [fermion] (e),
				};
			\end{feynman}
		\end{tikzpicture}$iC^{ijlk}_{\ell edq}\delta_{IK} P_R \otimes P_L$
	\end{subfigure}
	\begin{subfigure}{.4\textwidth}
		\centering
		\begin{tikzpicture}[scale=1.2, baseline={([yshift=-.5ex]current bounding box.center)}]
			\begin{feynman}[small]
				\vertex (a) at (-1,1){\(e_{j}\)};
				\vertex[style=crossed dot] (b) at (0,0) {};
				\vertex (c) at (1,1) {\(\ell_{iI}\)};
				\vertex (d) at (-1,-1) {\(u_{k}\)};
				\vertex (e) at (1,-1){\(q_{lL}\)};
				
				\diagram* {
					(a)  --  [fermion] (b)  -- [fermion] (c),
					(d) --  [fermion] (b) -- [fermion] (e),
				};
			\end{feynman}
		\end{tikzpicture}$\begin{array}{c}i(C^{(1)ijlk}_{\ell equ}\epsilon_{IL} P_R \otimes P_R+\\+C^{(3)ijlk}_{\ell equ}\epsilon_{IL} \sigma^{\alpha\beta} P_R \otimes \sigma_{\alpha\beta}P_R)\end{array}$
	\end{subfigure} \\ 
	\caption{Feynman rules for the dimension six SMEFT four-fermion interaction $4f_6$ of section \ref{sec:Operators}. In the product $\Gamma_1\otimes \Gamma_2$ the left matrix $\Gamma_1$ multiplies the lepton bilinear. Scalar and tensor with opposite chiralities have the same Feynman rules with conjugate coefficients and exchanged flavour indices within lepton and quark bilinears.}\label{fig:FeynRules4f6}
\end{figure}
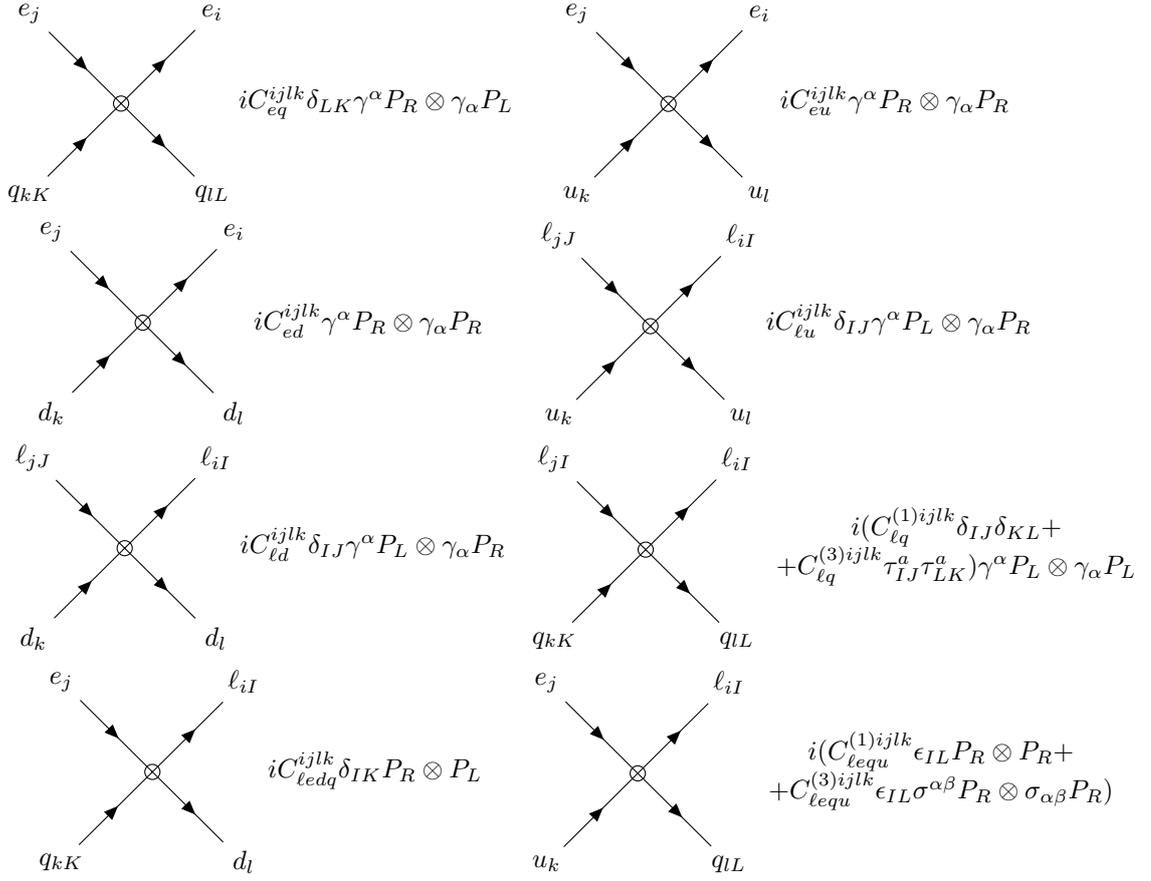

\begin{figure}[!h]
	\centering
	\begin{subfigure}{.45\textwidth}
		\centering
		\begin{tikzpicture}[scale=1.2, baseline={([yshift=-.5ex]current bounding box.center)}]
			\begin{feynman}[small]
				\vertex (a) at (-1,1){\(e_j\)};
				\vertex[style=crossed dot] (b) at (0,0) {};
				\vertex (c) at (1,1) {\(\ell_{iI}\)};
				\vertex (d) at (0,1) {\(H_J\)};
				\vertex (e) at (-1,-1){\(H_K\)};
				\vertex (f) at (1,-1){\(H_L\)};
				
				\diagram* {
					(a)  --  [fermion] (b)  -- [fermion] (c),
					(b) --  [anti charged scalar] (d),
					(b) -- [anti charged scalar] (e),
					(b) -- [ charged scalar] (f),
				};
			\end{feynman}
		\end{tikzpicture}$iC_{eH}(\delta_{IJ}\delta_{LK}+\delta_{IK}\delta_{LJ})P_R$
	\end{subfigure}
	\begin{subfigure}{.45\textwidth}
		\centering
		\begin{tikzpicture}[scale=1.2, baseline={([yshift=-.5ex]current bounding box.center)}]
			\begin{feynman}[small]
				\vertex (a) at (-1,-1){\(\ell_{jJ}\)};
				\vertex[style=crossed dot] (b) at (0,0) {};
				\vertex (c) at (1,-1) {\(\ell_{iI}\)};
				\vertex (d) at (0,1) {\(B\)};
				\vertex (e) at (-1,1){\(H_K\)};
				\vertex (f) at (1,1){\(H_L\)};
				
				\diagram* {
					(a)  --  [fermion] (b)  -- [fermion] (c),
					(b) --  [photon] (d),
					(b) -- [anti charged scalar] (e),
					(b) -- [ charged scalar] (f),
				};
			\end{feynman}
		\end{tikzpicture}$\begin{array}{c}-i2g'Y(H)(C^{ij}_{H\ell(1)}\delta_{IJ}+\\+C^{ij}_{H\ell(3)}\tau^b_{IJ}\tau^b_{LK})\gamma^\alpha P_L\end{array}$
	\end{subfigure}\\
	\begin{subfigure}{.45\textwidth}
		\centering
		\begin{tikzpicture}[scale=1.2, baseline={([yshift=-.5ex]current bounding box.center)}]
			\begin{feynman}[small]
				\vertex (a) at (-1,-1){\(e_{j}\)};
				\vertex[style=crossed dot] (b) at (0,0) {};
				\vertex (c) at (1,-1) {\(e_{i}\)};
				\vertex (d) at (0,1) {\(B\)};
				\vertex (e) at (-1,1){\(H_K\)};
				\vertex (f) at (1,1){\(H_L\)};
				
				\diagram* {
					(a)  --  [fermion] (b)  -- [fermion] (c),
					(b) --  [photon] (d),
					(b) -- [anti charged scalar] (e),
					(b) -- [ charged scalar] (f),
				};
			\end{feynman}
		\end{tikzpicture}$-i2g'Y(H)C^{ij}_{He}\delta_{LK} \gamma^\alpha P_R$
	\end{subfigure}
	\begin{subfigure}{.45\textwidth}
		\centering
		\begin{tikzpicture}[scale=1.2, baseline={([yshift=-.5ex]current bounding box.center)}]
			\begin{feynman}[small]
				\vertex (a) at (-1,-1){\(e_{j}\)};
				\vertex[style=crossed dot] (b) at (0,0) {};
				\vertex (c) at (1,-1) {\(e_{i}\)};
				\vertex (e) at (-1,1){\(H_K\)};
				\vertex (f) at (1,1){\(H_L\)};
				
				\diagram* {
					(a)  --  [fermion] (b)  -- [fermion] (c),
					(b) -- [anti charged scalar] (e),
					(b) -- [ charged scalar] (f),
				};
			\end{feynman}
		\end{tikzpicture}$iC^{ij}_{He}\delta_{LK}(\slashed{p}_{H_K}+\slashed{p}_{H_L})P_R$
	\end{subfigure}\\
	\begin{tikzpicture}[scale=1.2, baseline={([yshift=-.5ex]current bounding box.center)}]
		\begin{feynman}[small]
			\vertex (a) at (-1,-1){\(\ell_{jJ}\)};
			\vertex[style=crossed dot] (b) at (0,0) {};
			\vertex (c) at (1,-1) {\(\ell_{iI}\)};
			\vertex (e) at (-1,1){\(H_K\)};
			\vertex (f) at (1,1){\(H_L\)};
			
			\diagram* {
				(a)  --  [fermion] (b)  -- [fermion] (c),
				(b) -- [anti charged scalar] (e),
				(b) -- [ charged scalar] (f),
			};
	\end{feynman}\end{tikzpicture}$\begin{array}{c}i(C^{ij}_{H\ell(1)}\delta_{IJ}\delta_{LK}\\+C^{ij}_{H\ell(3)}\tau^a_{IJ}\tau^a_{LK})(\slashed{p}_{H_K}+\slashed{p}_{H_L})P_L\end{array}$
	\caption{Feynman rules for the dimension six SMEFT two fermion operators $Y_6,P_6$ of section \ref{sec:Operators}. The Higgs momenta follow the hypercharge arrows.}\label{fig:FeynRules2f6}
\end{figure}
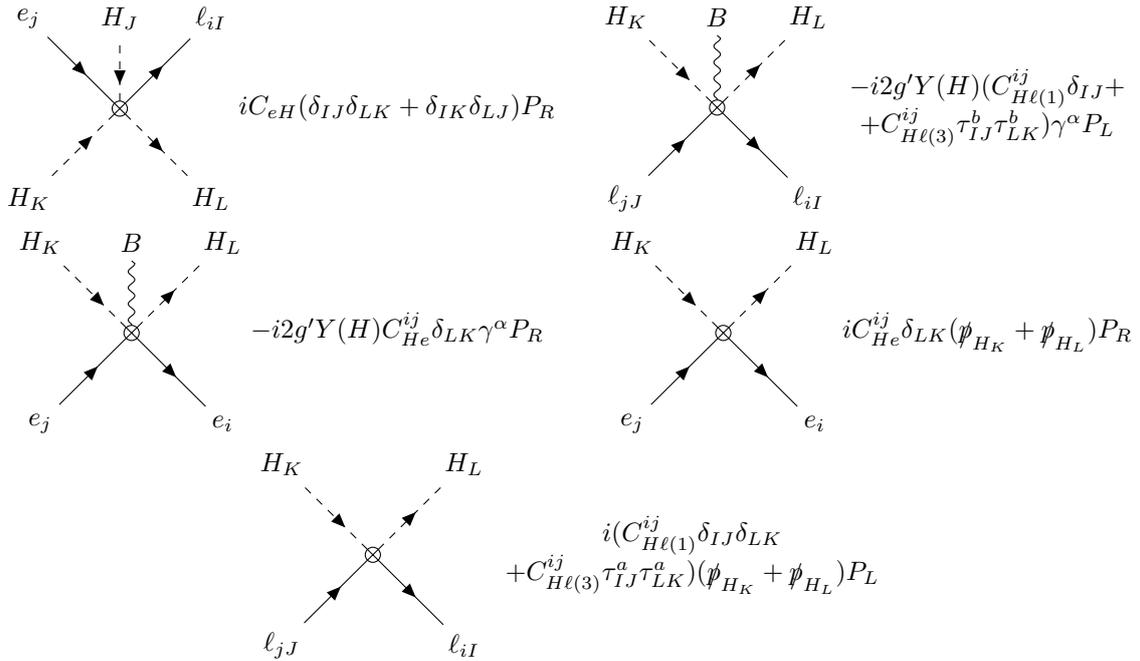

\begin{figure}[!h]
	\centering
	\begin{subfigure}{.45\textwidth}
		\centering
		\begin{tikzpicture}[scale=1.2, baseline={([yshift=-.5ex]current bounding box.center)}]
			\begin{feynman}[small]
				\vertex (a) at (-1,-1){\(e_{j}\)};
				\vertex[style=crossed dot] (b) at (0,0) {};
				\vertex (c) at (1,-1) {\(\ell_{iI}\)};
				\vertex (d) at (0,1) {\(B(q)\)};
				\vertex (e) at (-1,1){\(H_K\)};
				\vertex (f) at (1,1){\(H_L\)};
				\vertex (g) at (0,-1) {\(H_J\)};
				
				\diagram* {
					(a)  --  [fermion] (b)  -- [fermion] (c),
					(b) --  [photon] (d),
					(b) -- [anti charged scalar] (e),
					(b) -- [ charged scalar] (f),
					(b) -- [anti charged scalar] (g),
				};
			\end{feynman}
		\end{tikzpicture}$\begin{array}{c}-2C^{ij}_{\ell eBH^3}q_\alpha \sigma^{\alpha\beta}P_R(\delta_{IJ}\delta_{LK}\\+\delta_{IK}\delta_{LJ})\end{array}$
	\end{subfigure}\qquad
	\begin{subfigure}{.45\textwidth}
		\centering
		\begin{tikzpicture}[scale=1.2, baseline={([yshift=-.5ex]current bounding box.center)}]
			\begin{feynman}[small]
				\vertex (a) at (-1,-1){\(e_{j}\)};
				\vertex[style=crossed dot] (b) at (0,0) {};
				\vertex (c) at (1,-1) {\(\ell_{iI}\)};
				\vertex (d) at (0,1) {\(W^a(q)\)};
				\vertex (e) at (-1,1){\(H_K\)};
				\vertex (f) at (1,1){\(H_L\)};
				\vertex (g) at (0,-1) {\(H_J\)};
				
				\diagram* {
					(a)  --  [fermion] (b)  -- [fermion] (c),
					(b) --  [photon] (d),
					(b) -- [anti charged scalar] (e),
					(b) -- [ charged scalar] (f),
					(b) -- [anti charged scalar] (g),
				};
			\end{feynman}
		\end{tikzpicture}$\begin{array}{c}-2 q_\alpha\sigma^{\alpha\beta}P_R[C^{(1)ij}_{\ell eWH^3}(\tau^a_{IJ}\delta_{LK}+\tau^a_{IK}\delta_{LJ})\\+C^{(2)ij}_{\ell eWH^3}(\delta_{IJ}\tau^a_{LK}+\delta_{IK}\tau^a_{LJ})]\end{array}$
	\end{subfigure}\\
	\begin{subfigure}{.45\textwidth}
		\centering
		\begin{tikzpicture}[scale=1.2, baseline={([yshift=-.5ex]current bounding box.center)}]
			\begin{feynman}[small]
				\vertex (a) at (-1,-1){\(e_{j}\)};
				\vertex[style=crossed dot] (b) at (0,0) {};
				\vertex (c) at (1,-1) {\(e_{i}\)};
				\vertex (d) at (0,1) {\(B\)};
				\vertex (e) at (-1,1){\(H_K\)};
				\vertex (f) at (1,1){\(H_L\)};
				\vertex (g) at (-1,0) {\(H_N\)};
				\vertex (h) at (1,0) {\(H_M\)};
				
				\diagram* {
					(a)  --  [fermion] (b)  -- [fermion] (c),
					(b) --  [photon] (d),
					(b) -- [anti charged scalar] (e),
					(b) -- [ charged scalar] (f),
					(g) -- [charged scalar] (b) -- [charged scalar] (h),
				};
			\end{feynman}
		\end{tikzpicture}$\begin{array}{c}-4ig'Y(H)C^{ij}_{e^2H^4D}(\delta_{LN}\delta_{MK}+\\+\delta_{LK}\delta_{MN})\gamma^\alpha P_R\end{array}$
	\end{subfigure}\qquad
	\begin{subfigure}{.45\textwidth}
		\centering
		\begin{tikzpicture}[scale=1.2, baseline={([yshift=-.5ex]current bounding box.center)}]
			\begin{feynman}[small]
				\vertex (a) at (-1,-1){\(e_{j}\)};
				\vertex[style=crossed dot] (b) at (0,0) {};
				\vertex (c) at (1,-1) {\(e_{i}\)};
				\vertex (d) at (0,1) {\(W^a\)};
				\vertex (e) at (-1,1){\(H_K\)};
				\vertex (f) at (1,1){\(H_L\)};
				\vertex (g) at (-1,0) {\(H_N\)};
				\vertex (h) at (1,0) {\(H_M\)};
				
				\diagram* {
					(a)  --  [fermion] (b)  -- [fermion] (c),
					(b) --  [photon] (d),
					(b) -- [anti charged scalar] (e),
					(b) -- [ charged scalar] (f),
					(g) -- [charged scalar] (b) -- [charged scalar] (h),
				};
			\end{feynman}
		\end{tikzpicture}$\begin{array}{c}-igC^{ij}_{e^2H^4D}\bigg(\tau^a_{LN}\delta_{MK}+\\+\tau^a_{MN}\delta_{LK}+ (N\leftrightarrow K)\bigg)\gamma^\alpha P_R\end{array}$
	\end{subfigure}\\
	\begin{subfigure}{.45\textwidth}
		\centering
		\begin{tikzpicture}[scale=1.2, baseline={([yshift=-.5ex]current bounding box.center)}]
			\begin{feynman}[small]
				\vertex (a) at (-1,-1){\(\ell_{jJ}\)};
				\vertex[style=crossed dot] (b) at (0,0) {};
				\vertex (c) at (1,-1) {\(\ell_{iI}\)};
				\vertex (d) at (0,1) {\(B\)};
				\vertex (e) at (-1,1){\(H_K\)};
				\vertex (f) at (1,1){\(H_L\)};
				\vertex (g) at (-1,0) {\(H_N\)};
				\vertex (h) at (1,0) {\(H_M\)};
				
				\diagram* {
					(a)  --  [fermion] (b)  -- [fermion] (c),
					(b) --  [photon] (d),
					(b) -- [anti charged scalar] (e),
					(b) -- [ charged scalar] (f),
					(g) -- [charged scalar] (b) -- [charged scalar] (h),
				};
			\end{feynman}
		\end{tikzpicture}$\begin{array}{c}-i4g'Y(H)\bigg[C^{(1)ij}_{\ell^2H^4D}\delta_{IJ}\bigg(\delta_{LN}\delta_{MK}+\delta_{LK}\delta_{MN}\bigg)+\\+C^{(2)ij}_{\ell^2H^4D}\tau^b_{IJ}\bigg(\tau^b_{LN}\delta_{MK}+\tau^b_{MN}\delta_{LK}+ (N\leftrightarrow K)\bigg)\bigg]\gamma^\alpha P_R\end{array}$
	\end{subfigure}\\
	\begin{subfigure}{.45\textwidth}
		\centering
		\begin{tikzpicture}[scale=1.2, baseline={([yshift=-.5ex]current bounding box.center)}]
			\begin{feynman}[small]
				\vertex (a) at (-1,-1){\(\ell_{jJ}\)};
				\vertex[style=crossed dot] (b) at (0,0) {};
				\vertex (c) at (1,-1) {\(\ell_{iI}\)};
				\vertex (d) at (0,1) {\(W^a\)};
				\vertex (e) at (-1,1){\(H_K\)};
				\vertex (f) at (1,1){\(H_L\)};
				\vertex (g) at (-1,0) {\(H_N\)};
				\vertex (h) at (1,0) {\(H_M\)};
				
				\diagram* {
					(a)  --  [fermion] (b)  -- [fermion] (c),
					(b) --  [photon] (d),
					(b) -- [anti charged scalar] (e),
					(b) -- [ charged scalar] (f),
					(g) -- [charged scalar] (b) -- [charged scalar] (h),
				};
			\end{feynman}
		\end{tikzpicture}$\begin{array}{c}-ig\bigg[C^{(1)ij}_{\ell^2H^4D}\delta_{IJ}\bigg(\tau^a_{LN}\delta_{MK}+\tau^a_{MN}\delta_{LK}+ (N\leftrightarrow K)\bigg)+\\+C^{(2)ij}_{\ell^2H^4D}\tau^b_{IJ}\bigg(\delta^{ab}(\delta_{LN}\delta_{MK}+\delta_{LK}\delta_{MN})+\\+\tau^a_{LN}\tau^b_{MK}+\tau^a_{MN}\tau^b_{LK}+(N\leftrightarrow K)\bigg)\bigg]\gamma^\alpha P_R\end{array}$
	\end{subfigure}\\
	\begin{subfigure}{.45\textwidth}
		\centering
		\begin{tikzpicture}[scale=1.2, baseline={([yshift=-.5ex]current bounding box.center)}]
			\begin{feynman}[small]
				\vertex (a) at (-1,-1){\(e_{j}\)};
				\vertex[style=crossed dot] (b) at (0,0) {};
				\vertex (c) at (1,-1) {\(e_{i}\)};
				\vertex (e) at (-1,1){\(H_K\)};
				\vertex (f) at (1,1){\(H_L\)};
				\vertex (g) at (-1,0) {\(H_N\)};
				\vertex (h) at (1,0) {\(H_M\)};
				
				\diagram* {
					(a)  --  [fermion] (b)  -- [fermion] (c),
					(b) -- [anti charged scalar] (e),
					(b) -- [ charged scalar] (f),
					(g) -- [charged scalar] (b) -- [charged scalar] (h),
				};
			\end{feynman}
		\end{tikzpicture}$\begin{array}{c}iC^{ij}_{e^2H^4D}(\slashed{p}_{H_L}+\slashed{p}_{H_M}+\slashed{p}_{H_N}+\slashed{p}_{H_K})(\delta_{LN}\delta_{MK}+\\+\delta_{LK}\delta_{MN}) P_R\end{array}$
	\end{subfigure}\\
	\begin{subfigure}{.45\textwidth}
		\centering
		\begin{tikzpicture}[scale=1.2, baseline={([yshift=-.5ex]current bounding box.center)}]
			\begin{feynman}[small]
				\vertex (a) at (-1,-1){\(\ell_{jJ}\)};
				\vertex[style=crossed dot] (b) at (0,0) {};
				\vertex (c) at (1,-1) {\(\ell_{iI}\)};
				\vertex (e) at (-1,1){\(H_K\)};
				\vertex (f) at (1,1){\(H_L\)};
				\vertex (g) at (-1,0) {\(H_N\)};
				\vertex (h) at (1,0) {\(H_M\)};
				
				\diagram* {
					(a)  --  [fermion] (b)  -- [fermion] (c),
					(b) -- [anti charged scalar] (e),
					(b) -- [ charged scalar] (f),
					(g) -- [charged scalar] (b) -- [charged scalar] (h),
				};
			\end{feynman}
		\end{tikzpicture}$\begin{array}{c}i(\slashed{p}_{H_L}+\slashed{p}_{H_M}+\slashed{p}_{H_N}+\slashed{p}_{H_K})\bigg[C^{(1)ij}_{\ell^2H^4D}\delta_{IJ}(\delta_{LN}\delta_{MK}+\delta_{LK}\delta_{MN})+\\+C^{(2)ij}_{\ell^2H^4D}\tau_{IJ}^a\bigg(\tau^a_{LN}\delta_{MK}+\tau^a_{MN}\delta_{LK}+(N\leftrightarrow K)\bigg)\bigg] P_L\end{array}$
	\end{subfigure}\\
	\caption{Feynman rules for the dimension eight SMEFT two fermion operators $D_8,P_8$ of section \ref{sec:Operators}. The Higgs momenta directions follow the hypercharge arrow, while the bosons momentum $q$ is outgoing}\label{fig:FeynRules2f8}
\end{figure}

\begin{figure}[!h]
	\centering
	\begin{subfigure}{.45\textwidth}
		\centering
		\begin{tikzpicture}[scale=1.2, baseline={([yshift=-.5ex]current bounding box.center)}]
			\begin{feynman}[small]
				\vertex (a) at (-1,1){\(e_j\)};
				\vertex[style=crossed dot] (b) at (0,0) {};
				\vertex (c) at (1,1) {\(e_i\)};
				\vertex (f) at (-1,0) {\(H_K\)};
				\vertex (g) at (1,0) {\(H_L\)};
				\vertex (d) at (-1,-1) {\(u_k\)};
				\vertex (e) at (1,-1){\(u_l\)};
				
				\diagram* {
					(a)  --  [fermion] (b)  -- [fermion] (c),
					(d) --  [fermion] (b) -- [fermion] (e),
					(f) -- [charged scalar] (b),
					(b) -- [charged scalar] (g),
				};
			\end{feynman}
		\end{tikzpicture}\ $iC^{ijlk}_{e^2u^2H^2}\delta_{LK}\gamma^\alpha P_R \otimes \gamma_\alpha P_R$
	\end{subfigure}\qquad
	\begin{subfigure}{.45\textwidth}
		\centering
		\begin{tikzpicture}[scale=1.2, baseline={([yshift=-.5ex]current bounding box.center)}]
			\begin{feynman}[small]
				\vertex (a) at (-1,1){\(e_j\)};
				\vertex[style=crossed dot] (b) at (0,0) {};
				\vertex (c) at (1,1) {\(e_i\)};
				\vertex (f) at (-1,0) {\(H_K\)};
				\vertex (g) at (1,0) {\(H_L\)};
				\vertex (d) at (-1,-1) {\(d_k\)};
				\vertex (e) at (1,-1){\(d_l\)};
				
				\diagram* {
					(a)  --  [fermion] (b)  -- [fermion] (c),
					(d) --  [fermion] (b) -- [fermion] (e),
					(f) -- [charged scalar] (b),
					(b) -- [charged scalar] (g),
				};
			\end{feynman}
		\end{tikzpicture}\ $iC^{ijlk}_{e^2d^2H^2}\delta_{LK}\gamma^\alpha P_R \otimes \gamma_\alpha P_R$
	\end{subfigure}\\
	\begin{subfigure}{.45\textwidth}
		\centering
		\begin{tikzpicture}[scale=1.2, baseline={([yshift=-.5ex]current bounding box.center)}]
			\begin{feynman}[small]
				\vertex (a) at (-1,1){\(\ell_{jJ}\)};
				\vertex[style=crossed dot] (b) at (0,0) {};
				\vertex (c) at (1,1) {\(\ell_{iI}\)};
				\vertex (f) at (-1,0) {\(H_K\)};
				\vertex (g) at (1,0) {\(H_L\)};
				\vertex (d) at (-1,-1) {\(u_k\)};
				\vertex (e) at (1,-1){\(u_l\)};
				
				\diagram* {
					(a)  --  [fermion] (b)  -- [fermion] (c),
					(d) --  [fermion] (b) -- [fermion] (e),
					(f) -- [charged scalar] (b),
					(b) -- [charged scalar] (g),
				};
			\end{feynman}
		\end{tikzpicture}$\begin{array}{c}
			i\big(C^{(1)ijlk}_{\ell^2u^2H^2}\delta_{IJ}\delta_{LK}+\\+C^{(2)ijlk}_{\ell^2u^2H^2}\tau^a_{IJ}\tau^a_{LK}\big)\gamma^\alpha P_L \otimes \gamma_\alpha P_R \end{array}$
	\end{subfigure}\qquad
	\begin{subfigure}{.45\textwidth}
		\centering
		\begin{tikzpicture}[scale=1.2, baseline={([yshift=-.5ex]current bounding box.center)}]
			\begin{feynman}[small]
				\vertex (a) at (-1,1){\(\ell_{jJ}\)};
				\vertex[style=crossed dot] (b) at (0,0) {};
				\vertex (c) at (1,1) {\(\ell_{iI}\)};
				\vertex (f) at (-1,0) {\(H_K\)};
				\vertex (g) at (1,0) {\(H_L\)};
				\vertex (d) at (-1,-1) {\(d_k\)};
				\vertex (e) at (1,-1){\(d_l\)};
				
				\diagram* {
					(a)  --  [fermion] (b)  -- [fermion] (c),
					(d) --  [fermion] (b) -- [fermion] (e),
					(f) -- [charged scalar] (b),
					(b) -- [charged scalar] (g),
				};
			\end{feynman}
		\end{tikzpicture}$\begin{array}{c}
			i\big(C^{(1)ijlk}_{\ell^2d^2H^2}\delta_{IJ}\delta_{LK}+\\+C^{(2)ijlk}_{\ell^2d^2H^2}\tau^a_{IJ}\tau^a_{LK}\big)\gamma^\alpha P_L \otimes \gamma_\alpha P_R \end{array}$
	\end{subfigure}\\
	\begin{subfigure}{.9\textwidth}
		\centering
		\begin{tikzpicture}[scale=1.2, baseline={([yshift=-.5ex]current bounding box.center)}]
			\begin{feynman}[small]
				\vertex (a) at (-1,1){\(\ell_{jJ}\)};
				\vertex[style=crossed dot] (b) at (0,0) {};
				\vertex (c) at (1,1) {\(\ell_{iI}\)};
				\vertex (f) at (-1,0) {\(H_N\)};
				\vertex (g) at (1,0) {\(H_M\)};
				\vertex (d) at (-1,-1) {\(q_{kK}\)};
				\vertex (e) at (1,-1){\(q_{lL}\)};
				
				\diagram* {
					(a)  --  [fermion] (b)  -- [fermion] (c),
					(d) --  [fermion] (b) -- [fermion] (e),
					(f) -- [charged scalar] (b),
					(b) -- [charged scalar] (g),
				};
			\end{feynman}
		\end{tikzpicture}$\begin{array}{c}
			i\big(C^{(1)ijlk}_{\ell^2q^2H^2}\delta_{IJ}\delta_{LK}\delta_{MN}+C^{(2)ijlk}_{\ell^2q^2H^2}\tau^a_{IJ}\tau^a_{MN}\delta_{LK}+\\+C^{(3)ijlk}_{\ell^2q^2H^2}\tau^a_{IJ}\tau^a_{LK}\delta_{MN}+C^{(4)ijlk}_{\ell^2q^2H^2}\tau^a_{LK}\tau^a_{MN}\delta_{IJ}\big)\gamma^\alpha P_L \otimes \gamma_\alpha P_L \end{array}$
	\end{subfigure}\\
	\begin{subfigure}{.9\textwidth}
		\centering
		\begin{tikzpicture}[scale=1.2, baseline={([yshift=-.5ex]current bounding box.center)}]
			\begin{feynman}[small]
				\vertex (a) at (-1,1){\(e_j\)};
				\vertex[style=crossed dot] (b) at (0,0) {};
				\vertex (c) at (1,1) {\(\ell_{iI}\)};
				\vertex (f) at (-1,0) {\(H_N\)};
				\vertex (g) at (1,0) {\(H_M\)};
				\vertex (d) at (-1,-1) {\(q_{kK}\)};
				\vertex (e) at (1,-1){\(d_l\)};
				
				\diagram* {
					(a)  --  [fermion] (b)  -- [fermion] (c),
					(d) --  [fermion] (b) -- [fermion] (e),
					(f) -- [charged scalar] (b),
					(b) -- [charged scalar] (g),
				};
			\end{feynman}
		\end{tikzpicture}\ $i(C^{(1)ijlk}_{\ell edqH^2}\delta_{IK}\delta_{MN}+C^{(2)ijlk}_{\ell edqH^2}\tau^a_{IK}\tau^a_{MN}) P_R \otimes  P_L$
	\end{subfigure}\\
	\begin{subfigure}{.9\textwidth}
		\centering
		\begin{tikzpicture}[scale=1.2, baseline={([yshift=-.5ex]current bounding box.center)}]
			\begin{feynman}[small]
				\vertex (a) at (-1,1){\(e_j\)};
				\vertex[style=crossed dot] (b) at (0,0) {};
				\vertex (c) at (1,1) {\(\ell_{iI}\)};
				\vertex (f) at (-1,0) {\(H_N\)};
				\vertex (g) at (1,0) {\(H_M\)};
				\vertex (d) at (-1,-1) {\(u_{k}\)};
				\vertex (e) at (1,-1){\(q_{lL}\)};
				
				\diagram* {
					(a)  --  [fermion] (b)  -- [fermion] (c),
					(d) --  [fermion] (b) -- [fermion] (e),
					(f) -- [charged scalar] (b),
					(b) -- [charged scalar] (g),
				};
			\end{feynman}
		\end{tikzpicture}$\begin{array}{c}i\bigg[(C^{(1)ijlk}_{\ell equH^2}\epsilon_{IK}\delta_{MN}+C^{(2)ijlk}_{\ell equH^2}(\tau^a\epsilon)_{IK}\tau^a_{MN}) P_R \otimes  P_R+\\+(C^{(3)ijlk}_{\ell equH^2}\epsilon_{IK}\delta_{MN}+C^{(4)ijlk}_{\ell equH^2}(\tau^a\epsilon)_{IK}\tau^a_{MN}) \sigma^{\alpha\beta}P_R \otimes  \sigma_{\alpha\beta}P_R\bigg]\end{array}$
	\end{subfigure}\\
	\caption{Feynman rules for the dimension eight SMEFT four-fermion interactions $4f_8$ of section \ref{sec:Operators}. We consider only the dimension eight operators involved in the diagrams of section \ref{ssec:SMEFTrunningsub}. In the product $\Gamma_1\otimes \Gamma_2$ the left matrix $\Gamma_1$ multiplies the lepton bilinear. Scalar and tensor with opposite chiralities have the same Feynman rules with conjugate coefficients and exchanged flavour indices within lepton and quark bilinears.}\label{fig:FeynRules4f8}
\end{figure}
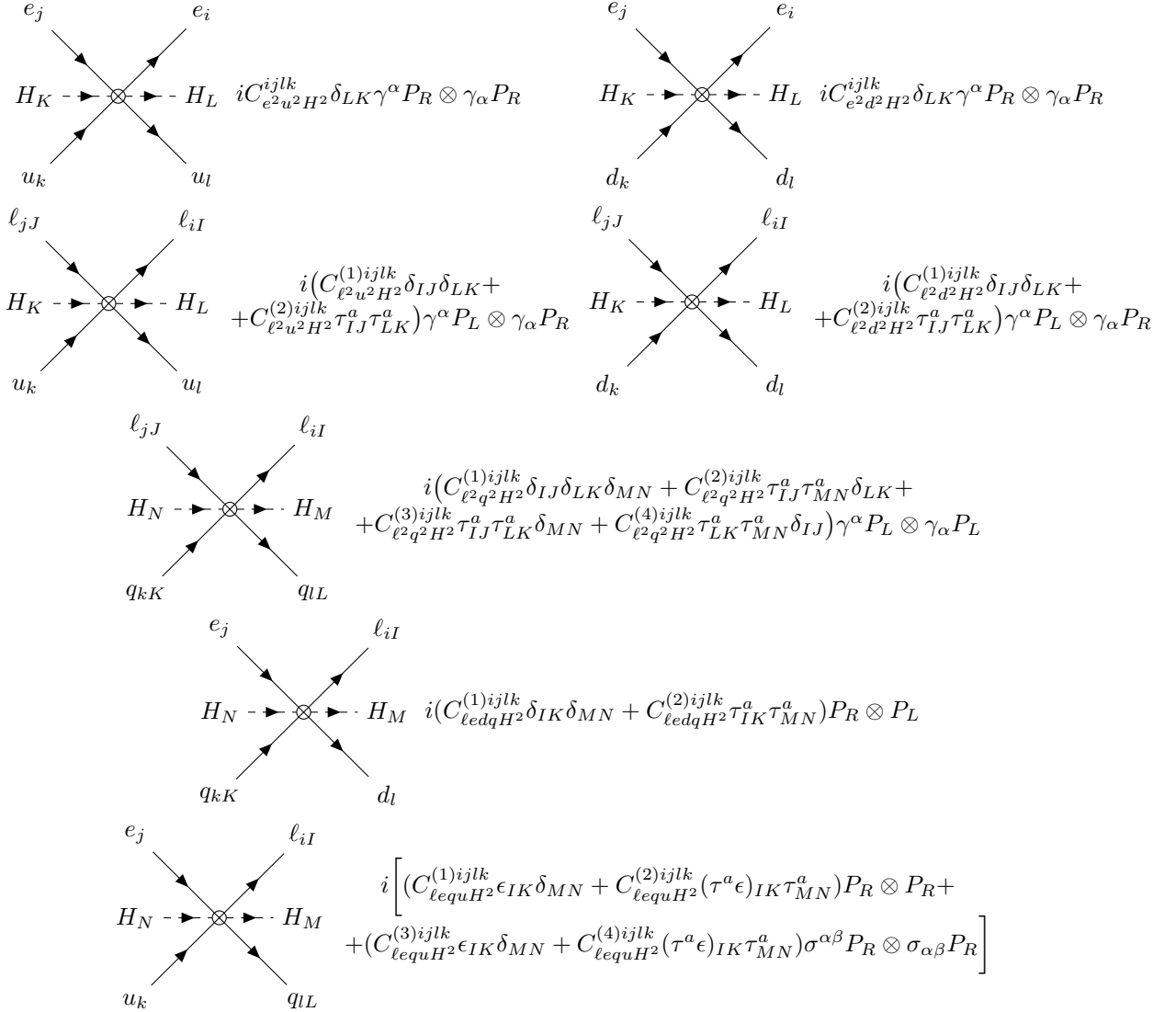

\clearpage

\section{Anomalous Dimensions}\label{appendix:AnomalousDimensions}

In this section we write the renormalization group equations for the mixing of
{ a $\mu\to \tau$ dimension six  operator, multiplied by a
	$\tau\to e$ dimension six operator, into  a dimension eight $\mu\to e$ operator. These anomalous dimensions are generated } by the diagrams of section \ref{ssec:SMEFTrunningsub}. We conveniently present the RGEs divided in the ``classes" introduced in the same section. The operator definitions can be found in section \ref{sec:Operators}. The upper dot $\dot{C}$ on the Wilson coefficient indicate the logarithmic derivative with respect to the renormalization scale $M$.
{ The anomalous dimensions are written for the dimension eight operators of Section \ref{sec:Operators}, which are relevant for $\mu\to e$ processes that are otherwise flavour diagonal, although more general flavour structures can be obtained with the appropriate substitutions. For non-Hermitian operators such as $\mathcal{O}^{(1)}_{\ell e quH^2}$, we write the RGEs for the $\mu\to e$ operators $\mathcal{O}^{(1)e\mu ii}_{\ell e quH^2}, \mathcal{O}^{*(1)\mu e ii}_{\ell e quH^2}$.  This is to more explicitly show  the $\tl$ operator pairs upon which we obtain  limits  in section \ref{sec:Pheno}.     }

\subsection{$4f_6\times4f_6\to 4f_8$}
Figure \ref{subfig:6} shows the mixing $\propto y_t y_\tau$ of pairs of  dimension six $\tau\to l$ operators into the dimension eight $\mu\to e$ tensor with top legs. We align $\te,\tm$ Wilson coefficients respectively in row and column vectors to write the following anomalous dimensions, relevant for the $B_{\mue}$ sensitivity of Table \ref{tab:toptensors}. 
\begin{align}
	16\pi^2 \dot{C}^{(3)e\mu 3t}_{\ell equH^2}=\begin{pmatrix}
		C^{e\tau tt}_{\ell u} & C^{(1)e\tau 33}_{\ell q} & C^{(3)e\tau 33}_{\ell q} & C^{(1)e\tau 3t}_{\ell equ} & C^{(3)e\tau 3t}_{\ell equ}\end{pmatrix} \nonumber\\ \begin{pmatrix}
		y_\tau y_t & 0 & 0 & 0\\
		0 & -y_\tau y_t & 0 & 0\\
		0 & y_\tau y_t & 0 & 0\\
		0 & 0 & 0 & 3y_\tau y_t\\
		0 & 0 & 3y_\tau y_t & -8y_\tau y_t\\
	\end{pmatrix} \begin{pmatrix}
		C^{\tau\mu 33}_{eq} \\ 
		C^{\tau\mu tt}_{eu} \\
		C^{(1)\tau\mu 3t}_{\ell equ} \\
		C^{(3)\tau\mu 3t}_{\ell equ}
	\end{pmatrix}
\end{align}
\begin{align}
	16\pi^2 \dot{C}^{*(3)\mu e 3t}_{\ell equH^2}=\begin{pmatrix}
		C^{\tau \mu tt}_{\ell u} & C^{(1)\tau\mu 33}_{\ell q} & C^{(3)\tau\mu 33}_{\ell q} & C^{*(1)\mu\tau 3t}_{\ell equ} & C^{*(3)\mu\tau 3t}_{\ell equ}\end{pmatrix} \nonumber\\ \begin{pmatrix}
		y_\tau y_t & 0 & 0 & 0\\
		0 & -y_\tau y_t & 0 & 0\\
		0 & y_\tau y_t & 0 & 0\\
		0 & 0 & 0 & 3y_\tau y_t\\
		0 & 0 & 3y_\tau y_t & -8y_\tau y_t\\
	\end{pmatrix} \begin{pmatrix}
		C^{e \tau 33}_{eq} \\ 
		C^{e\tau tt}_{eu} \\
		C^{*(1)\tau e 3t}_{\ell equ} \\
		C^{*(3)\tau e 3t}_{\ell equ}
	\end{pmatrix}
\end{align}
\begin{align}
	16\pi^2 \dot{C}^{(4)e\mu 3t}_{\ell equH^2}=\begin{pmatrix}
		C^{e\tau tt}_{\ell u} & C^{(1)e\tau 33}_{\ell q} & C^{(3)e\tau 33}_{\ell q} & C^{(1)e\tau 3t}_{\ell equ} & C^{(3)e\tau 3t}_{\ell equ}\end{pmatrix} \nonumber\\ \begin{pmatrix}
		y_\tau y_t & 0 & 0 & 0\\
		0 & -y_\tau y_t & 0 & 0\\
		0 & -y_\tau y_t & 0 & 0\\
		0 & 0 & 0 & y_\tau y_t\\
		0 & 0 & y_\tau y_t & 8y_\tau y_t\\
	\end{pmatrix} \begin{pmatrix}
		C^{\tau\mu 33}_{eq} \\ 
		C^{\tau\mu tt}_{eu} \\
		C^{(1)\tau\mu 3t}_{\ell equ} \\
		C^{(3)\tau\mu 3t}_{\ell equ}
	\end{pmatrix}
\end{align}
\begin{align}
	16\pi^2 \dot{C}^{*(4)\mu e 3t}_{\ell equH^2}=\begin{pmatrix}
		C^{\tau \mu tt}_{\ell u} & C^{(1)\tau\mu 33}_{\ell q} & C^{(3)\tau\mu 33}_{\ell q} & C^{*(1)\mu\tau 3t}_{\ell equ} & C^{*(3)\mu\tau 3t}_{\ell equ}\end{pmatrix} \nonumber\\ \begin{pmatrix}
		y_\tau y_t & 0 & 0 & 0\\
		0 & -y_\tau y_t & 0 & 0\\
		0 & -y_\tau y_t & 0 & 0\\
		0 & 0 & 0 & y_\tau y_t\\
		0 & 0 & y_\tau y_t & 8y_\tau y_t\\
	\end{pmatrix} \begin{pmatrix}
		C^{e \tau 33}_{eq} \\ 
		C^{e\tau tt}_{eu} \\
		C^{*(1)\tau e 3t}_{\ell equ} \\
		C^{*(3)\tau e 3t}_{\ell equ}
	\end{pmatrix}
\end{align}
In Figure \ref{subfig:5} we show a representative diagram with the  double insertion of two-lepton two-quark $\tau\to l$ operators  of  dimension six, which renormalizes the coefficient of $\mu\to e$ dimension eight four fermion operators. The mixing is proportional to the square of the top Yukawa $y^2_t$. The RGEs for scalar and tensor with a up-singlet quark (the sensitivities of $\mu\to e$ processes that we obtain from this mixing are summarized in Tables \ref{tab:Bdecaysscalarup} and \ref{tab:Bdecaystensorcharm}) read 
\begin{align}
	16\pi^2 \dot{C}^{(1)e\mu ii}_{\ell equH^2}=\begin{pmatrix}
		C^{e\tau ti}_{\ell u} & C^{(1)e\tau ti}_{\ell equ} & C^{(3)e\tau 3i}_{\ell equ} & C^{(1)e\tau it}_{\ell equ} & C^{(3)e\tau it}_{\ell equ} & C^{(1)e\tau i3}_{\ell q} & C^{(3)e\tau i3}_{\ell q} \end{pmatrix} \nonumber \\
	\begin{pmatrix}
		-2y^2_t & -24y^2_t & 0 & 0 & 0 & 0\\
		0 & 0& -y^2_t & 0 & 0 & 0\\
		0 & 0& -12y^2_t & 0 & 0 & 0\\
		0 & 0& 0 & 2y^2_t & 0 & 0\\
		0 & 0& 0 & -24y^2_t & 0 & 0\\
		0 & 0& 0 & 0 & y^2_t & -12y^2_t\\
		0 & 0& 0 & 0 & -3y^2_t & 36y^2_t\\
	\end{pmatrix} \begin{pmatrix}
		C^{(1)\tau\mu it}_{\ell equ} \\ 
		C^{(3)\tau\mu it}_{\ell equ} \\
		C^{\tau\mu i3 }_{eq} \\
		C^{\tau\mu ti }_{eu} \\
		C^{(1)\tau\mu 3i}_{\ell equ}\\
		C^{(3)\tau\mu 3i}_{\ell equ}\\
	\end{pmatrix}
\end{align}
\begin{align}
	16\pi^2 \dot{C}^{*(1)\mu e ii}_{\ell equH^2}=\begin{pmatrix}
		C^{\tau\mu it}_{\ell u} & C^{*(1)\mu\tau 3i}_{\ell equ} & C^{*(3)\mu\tau 3i}_{\ell equ} & C^{*(1)\mu\tau it}_{\ell equ} & C^{*(3)\mu\tau it}_{\ell equ} & C^{(1)\tau\mu 3i}_{\ell q} & C^{(3)\tau\mu 3i}_{\ell q} \end{pmatrix} \nonumber \\
	\begin{pmatrix}
		-2y^2_t & -24y^2_t & 0 & 0 & 0 & 0\\
		0 & 0& -y^2_t & 0 & 0 & 0\\
		0 & 0& -12y^2_t & 0 & 0 & 0\\
		0 & 0& 0 & 2y^2_t & 0 & 0\\
		0 & 0& 0 & -24y^2_t & 0 & 0\\
		0 & 0& 0 & 0 & y^2_t & -12y^2_t\\
		0 & 0& 0 & 0 & -3y^2_t & 36y^2_t\\
	\end{pmatrix} \begin{pmatrix}
		C^{*(1)\tau e it}_{\ell equ} \\ 
		C^{*(3)\tau e it}_{\ell equ} \\
		C^{e\tau 3i}_{eq} \\
		C^{e\tau it }_{eu} \\
		C^{*(1)\tau e 3i}_{\ell equ}\\
		C^{*(3)\tau e 3i}_{\ell equ}\\
	\end{pmatrix}
\end{align}
\begin{align}
	16\pi^2 \dot{C}^{(2)e\mu ii}_{\ell equH^2}=\begin{pmatrix}
		C^{(1)e\tau 3i}_{\ell equ} & C^{(3)e\tau 3i}_{\ell equ} &  C^{(1)e\tau i3}_{\ell q} & C^{(3)e\tau i3}_{\ell q} \end{pmatrix} \nonumber \\
	\begin{pmatrix}
		-y^2_t  & 0 & 0\\
		-12y^2_t  & 0 & 0\\
		0  & y^2_t & -12y^2_t\\
		0  & y^2_t & -12y^2_t\\
	\end{pmatrix} \begin{pmatrix}
		C^{\tau\mu i3}_{eq} \\
		C^{(1)\tau\mu 3i}_{\ell equ}\\
		C^{(3)\tau\mu 3i}_{\ell equ}\\
	\end{pmatrix}
\end{align}
\begin{align}
	16\pi^2 \dot{C}^{*(2)\mu e ii}_{\ell equH^2}=\begin{pmatrix}
		C^{*(1)\mu\tau 3i}_{\ell equ} & C^{*(3)\mu\tau 3i}_{\ell equ} &  C^{(1)\tau\mu 3i}_{\ell q} & C^{(3)\tau\mu 3i}_{\ell q} \end{pmatrix} \nonumber \\
	\begin{pmatrix}
		-y^2_t  & 0 & 0\\
		-12y^2_t  & 0 & 0\\
		0  & y^2_t & -12y^2_t\\
		0  & y^2_t & -12y^2_t\\
	\end{pmatrix} \begin{pmatrix}
		C^{e\tau 3i}_{eq} \\
		C^{*(1)\tau e 3i}_{\ell equ}\\
		C^{*(3)\tau e 3i}_{\ell equ}\\
	\end{pmatrix}
\end{align}
\begin{align}
	16\pi^2 \dot{C}^{(3)e\mu ii}_{\ell equH^2}=\begin{pmatrix}
		C^{e\tau ti}_{\ell u} & C^{(1)e\tau 3i}_{\ell equ} & C^{(3)e\tau 3i}_{\ell equ} & C^{(1)e\tau it}_{\ell equ} & C^{(3)e\tau it}_{\ell equ} & C^{(1)e\tau i3}_{\ell q} & C^{(3)e\tau i3}_{\ell q} \end{pmatrix} \nonumber \\
	\begin{pmatrix}
		-y^2_t/2 & -6y^2_t & 0 & 0 & 0 & 0\\
		0 & 0& -y^2_t/4 & 0 & 0 & 0\\
		0 & 0& -3y^2_t & 0 & 0 & 0\\
		0 & 0& 0 & -y^2_t/2 & 0 & 0\\
		0 & 0& 0 & 6y^2_t & 0 & 0\\
		0 & 0& 0 & 0 & -y^2_t/4 & 3y^2_t\\
		0 & 0& 0 & 0 & 3y^2_t/4 & -9y^2_t\\
	\end{pmatrix} \begin{pmatrix}
		C^{(1)\tau\mu it}_{\ell equ} \\ 
		C^{(3)\tau\mu it}_{\ell equ} \\
		C^{\tau\mu i3}_{eq} \\
		C^{\tau\mu ti }_{eu} \\
		C^{(1)\tau\mu 3i}_{\ell equ}\\
		C^{(3)\tau\mu 3i}_{\ell equ}\\
	\end{pmatrix}
\end{align}
\begin{align}
	16\pi^2 \dot{C}^{*(3)\mu e ii}_{\ell equH^2}=\begin{pmatrix}
		C^{\tau\mu it}_{\ell u} & C^{*(1)\mu\tau 3i}_{\ell equ} & C^{*(3)\mu\tau 3i}_{\ell equ} & C^{*(1)\mu\tau it}_{\ell equ} & C^{*(3)\mu\tau it}_{\ell equ} & C^{(1)\tau\mu 3i}_{\ell q} & C^{(3)\tau\mu 3i}_{\ell q} \end{pmatrix} \nonumber \\
	\begin{pmatrix}
		-y^2_t/2 & -6y^2_t & 0 & 0 & 0 & 0\\
		0 & 0& -y^2_t/4 & 0 & 0 & 0\\
		0 & 0& -3y^2_t & 0 & 0 & 0\\
		0 & 0& 0 & -y^2_t/2 & 0 & 0\\
		0 & 0& 0 & 6y^2_t & 0 & 0\\
		0 & 0& 0 & 0 & -y^2_t/4 & 3y^2_t\\
		0 & 0& 0 & 0 & 3y^2_t/4 & -9y^2_t\\
	\end{pmatrix} \begin{pmatrix}
		C^{*(1)\tau e it}_{\ell equ} \\ 
		C^{*(3)\tau e it}_{\ell equ} \\
		C^{e\tau 3i}_{eq} \\
		C^{e\tau it }_{eu} \\
		C^{*(1)\tau e 3i}_{\ell equ}\\
		C^{*(3)\tau e 3i}_{\ell equ}\\
	\end{pmatrix}
\end{align}
\begin{align}
	16\pi^2 \dot{C}^{(4)e\mu ii}_{\ell equH^2}=\begin{pmatrix}
		C^{(1)e\tau 3i}_{\ell equ} & C^{(3)e\tau 3i}_{\ell equ} &  C^{(1)e\tau i3}_{\ell q} & C^{(3)e\tau i3}_{\ell q} \end{pmatrix} \nonumber \\
	\begin{pmatrix}
		-y^2_t/4  & 0 & 0\\
		-3y^2_t  & 0 & 0\\
		0  & -y^2_t/4 & 3y^2_t\\
		0  & -y^2_t/4 & 3y^2_t\\
	\end{pmatrix} \begin{pmatrix}
		C^{\tau\mu i3}_{eq} \\
		C^{(1)\tau\mu 3i}_{\ell equ}\\
		C^{(3)\tau\mu 3i}_{\ell equ}\\
	\end{pmatrix}
\end{align}
\begin{align}
	16\pi^2 \dot{C}^{*(4)\mu e ii}_{\ell equH^2}=\begin{pmatrix}
		C^{*(1)\mu\tau 3i}_{\ell equ} & C^{*(3)\mu\tau 3i}_{\ell equ} &  C^{(1)\tau\mu 3i}_{\ell q} & C^{(3)\tau\mu 3i}_{\ell q} \end{pmatrix} \nonumber \\
	\begin{pmatrix}
		-y^2_t/4  & 0 & 0\\
		-3y^2_t  & 0 & 0\\
		0  & -y^2_t/4 & 3y^2_t\\
		0  & -y^2_t/4 & 3y^2_t\\
	\end{pmatrix}  \begin{pmatrix}
		C^{e\tau 3i}_{eq} \\
		C^{*(1)\tau e 3i}_{\ell equ}\\
		C^{*(3)\tau e 3i}_{\ell equ}\\
	\end{pmatrix}.
\end{align}
For scalars with a singlet down-quark (sensitivities in Table \ref{tab:Bdecaysscalardown}), the mixing is 
\begin{align}
	16\pi^2 \dot{C}^{(1)e\mu ii}_{\ell edqH^2}=\begin{pmatrix}
		C^{(1)e\tau i3}_{\ell q} & C^{(3)e\tau i3}_{\ell q} & C^{e\tau i3}_{ledq} \end{pmatrix} \nonumber \\
	\begin{pmatrix}
		-y^2_t  & 0 \\
		-3y^2_t  & 0 \\
		0  & y^2_t & \\
	\end{pmatrix} \begin{pmatrix}
		C^{\tau\mu i3 }_{\ell edq} \\
		C^{\tau \mu 3i}_{eq}\\
	\end{pmatrix}
\end{align}
\begin{align}
	16\pi^2 \dot{C}^{*(1)\mu e ii}_{\ell edqH^2}=\begin{pmatrix}
		C^{(1)\tau\mu 3i}_{\ell q} & C^{(3)\tau\mu 3i}_{\ell q} & C^{*\mu\tau i3}_{\ell edq} \end{pmatrix} \nonumber \\
	\begin{pmatrix}
		-y^2_t  & 0 \\
		-3y^2_t  & 0 \\
		0  & y^2_t & \\
	\end{pmatrix} \begin{pmatrix}
		C^{*\tau e i3 }_{\ell edq} \\
		C^{e\tau i3}_{eq}\\
	\end{pmatrix}
\end{align}
\begin{align}
	16\pi^2 \dot{C}^{(2)e\mu ii}_{\ell edqH^2}=\begin{pmatrix}
		C^{(1)e\tau i3}_{\ell q} & C^{(3)e\tau i3}_{\ell q} & C^{e\tau i3}_{\ell edq} \end{pmatrix} \nonumber \\
	\begin{pmatrix}
		y^2_t  & 0 \\
		-y^2_t  & 0 \\
		0  & -y^2_t & \\
	\end{pmatrix} \begin{pmatrix}
		C^{\tau\mu i3 }_{\ell edq} \\
		C^{\tau \mu 3i}_{eq}\\
	\end{pmatrix}
\end{align}
\begin{align}
	16\pi^2 \dot{C}^{*(2)\mu e ii}_{\ell edqH^2}=\begin{pmatrix}
		C^{(1)\tau\mu 3i}_{\ell q} & C^{(3)\tau\mu 3i}_{\ell q} & C^{*\mu\tau i3}_{\ell edq} \end{pmatrix} \nonumber \\
	\begin{pmatrix}
		y^2_t  & 0 \\
		-y^2_t  & 0 \\
		0  & -y^2_t & \\
	\end{pmatrix} \begin{pmatrix}
		C^{*\tau e i3 }_{\ell edq} \\
		C^{ e\tau i3}_{eq}\\
	\end{pmatrix}
\end{align}
The anomalous dimensions for the mixing into $\mu\to e$ vectors with SU(2) lepton singlets are (sensitivities in Table \ref{tab:BdecaysvectR})
\begin{align}
	16\pi^2 \dot{C}^{(1) e \mu ii}_{e^2q^2H^2}=\begin{pmatrix}
		C^{e \tau 3 i}_{eq} & C^{e \tau i3}_{eq} & C^{*(1)\tau e it}_{\ell equ} & C^{*(3)\tau e it}_{\ell equ} \end{pmatrix} \nonumber \\
	\begin{pmatrix}
		-4y^2_t  & 0 & 0 & 0 \\
		0  & y^2_t & 0 & 0 \\
		0  & 0 & -y^2_t/2 & -6y^2_t \\
		0  & 0 & -6y^2_t & -72y^2_t \\
	\end{pmatrix} \begin{pmatrix}
		C^{\tau \mu i3 }_{eq} \\
		C^{\tau \mu 3i }_{eq}\\
		C^{(1)\tau \mu it }_{\ell equ} \\
		C^{(3)\tau \mu it }_{\ell equ}\\
	\end{pmatrix}
\end{align}
\begin{align}
	16\pi^2 \dot{C}^{(2) e \mu ii}_{e^2q^2H^2}=\begin{pmatrix}
		C^{e \tau 3 i}_{eq} & C^{e \tau i3}_{eq} \end{pmatrix} \nonumber \\
	\begin{pmatrix}
		4y^2_t  & 0 \\
		0  & -y^2_t 
	\end{pmatrix} \begin{pmatrix}
		C^{\tau \mu i3 }_{eq} \\
		C^{\tau \mu 3i }_{eq}\\
	\end{pmatrix}
\end{align}
\begin{align}
	16\pi^2 \dot{C}^{e \mu ii}_{e^2u^2H^2}=\begin{pmatrix}
		C^{e \tau t i}_{eu} & C^{e \tau it}_{eu} & C^{*(1)\tau e 3i}_{\ell equ} & C^{*(3)\tau e 3i}_{\ell equ} \end{pmatrix} \nonumber \\
	\begin{pmatrix}
		-2y^2_t  & 0 & 0 & 0 \\
		0  & 8y^2_t & 0 & 0 \\
		0  & 0 & y^2_t/2 & -6y^2_t \\
		0  & 0 & -6y^2_t & 72y^2_t \\
	\end{pmatrix} \begin{pmatrix}
		C^{\tau \mu it }_{eu} \\
		C^{\tau \mu ti }_{eu}\\
		C^{(1)\tau \mu 3i }_{\ell equ} \\
		C^{(3)\tau \mu 3i }_{\ell equ}\\
	\end{pmatrix}
\end{align}
\begin{align}
	16\pi^2 \dot{C}^{e \mu ii}_{e^2d^2H^2}=-\frac{y_t^2}{2}C^{*\tau e i3}_{\ell edq}C^{\tau \mu i3}_{\ell edq}
\end{align}
while for vectors with lepton doublets (sensitivities in Table \ref{tab:BdecaysvectL}) these are
\begin{align}
	16\pi^2 \dot{C}^{(1)e \mu ii}_{\ell ^2q^2H^2}=\begin{pmatrix}
		C^{(1)e \tau 3 i}_{\ell q} & C^{(3)e \tau 3 i}_{\ell q} & C^{(1)e \tau i3 }_{\ell q} & C^{(3)e \tau i3 }_{\ell q} & C^{(1)e\tau it}_{\ell equ} &  C^{(3)e\tau it}_{\ell equ}\end{pmatrix} \nonumber \\
	\begin{pmatrix}
		-y^2_t  & 0 & 0 & 0 & 0 & 0\\
		0  & -3y^2_t & 0 & 0 & 0 & 0\\
		0  & 0 & 4y^2_t & 0 & 0 & 0\\
		0  & 0 & 0 & 12y^2_t & 0 & 0\\
		0  & 0 & 0 & 0 & y^2_t/4 & -3y^2_t\\
		0  & 0 & 0 & 0 & -3y^2_t & 36y^2_t\\
	\end{pmatrix} \begin{pmatrix}
		C^{(1)\tau \mu i3}_{\ell q}\\
		C^{(3)\tau \mu i3}_{\ell q}\\
		C^{(1)\tau \mu 3i}_{\ell q}\\
		C^{(3)\tau \mu 3i}_{\ell q}\\
		C^{*(1)\mu \tau it}_{\ell equ}\\
		C^{*(3)\mu \tau it}_{\ell equ}\\
	\end{pmatrix}
\end{align}
\begin{align}
	16\pi^2 \dot{C}^{(2)e \mu ii}_{\ell^2q^2H^2}=\begin{pmatrix}
		C^{(1)e \tau t i}_{\ell q} & C^{(3)e \tau t i}_{\ell q} & C^{(1)e \tau it }_{\ell q} & C^{(3)e \tau it }_{\ell q} \end{pmatrix} \nonumber \\
	\begin{pmatrix}
		0  & y^2_t & 0 & 0 \\
		y^2_t & -2y^2_t & 0 & 0 \\
		0  & 0 & 0 & -4y^2_t \\
		0  & 0 & -4y^2_t & -8y^2_t \\
	\end{pmatrix} \begin{pmatrix}
		C^{(1)\tau \mu i3}_{\ell q}\\
		C^{(3)\tau \mu i3}_{\ell q}\\
		C^{(1)\tau \mu 3i}_{\ell q}\\
		C^{(3)\tau \mu 3i}_{\ell q}\\
	\end{pmatrix}
\end{align}
\begin{align}
	16\pi^2 \dot{C}^{(3)e \mu ii}_{\ell ^2q^2H^2}=\begin{pmatrix}
		C^{(1)e \tau 3 i}_{\ell q} & C^{(3)e \tau 3 i}_{\ell q} & C^{(1)e \tau i3 }_{\ell q} & C^{(3)e \tau i3 }_{\ell q} & C^{(1)e\tau it}_{\ell equ} &  C^{(3)e\tau it}_{\ell equ}\end{pmatrix} \nonumber \\
	\begin{pmatrix}
		0  & -y^2_t & 0 & 0 & 0 & 0\\
		-y^2_t  & -2y^2_t & 0 & 0 & 0 & 0\\
		0  & 0 & 0 & 4y^2_t & 0 & 0\\
		0  & 0 & 4y^2_t & -8y^2_t & 0 & 0\\
		0  & 0 & 0 & 0 & -y^2_t/4 & 3y^2_t\\
		0  & 0 & 0 & 0 & 3y^2_t & -36y^2_t\\
	\end{pmatrix} \begin{pmatrix}
		C^{(1)\tau \mu i3}_{\ell q}\\
		C^{(3)\tau \mu i3}_{\ell q}\\
		C^{(1)\tau \mu 3i}_{\ell q}\\
		C^{(3)\tau \mu 3i}_{\ell q}\\
		C^{*(1) \mu \tau it}_{\ell equ}\\
		C^{*(3) \mu \tau it}_{\ell equ}\\
	\end{pmatrix}
\end{align}
\begin{align}
	16\pi^2 \dot{C}^{(4)e \mu ii}_{\ell^2q^2H^2}=\begin{pmatrix}
		C^{(1)e \tau 3 i}_{\ell q} & C^{(3)e \tau 3 i}_{\ell q} & C^{(1)e \tau i3 }_{\ell q} & C^{(3)e \tau i3 }_{\ell q} \end{pmatrix} \nonumber \\
	\begin{pmatrix}
		y^2_t & 0 & 0 & 0 \\
		& -y^2_t & 0 & 0 \\
		0  & 0 & -4y^2_t & 0 \\
		0  & 0 & 0 & 4y^2_t \\
	\end{pmatrix} \begin{pmatrix}
		C^{(1)\tau \mu i3}_{\ell q}\\
		C^{(3)\tau \mu i3}_{\ell q}\\
		C^{(1)\tau \mu 3i}_{\ell q}\\
		C^{(3)\tau \mu 3i}_{\ell q}\\
	\end{pmatrix}
\end{align}
\begin{align}
	16\pi^2 \dot{C}^{(1)e \mu ii}_{\ell^2u^2H^2}=\begin{pmatrix}
		C^{e \tau t i}_{\ell u} & C^{e \tau  it}_{\ell u} & C^{(1)e\tau 3i}_{\ell equ} &  C^{(3)e\tau 3i}_{\ell equ}\end{pmatrix} \nonumber \\
	\begin{pmatrix}
		-8y_t^2  & 0 & 0 & 0 \\
		0 & 2y^2_t & 0 & 0\\
		0  & 0 & -y^2_t/4 & -3y^2_t \\
		0  & 0 & -3y^2_t & -36y^2_t\\
	\end{pmatrix} \begin{pmatrix}
		C^{\tau\mu it}_{\ell u}\\
		C^{\tau\mu ti}_{\ell u}\\
		C^{*(1)\mu \tau 3i}_{\ell equ}\\
		C^{*(3)\mu \tau 3i}_{\ell equ}\\
	\end{pmatrix}
\end{align}
\begin{align}
	16\pi^2 \dot{C}^{(2)e \mu ii}_{\ell^2u^2H^2}=\begin{pmatrix}
		C^{(1)e\tau 3i}_{\ell equ} &  C^{(3)e\tau 3i}_{\ell equ}\end{pmatrix} \nonumber \\
	\begin{pmatrix}
		-y^2_t/4 & -3y^2_t \\
		-3y^2_t & -36y^2_t\\
	\end{pmatrix} \begin{pmatrix}
		C^{*(1)\mu \tau 3i}_{\ell equ}\\
		C^{*(3)\mu \tau 3i}_{\ell equ}\\
	\end{pmatrix}
\end{align}
\begin{align}
	16\pi^2 \dot{C}^{(1)e \mu ii}_{\ell^2d^2H^2}=\frac{y_t^2}{4}C^{e\tau i 3}_{\ell edq}C^{*\mu\tau i3}_{\ell edq}\qquad 16\pi^2 \dot{C}^{(2)e \mu ii}_{\ell^2d^2H^2}=-\frac{y_t^2}{4}C^{e\tau i 3}_{\ell edq}C^{*\mu\tau i3}_{\ell edq}
\end{align}
\subsection{$P_6\times 4f_6\to 4f_8$}
Dimension six $\tau\to l$ four fermion interactions renormalise $\mu\to e$ dimension eight operators via gauge loops where one vertex is a flavour changing penguin (eq. (\ref{eq:penguinsdim6a})-(\ref{eq:penguinsdim6b})), as depicted in Figure \ref{subfig:4}. One-particle-irreducible vertex corrections and ``wavefunction-like" contributions (see section \ref{ssec:EoM} for a discussion) give the following gauge invariant anomalous dimensions, where we align four-fermion interactions and penguins respectively in row and column vectors:  
\begin{align}
	16\pi^2 \dot{C}^{(1)e \mu ii}_{\ell equH^2}=\begin{pmatrix}
		C^{(1)e\tau ii}_{\ell equ} &  C^{(3)e\tau ii}_{\ell equ} & C^{(1)\tau\mu ii}_{\ell equ} &  C^{(3)\tau\mu ii}_{\ell equ}\end{pmatrix} \nonumber \\
	\begin{pmatrix}
		3g'^2  & 0 & 0 \\
		-20g'^2 & 0 & 0\\
		0  & 6g'^2 & 0  \\
		0  &  -20g'^2 & 36g'^2\\
	\end{pmatrix} \begin{pmatrix}
		C^{\tau\mu}_{He}\\
		C^{(1)e\tau}_{H\ell}\\
		C^{(3)e\tau}_{H\ell}\\
	\end{pmatrix}
\end{align}
\begin{align}
	16\pi^2 \dot{C}^{*(1) \mu e ii}_{\ell equH^2}=\begin{pmatrix}
		C^{*(1)\mu\tau ii}_{\ell equ} &  C^{*(3)\mu\tau ii}_{\ell equ} & C^{*(1)\tau eii}_{\ell equ} &  C^{*(3)\tau e ii}_{\ell equ}\end{pmatrix} \nonumber \\
	\begin{pmatrix}
		3g'^2  & 0 & 0 \\
		-20g'^2 & 0 & 0\\
		0  & 6g'^2 & 0  \\
		0  &  -20g'^2 & 36g'^2\\
	\end{pmatrix} \begin{pmatrix}
		C^{e\tau}_{He}\\
		C^{(1)\tau\mu}_{H\ell }\\
		C^{(3)\tau\mu}_{H\ell}\\
	\end{pmatrix}
\end{align}
\begin{align}
	16\pi^2 \dot{C}^{(2)e \mu ii}_{\ell equH^2}=\begin{pmatrix}
		C^{(1)e\tau ii}_{\ell equ} &  C^{(3)e\tau ii}_{\ell equ} & C^{(1)\tau\mu ii}_{\ell equ} &  C^{(3)\tau\mu ii}_{\ell equ}\end{pmatrix} \nonumber \\
	\begin{pmatrix}
		-3g'^2  & 0 & 0 \\
		12g'^2 & 0 & 0\\
		0  & 0 & 6g'^2  \\
		0  &  12g^2 & -20g'^2\\
	\end{pmatrix} \begin{pmatrix}
		C^{\tau\mu}_{He}\\
		C^{(1)e\tau}_{H\ell}\\
		C^{(3)e\tau}_{H\ell}\\
	\end{pmatrix}
\end{align}
\begin{align}
	16\pi^2 \dot{C}^{*(2) \mu e ii}_{\ell equH^2}=\begin{pmatrix}
		C^{*(1)\mu\tau ii}_{\ell equ} &  C^{*(3)\mu\tau ii}_{\ell equ} & C^{*(1)\tau eii}_{\ell equ} &  C^{*(3)\tau e ii}_{\ell equ}\end{pmatrix} \nonumber \\
	\begin{pmatrix}
		-3g'^2  & 0 & 0 \\
		12g'^2 & 0 & 0\\
		0  & 0 & 6g'^2  \\
		0  &  12g^2 & -20g'^2\\
	\end{pmatrix} \begin{pmatrix}
		C^{e\tau}_{He}\\
		C^{(1)\tau\mu}_{H\ell }\\
		C^{(3)\tau\mu}_{H\ell}\\
	\end{pmatrix}
\end{align}
\begin{align}
	16\pi^2 \dot{C}^{(3)e \mu ii}_{\ell equH^2}=\begin{pmatrix}
		C^{(1)e\tau ii}_{\ell equ} &  C^{(3)e\tau ii}_{\ell equ} & C^{(1)\tau\mu ii}_{\ell equ} &  C^{(3)\tau\mu ii}_{\ell equ}\end{pmatrix} \nonumber \\
	\begin{pmatrix}
		-5g'^2/12  & 0 & 0 \\
		g'^2 & 0 & 0\\
		0  & -5g'^2/12 & 3g^2/4  \\
		0  &  -4g'^2 & -6g^2\\
	\end{pmatrix} \begin{pmatrix}
		C^{\tau\mu}_{He}\\
		C^{(1)e\tau}_{H\ell }\\
		C^{(3)e\tau}_{H\ell}\\
	\end{pmatrix}
\end{align}
\begin{align}
	16\pi^2 \dot{C}^{*(3) \mu e ii}_{\ell equH^2}=\begin{pmatrix}
		C^{*(1)\mu\tau ii}_{\ell equ} &  C^{*(3)\mu\tau ii}_{\ell equ} & C^{*(1)\tau eii}_{\ell equ} &  C^{*(3)\tau e ii}_{\ell equ}\end{pmatrix} \nonumber \\
	\begin{pmatrix}
		-5g'^2/12  & 0 & 0 \\
		g'^2 & 0 & 0\\
		0  & -5g'^2/12 & 3g^2/4  \\
		0  &  -4g'^2 & -6g^2\\
	\end{pmatrix} \begin{pmatrix}
		C^{e\tau}_{He}\\
		C^{(1)\tau\mu}_{H\ell}\\
		C^{(3)\tau\mu}_{H\ell}\\
	\end{pmatrix}
\end{align}
\begin{align}
	16\pi^2 \dot{C}^{(4)e \mu ii}_{\ell equH^2}=\begin{pmatrix}
		C^{(1)e\tau ii}_{\ell equ} &  C^{(3)e\tau ii}_{\ell equ} & C^{(1)\tau\mu ii}_{\ell equ} &  C^{(3)\tau\mu ii}_{\ell equ}\end{pmatrix} \nonumber \\
	\begin{pmatrix}
		g^2/4  & 0 & 0 \\
		3g^2 & 0 & 0\\
		0  & g^2/4 & -5g'^2/12  \\
		0  &  -2g^2 & -4g'^2\\
	\end{pmatrix} \begin{pmatrix}
		C^{\tau\mu}_{He}\\
		C^{(1)e\tau}_{H\ell}\\
		C^{(3)e\tau}_{H\ell}\\
	\end{pmatrix}
\end{align}
\begin{align}
	16\pi^2 \dot{C}^{*(4) \mu e ii}_{\ell equH^2}=\begin{pmatrix}
		C^{*(1)\mu\tau ii}_{\ell equ} &  C^{*(3)\mu\tau ii}_{\ell equ} & C^{*(1)\tau eii}_{\ell equ} &  C^{*(3)\tau e ii}_{\ell equ}\end{pmatrix} \nonumber \\
	\begin{pmatrix}
		g^2/4  & 0 & 0 \\
		3g^2 & 0 & 0\\
		0  & g^2/4 & -5g'^2/12  \\
		0  &  -2g^2 & -4g'^2\\
	\end{pmatrix} \begin{pmatrix}
		C^{e\tau}_{He}\\
		C^{(1)\tau\mu}_{H\ell}\\
		C^{(3)\tau\mu}_{H\ell}\\
	\end{pmatrix}
\end{align}
\begin{align}
	16\pi^2 \dot{C}^{(1)e \mu ii}_{\ell edqH^2}=\begin{pmatrix}
		C^{e\tau ii}_{\ell edq} &  C^{\tau\mu ii}_{\ell edq} \end{pmatrix} \nonumber \\
	\begin{pmatrix}
		3g'^2  & 0  \\
		0 & 6g'^2 \\
	\end{pmatrix} \begin{pmatrix}
		C^{\tau\mu}_{He}\\
		C^{(1)e\tau}_{H\ell}\\
	\end{pmatrix}
\end{align}
\begin{align}
	16\pi^2 \dot{C}^{*(1)\mu e ii}_{\ell edqH^2}=\begin{pmatrix}
		C^{*\mu\tau ii}_{\ell edq} &  C^{*\tau e ii}_{\ell edq} \end{pmatrix} \nonumber \\
	\begin{pmatrix}
		3g'^2  & 0  \\
		0 & 6g'^2 \\
	\end{pmatrix} \begin{pmatrix}
		C^{e\tau}_{He}\\
		C^{(1)\tau\mu}_{H\ell}\\
	\end{pmatrix}
\end{align}
\begin{align}
	16\pi^2 \dot{C}^{(2)e \mu ii}_{\ell edqH^2}=\begin{pmatrix}
		C^{e\tau ii}_{\ell edq} &  C^{\tau\mu ii}_{\ell edq} \end{pmatrix} \nonumber \\
	\begin{pmatrix}
		-3g^2  & 0  \\
		0 & 6g'^2 \\
	\end{pmatrix} \begin{pmatrix}
		C^{\tau\mu}_{He}\\
		C^{(3)e\tau}_{H\ell}\\
	\end{pmatrix}
\end{align}
\begin{align}
	16\pi^2 \dot{C}^{*(2)\mu e ii}_{\ell edqH^2}=\begin{pmatrix}
		C^{*\mu\tau ii}_{\ell edq} &  C^{*\tau e ii}_{\ell edq} \end{pmatrix} \nonumber \\
	\begin{pmatrix}
		-3g^2  & 0  \\
		0 & 6g'^2 \\
	\end{pmatrix} \begin{pmatrix}
		C^{e\tau}_{He}\\
		C^{(3)\tau\mu}_{H\ell}\\
	\end{pmatrix}
\end{align}
\begin{align}
	16\pi^2 \dot{C}^{(1)e \mu ii}_{\ell^2q^2H^2}=\begin{pmatrix}
		C^{(1)e\tau ii}_{\ell q} &  C^{(3)e\tau ii}_{\ell q} & C^{(1)\tau\mu ii}_{\ell q} &  C^{(3)\tau\mu ii}_{\ell q}\end{pmatrix} \nonumber \\
	\begin{pmatrix}
		g'^2  & 0 & 0 & 0 \\
		0 & 9g^2 & 0 & 0\\
		0 & 0 & g'^2  & 0  \\
		0& 0& 0 & 9g^2 \\
	\end{pmatrix} \begin{pmatrix}
		C^{(1)\tau\mu}_{H\ell }\\
		C^{(3)\tau\mu}_{H\ell}\\
		C^{(1)e\tau}_{H\ell}\\
		C^{(3)e\tau}_{H\ell}\\
	\end{pmatrix}
\end{align}
\begin{align}
	16\pi^2 \dot{C}^{(2)e \mu ii}_{\ell^2q^2H^2}=\begin{pmatrix}
		C^{(1)e\tau ii}_{\ell q} &  C^{(3)e\tau ii}_{\ell q} & C^{(1)\tau\mu ii}_{\ell q} &  C^{(3)\tau\mu ii}_{\ell q}\end{pmatrix} \nonumber \\
	\begin{pmatrix}
		0  & g'^2 & 0 & 0 \\
		3g^2 & 0 & 0 & 0\\
		0 & 0 &  0 & g'^2  \\
		0& 0& 3g^2 & 0 \\
	\end{pmatrix} \begin{pmatrix}
		C^{(1)\tau\mu}_{H\ell }\\
		C^{(3)\tau\mu}_{H\ell}\\
		C^{(1)e\tau}_{H\ell}\\
		C^{(3)e\tau}_{H\ell}\\
	\end{pmatrix}
\end{align}
\begin{align}
	16\pi^2 \dot{C}^{(3)e \mu ii}_{\ell^2q^2H^2}=\begin{pmatrix}
		C^{(1)e\tau ii}_{\ell q} &  C^{(3)e\tau ii}_{\ell q} & C^{(1)\tau\mu ii}_{\ell q} &  C^{(3)\tau\mu ii}_{\ell q}\end{pmatrix} \nonumber \\
	\begin{pmatrix}
		0  & 3g^2 & 0 & 0 \\
		g'^2 & -10g^2 & 0 & 0\\
		0 & 0 &  0 & 3g^2  \\
		0& 0& g'^2 & -10g^2 \\
	\end{pmatrix} \begin{pmatrix}
		C^{(1)\tau\mu}_{H\ell}\\
		C^{(3)\tau\mu}_{H\ell}\\
		C^{(1)e\tau}_{H\ell}\\
		C^{(3)e\tau}_{H\ell}\\
	\end{pmatrix}
\end{align}
\begin{align}
	16\pi^2 \dot{C}^{(4)e \mu ii}_{\ell^2q^2H^2}=\begin{pmatrix}
		C^{(1)e\tau ii}_{\ell q} &  C^{(3)e\tau ii}_{\ell q} & C^{(1)\tau\mu ii}_{\ell q} &  C^{(3)\tau\mu ii}_{\ell q}\end{pmatrix} \nonumber \\
	\begin{pmatrix}
		3g^2  & 0 & 0 & 0 \\
		0 & g'^2 & 0 & 0\\
		0 & 0 & 3g^2  & 0  \\
		0& 0& 0 & g'^2 \\
	\end{pmatrix} \begin{pmatrix}
		C^{(1)\tau\mu}_{H\ell }\\
		C^{(3)\tau\mu}_{H\ell}\\
		C^{(1)e\tau}_{H\ell}\\
		C^{(3)e\tau}_{H\ell}\\
	\end{pmatrix}
\end{align}
\begin{align}
	16\pi^2 \dot{C}^{e \mu ii}_{e^2u^2H^2}=\begin{pmatrix}
		C^{e\tau ii}_{eu} &  C^{\tau\mu ii}_{eu} \end{pmatrix} \nonumber \\
	\begin{pmatrix}
		4g'^2  & 0  \\
		0 & 4g'^2 \\
	\end{pmatrix} \begin{pmatrix}
		C^{\tau\mu}_{He}\\
		C^{e\tau}_{He}\\
	\end{pmatrix}
\end{align}
\begin{align}
	16\pi^2 \dot{C}^{e \mu ii}_{e^2d^2H^2}=\begin{pmatrix}
		C^{e\tau ii}_{ed} &  C^{\tau\mu ii}_{ed} \end{pmatrix} \nonumber \\
	\begin{pmatrix}
		-2g'^2  & 0  \\
		0 & -2g'^2 \\
	\end{pmatrix} \begin{pmatrix}
		C^{\tau\mu}_{He}\\
		C^{e\tau}_{He}\\
	\end{pmatrix}
\end{align}
\begin{align}
	16\pi^2 \dot{C}^{(1)e \mu ii}_{e^2q^2H^2}=\begin{pmatrix}
		C^{e\tau ii}_{eq} &  C^{\tau\mu ii}_{eq} \end{pmatrix} \nonumber \\
	\begin{pmatrix}
		-g'^2  & 0  \\
		0 & -g'^2 \\
	\end{pmatrix} \begin{pmatrix}
		C^{\tau\mu}_{He}\\
		C^{e\tau}_{He}\\
	\end{pmatrix}
\end{align}
\begin{align}
	16\pi^2 \dot{C}^{(2)e \mu ii}_{e^2q^2H^2}=\begin{pmatrix}
		C^{e\tau ii}_{eq} &  C^{\tau\mu ii}_{eq} \end{pmatrix} \nonumber \\
	\begin{pmatrix}
		-3g^2  & 0  \\
		0 & -3g^2 \\
	\end{pmatrix} \begin{pmatrix}
		C^{\tau\mu}_{He}\\
		C^{e\tau}_{He}\\
	\end{pmatrix}
\end{align}
\begin{align}
	16\pi^2 \dot{C}^{(1)e \mu ii}_{\ell^2u^2H^2}=\begin{pmatrix}
		C^{e\tau ii}_{\ell u} &  C^{\tau\mu ii}_{\ell u} \end{pmatrix} \nonumber \\
	\begin{pmatrix}
		-4g'^2  & 0  \\
		0 & -4g'^2 \\
	\end{pmatrix} \begin{pmatrix}
		C^{(1)\tau\mu}_{H\ell }\\
		C^{(1)e\tau}_{H\ell}\\
	\end{pmatrix}
\end{align}
\begin{align}
	16\pi^2 \dot{C}^{(2)e \mu ii}_{\ell^2u^2H^2}=\begin{pmatrix}
		C^{e\tau ii}_{\ell u} &  C^{\tau\mu ii}_{\ell u} \end{pmatrix} \nonumber \\
	\begin{pmatrix}
		-4g'^2  & 0  \\
		0 & -4g'^2 \\
	\end{pmatrix} \begin{pmatrix}
		C^{(3)\tau\mu}_{H\ell}\\
		C^{(3)e\tau}_{H\ell}\\
	\end{pmatrix}
\end{align}
\begin{align}
	16\pi^2 \dot{C}^{(1)e \mu ii}_{\ell^2d^2H^2}=\begin{pmatrix}
		C^{e\tau ii}_{\ell d} &  C^{\tau\mu ii}_{\ell d} \end{pmatrix} \nonumber \\
	\begin{pmatrix}
		2g'^2  & 0  \\
		0 & 2g'^2 \\
	\end{pmatrix} \begin{pmatrix}
		C^{(1)\tau\mu}_{H\ell}\\
		C^{(1)e\tau}_{H\ell}\\
	\end{pmatrix}
\end{align}
\begin{align}
	16\pi^2 \dot{C}^{(2)e \mu ii}_{\ell^2d^2H^2}=\begin{pmatrix}
		C^{e\tau ii}_{\ell d} &  C^{\tau\mu ii}_{\ell d} \end{pmatrix} \nonumber \\
	\begin{pmatrix}
		2g'^2  & 0  \\
		0 & 2g'^2 \\
	\end{pmatrix} \begin{pmatrix}
		C^{(3)\tau\mu}_{H\ell}\\
		C^{(3)e\tau}_{H\ell}\\
	\end{pmatrix}
\end{align}
\subsection{$Y_6\times Y_6\to P_8$}
We here write the RGEs for the mixing of two dimension six $\tau\to l$ Yukawa (eq. (\ref{eq:yukdim6})) into the the dimension eight $\mu\to e$ penguins (eq. (\ref{eq:dim8peng})). More details can be found in section \ref{ssec:SMEFTrunningsub} of the text.
\begin{align}
	16\pi^2 \dot{C}^{e \mu }_{e^2H^4D}=-C^{\tau\mu}_{eH}C^{*\tau e}_{eH}
\end{align}
\begin{align}
	16\pi^2 \dot{C}^{(1)e \mu }_{\ell^2H^4D}=\frac{1}{2}C^{*\mu\tau }_{eH}C^{e\tau}_{eH}\qquad 	16\pi^2 \dot{C}^{(2)e \mu }_{\ell^2H^4D}=\frac{1}{4}C^{*\mu\tau }_{eH}C^{e\tau}_{eH}
\end{align}

\section{Limits from B Decays}\label{appendix:Bdecays}

In the body of the paper, we saw
that  $\me$ processes
have  a good sensitivity to  products of
$\tl$ coefficients which both involve a top quark, via  
the fish diagram of Figure \ref{fig:diagramsRunning} e). 
When the top quark is in  an doublet, these same  $\tl$ coefficients
mediate   B decays, which is discussed in   this section.

We set limits on  the $\tl$ coefficients from their contributions to
leptonic and semi-leptonic $B$ decays.   They can induce ``neutral current'' processes, such as $B_d\to \tau^\pm l^\mp$, which are absent in the SM,   and also contribute to ``charged current''  decays such as
$B^+\to \bar{\tau}\nu$, to which the SM does contribute but with a  different-flavoured neutrino.  Since our coefficients are lepton-flavour-changing, they cannot interfere with the  Standard Model, so neccessarily increase the Branching Ratios with respect to their SM  expectation.  This makes it difficult to  fit the current $B$ anomalies with LFV operators, because  many of the anomalies are experimental deficits with respect to the SM  predictions.

The list of decays that are included is given in table \ref{tab:tetu}, along with the  value of  the Branching Ratio(BR)  which we use to extract limits (A coefficient at its  upper limit  gives this  BR).
For processes where the SM contribution is negligeable, this value  is the  the experimental 95\% C.L. upper bound on the BR.
In the case  of SM processes  where prediction $\approx $  observation,
this value  is  the SM prediction + theory uncertainty + 2$\sigma$ experimental uncertainty.   This definition is used because we would like to  remove the SM  part and require that   the  flavour-changing  interactions contribute less than the remainder.  However, it can occur that the SM  prediction exceeds the experimental observation (as in some  ``B anomalies'').

To extrapolate the limits we obtain from current experimental constraints into  the future, we suppose a factor of 10 improvement in the experimental sensitivity  (and  in the theoretical precision), such that the future limits will be a factor of $\sim$ 3 better.

Our limits are obtained using Flavio\cite{Straub:2018kue}.
The limits  obtained from two-body leptonic decays were checked analytically,
using the well-known formula for  the rate  as a function of operator coefficients at  the
experimental scale $m_b$: 
\bea
\!\!\!\!\Gamma(B_0\to\bar{\tau} \mu)\!\!\!\! &=&\!\!\!\!\frac{E^2_\mu f_B^2 }{16\pi v^4 } {\Big \{}
(|C_{V,LX}^{db\mu \tau}|^2+ |C_{V,RX}^{db\mu \tau}|^2) (E_\tau\!\! -\!\! E_\mu)
+ (|C_{S,RX}^{db\mu \tau}|^2+ |C_{S,LX}^{db\mu \tau}|^2)
\frac{ m_B^2  }{m_b^2}(E_\tau\!\! +\!\! E_\mu)
+ ... {\Big \} } 
\eea
where ``...'' are cross-terms and $m_\mu$ is neglected.
A numerical limit can be obtained by, for instance, comparing to the experimental  rate for $B^+\to\bar{\tau} \nu$.

The coefficients are run from  $m_b \to \LNP=4$ TeV with the one-loop RGEs of QCD (which shrinks scalar coefficients by a factor $\sim 3/5$),  with tree-level matching to SMEFT  operators when passing $m_W$.  Electroweak running is neglected, except in the case of tensor to scalar mixing in
SMEFT\footnote{The tensor to scalar mixing  below $m_W$ in QED  is negligeable  for ``charged-current'' tensors involving a $b$ and a $\nu$.} (where
$C_S(m_W) \sim 0.3 C_T(\LNP)$), which  for instance,  mixes
single-top tensors ${\cal O}_{\ell e q u}^{(3)\tau e 3 u}$  into  scalars that induce
$B^+\to \bar{e}\nu$.

\begin{table}[!h]
	\begin{center}
		\begin{tabular}{|l|l|l|l|}
			\hline coefficient & limit & process & BR\\
			\hline
			$C^{e\tau 32}_{eq}$,$C^{(1)e\tau 32}_{\ell q }+ C^{(3)e\tau 32}_{\ell q}$ &$2.3 \times 10^{-3}$(c) &$B^+ \to K+ \tau^\pm e^\mp $ & $<4.4\times 10^{-5} $ \cite{BaBar:2012azg}\\
			$C^{e\tau 31}_{eq}$,$C^{(1)e\tau 31}_{\ell q}+ C^{(3)e\tau 31}_{\ell q}$ &$2.3 \times 10^{-3}$(c) &$B_d^0 \to \tau^\pm e^\mp$ & $<3.0\times 10^{-5} $\cite{BaBar:2008pet}\\
			$C^{\mu\tau 32}_{eq}$,$C^{(1)\mu\tau 32}_{\ell q}+ C^{(3)\mu\tau 32}_{\ell q}$ &$2.3 \times 10^{-3}$(c) &$B_s^0 \to \tau^\pm \mu^\mp$& $<4.3\times 10^{-5} $ \cite{LHCb:2019ujz}\\
			$C^{\mu\tau 31}_{eq}$,$C^{(1)\mu\tau 31}_{\ell q} + C^{(3)\mu\tau 31}_{\ell q}$ &$ 1.5\times 10^{-3}$(c) &$B_d^0 \to \tau^\pm \mu^\mp$ & $<1.2\times 10^{-5} $ \cite{LHCb:2019ujz}\\
			\hline
			$C^{e\tau d3}_{\ell edq}, C^{\tau e d3}_{
				\ell edq}$&$ 3.4\times 10^{-4}$(c)  &$B_d^0 \to e^\pm \tau^\mp$ & $<3.0\times 10^{-5} $ \cite{BaBar:2008pet} \\
			$C^{\mu\tau d3}_{\ell edq}, C^{\tau \mu d3}_{\ell edq}$&$ 2.2\times 10^{-4}$(c)  &$B_d^0 \to \mu^\pm \tau^\mp$ & $<1.2\times 10^{-5} $ \cite{LHCb:2019ujz}\\
			$C^{\mu\tau s3}_{\ell edq}, C^{\tau \mu s3}_{\ell edq}$&$ 3.3\times 10^{-4}$(c)  &$B_s^0 \to \mu^\pm \tau^\mp$ & $<4.3\times 10^{-5} $ \cite{LHCb:2019ujz}\\
			\hline
			$C^{(1)l\tau 3u}_{\ell equ}$ &$ 4.5 \times 10^{-4}$(c)& $B^-\to \tau  \bar{\nu}$ & $1.4\times 10^{-4} $\cite{Zyla:2020zbs,Straub:2018kue}\\
			$C^{(1)\tau  e 3u}_{\ell equ}$ &$5.8 \times 10^{-5}$(c)& $B^-\to e \bar{\nu}$ &  $\leq 1.2\times 10^{-6} $ \cite{Belle:2006tbq}  \\
			$C^{(1)\tau  \mu 3u}_{\ell equ}$ &$4.3 \times 10^{-5}$(c)& $B^-\to \mu  \bar{\nu}$ & $\leq 1.0\times 10^{-6} $ \cite{Belle:2019iji} \\
			%
			$C^{(1)l\tau 3c}_{\ell equ}$ &$ 1.0 \times 10^{-2}$(c)& $B^-_c\to \tau  \bar{\nu}$ & $0.1 $\cite{Straub:2018kue}\\
			$C^{(1)\tau  e 3c}_{\ell equ}$ &$9.0 \times 10^{-3}$(c)& $B_d^0\to D e \bar{\nu}$ &  $\leq 3.0\times 10^{-2} $ \cite{Straub:2018kue}  \\
			$C^{(1)\tau  \mu 3c}_{\ell equ}$ &$9.0 \times 10^{-3}$(c)& $B_d^0\to  D \mu  \bar{\nu}$ & $\leq 3.1\times 10^{-2} $ \cite{Straub:2018kue} \\
			$C^{(3)l\tau 3u}_{\ell equ}$  &$1.8 \times 10^{-3}$(c)& $B^-\to \tau \bar{\nu}$ & $1.4\times 10^{-4} $ \cite{Zyla:2020zbs,Straub:2018kue}\\
			$C^{(3)\tau e 3u}_{\ell equ}$  &$2.4 \times 10^{-4}$(c)& $B^-\to e \bar{\nu}$&  $\leq 1.2\times 10^{-6} $ \cite{Belle:2006tbq}\\
			$C^{(3)\tau \mu 3u}_{\ell equ}$  &$1.8 \times 10^{-4}$(c)& $B^-\to \mu \bar{\nu}$&
			$\leq 1.0\times 10^{-6} $ \cite{Belle:2019iji} \\
			%
			$C^{(3)l\tau 3c}_{\ell equ}$  &$5.0 \times 10^{-3}$(c)& $R_{\tau/l}(B\to D^* l \bar{\nu})$ & $0.28 $ \cite{Straub:2018kue}\\
			$C^{(3)\tau e 3c}_{\ell equ}$  &$5.3 \times 10^{-3}$(c)& $B_d^0\to D^* e \bar{\nu}$&  $\leq 7.3\times 10^{-2} $  \cite{Straub:2018kue}\\
			$C^{(3)\tau \mu 3c}_{\ell equ}$  &$6.4 \times 10^{-3}$(c)& $B_d^0\to D^*\mu \bar{\nu}$&
			$\leq 7.7\times 10^{-2} $  \cite{Straub:2018kue} \\
			\hline			\hline
		\end{tabular}
		\caption{ Current limits (c)  on
			$\te$ and $\tm$  coefficients of   SMEFT operators, at 4 TeV, arising from the
			$B$ decays given in the third column.  The limits saturate the Branching Ratio given in the last column (which may not be the cited experimental limit, see discussion in Appendix \ref{appendix:Bdecays}).  Limits on vector coefficients apply  for permuted lepton and quark flavour indices, scalars apply as given.
			\label{tab:tetu}}
	\end{center}
\end{table}
\clearpage

\section{Table of Sensitivities}\label{appendix:Sensitivities}
\begin{table}[!h]
	\begin{center}
		\begin{tabular}{|l|c|c|}
			\hline
			coefficients &     $B^{(f)}_{\te} B^{(f)}_{\tm}$  & $B^{(f)}_{\mue}$  \\
			\hline
			$ C^{(1)e \tau 1u}_{\ell equ } C^{\tau \mu}_{He} $ &  $8.3 \times 10^{-5}(f) \times 1.2 \times 10^{-4} (f)  $
			&$5 \times 10^{-9}  $ \\
			$ \bm{C^{(3)e \tau 1u}_{\ell equ } C^{\tau \mu}_{He}} $ &  $7.7 \times 10^{-5}(f) \times 1.2 \times 10^{-4}(f)  $
			&$2 \times 10^{-9}  $ \\
			$ (C^{(1) \tau e  1 u}_{\ell equ})^* C^{ \tau\mu }_{H\ell(1) }$ & $8.3 \times 10^{-5}(f) \times1.0 \times 10^{-4}(f)  $
			&$1 \times 10^{-8}  $ \\
			$ \bm{(C^{ (3)\tau e  1 u}_{\ell equ})^* C^{ \tau\mu }_{H\ell (1)}}$ & $7.7 \times 10^{-5}(f) \times1.0 \times 10^{-4}(f)  $
			&$2 \times 10^{-9}  $ \\
			$ (C^{ (1)\tau e  1 u}_{\ell equ})^* C^{ \tau\mu }_{H\ell (3) }$ & $8.3 \times 10^{-5}(f) \times1.0 \times 10^{-4}(f)  $
			&$1 \times 10^{-8}  $ \\
			$ \bm{(C^{(3) \tau e  1 u}_{\ell equ})^* C^{ \tau\mu }_{H\ell (3) }}$ & $7.7 \times 10^{-5}(f) \times1.0 \times 10^{-4}(f)  $
			&$3\times 10^{-10}  $ \\
			\hline
			$ C^{e \tau d1}_{\ell edq } C^{\tau \mu}_{He} $ &  $8.3 \times 10^{-5}(f) \times 1.2 \times 10^{-4}(f)  $
			&$5 \times 10^{-9}  $ \\
			$ (C^{ \tau e d 1 }_{\ell edq})^* C^{ \tau\mu }_{H\ell(1) }$ & $8.3 \times 10^{-5}(f) \times1.0 \times 10^{-4}(f)  $
			&$1 \times 10^{-8}  $ \\
			$ (C^{ \tau e  d1 }_{\ell edq})^* C^{ \tau\mu }_{H\ell (3) }$ & $8.3 \times 10^{-5}(f) \times1.0 \times 10^{-4}(f)  $
			&$1 \times 10^{-8}  $ \\
			\hline
		\end{tabular}
		\caption{  
			Pair of $\tl$ penguin and four fermion dimension six operators  that generate  $\mue$  scalar/tensor  dimension eight operators with a  singlet $u$ and $d$ quark. The future (f) ``limits" $B^{(f)}_{\tl}$ on $\tl$ vectors and scalars are from the upper bounds on the LFV decays $\tau\to l\rho (\eta)$ and $\tau\to \pi l$ respectively (adapted from \cite{Davidson:2020hkf}). The limits on penguins follow from their contribution to four-lepton vector interactions $\tau\to 3l$.  The same bound applies to the dimension six operators with $\mu\leftrightarrow e$ interchanged. The sensitivities $B^{(f)}_{\mue}$ arise from future $\mu\to e$ conversion. Bolded pairs indicate that the sensitivity of $\mu\to e$ is better than the one arising from direct $\tl$ searches (see eq.~(\ref{eq:hyperbolevsellipse})).
			\label{tab:peng4fscalar}} 
	\end{center}
\end{table}

\begin{table}[!h]
	\begin{center}
		\begin{tabular}{|l|c|c|}
			\hline
			coefficients &     $B^{(f)}_{\te} B^{(f)}_{\tm}$  & $B^{(f)}_{\mue}$  \\
			\hline
			$C^{e \tau uu}_{eu } C^{\tau \mu}_{He} $ &  $2.4 \times 10^{-4}(f) \times 1.1 \times 10^{-4}(f)  $
			&$4.6 \times 10^{-8}  $ \\
			$C^{e \tau dd}_{ed } C^{\tau \mu}_{He}$ &  $2.4 \times 10^{-4}(f) \times 1.1 \times 10^{-4}(f)  $
			&$8.2 \times 10^{-8}  $ \\
			$ C^{(1)e \tau 11}_{\ell q } C^{\tau \mu}_{H\ell(1)} $ &  $7.0 \times 10^{-4}(f) \times 1. \times 10^{-4}(f)  $ &$1 \times 10^{-7}  $ \\
			$ C^{(1)e \tau 11}_{\ell q } C^{\tau \mu}_{H\ell(3)} $ &  $7.0 \times 10^{-4}(f) \times 1. \times 10^{-4}(f)  $ &$8.5 \times 10^{-8}  $ \\
			$ C^{(3)e \tau 11}_{\ell q } C^{\tau \mu}_{H\ell(1)} $ &  $1.2 \times 10^{-4}(f) \times 1. \times 10^{-4}(f)  $ &$1 \times 10^{-8}  $ \\
			$ \bm{C^{(3)e \tau 11}_{\ell q } C^{\tau \mu}_{H\ell(3)}} $ &  $1.2 \times 10^{-4}(f) \times 1. \times 10^{-4}(f)  $ &$3.2 \times 10^{-9}  $ \\
			\hline
		\end{tabular}
		\caption{  
			Similar to Table \ref{tab:peng4fscalar} but with product of  penguin and four-fermion  dimension six operators that mix into $\mu\to e$ vectors at dimension eight.
			\label{tab:peng4fvector}} 
	\end{center}
\end{table}

\nocite{Davidson:2018zuo}

\pagebreak

\bibliography{references.bib}{}

\bibliographystyle{unsrt}

\end{document}